\documentclass[twocolumn]{aastex631}

\usepackage{graphicx}
\usepackage{amsmath}
\usepackage{amssymb}
\usepackage{bm}
\usepackage{verbatim}
\usepackage{xcolor}
\usepackage{mathtools}
\usepackage{amsfonts}
\usepackage{graphics}
\usepackage{listings}
\usepackage{tabularx}
\usepackage{soul}
\usepackage{float}
\usepackage{placeins}
\graphicspath{{apj_template/images_final}}

\usepackage[T1]{fontenc}

\newcommand{\mphi}{\boldsymbol{\Phi}}
\newcommand{\rmax}{r_{\rm max}}
\newcommand{\phibh}{\Phi_{\rm BH}}
\newcommand{\Bpol}{B_{\rm pol}}
\newcommand{\EQ}{\begin{equation}}
\newcommand{\EN}{\end{equation}}
\newcommand{\EQA}{\begin{eqnarray}}
\newcommand{\ENA}{\end{eqnarray}}

\newcommand{\Eq}[1]{Eq.~(\ref{#1})}
\newcommand{\Eqs}[2]{Eqs.~(\ref{#1}) and~(\ref{#2})}
\newcommand{\Fig}[1]{Fig.~\ref{#1}}

\begin{document}

\title{Dynamo and Jet interconnections in GRMHD simulations of black hole accretion disks}

\author{P. S. Santhiya}
\affiliation{International Centre for Theoretical Sciences, 
Bangalore, 560089, India}
\email{santhiya.ps@icts.res.in}

\author{Pallavi Bhat}
\affiliation{International Centre for Theoretical Sciences, Bangalore, 560089, India}
\email{pallavi.bhat@icts.res.in}

\author{Prayush Kumar}
\affiliation{International Centre for Theoretical Sciences, 
Bangalore, 560089, India}

\author{Tushar Mondal}
\affiliation{International Centre for Theoretical Sciences, Bangalore, 560089, India}

\author{Indu K. Dihingia}
\affiliation{Tsung-Dao Lee Institute, Shanghai Jiao Tong University, 1 Lisuo Road, Shanghai, 201210, People’s Republic of China}

\begin{abstract}
We present global 3D GRMHD simulations of black hole (BH) accretion disks designed to investigate how MRI-driven dynamo action regulates jet formation and evolution. Unlike standard SANE/MAD setups that impose a coherent large-scale poloidal loop, our ``sub-SANE'' initial conditions use multiple same-polarity small-scale magnetic loops. Rapid reconnection erases magnetic memory and enables large-scale dynamo to emerge early from MRI turbulence. We perform two such sub-SANE simulations at different BH spins ($a = 0.5, 0.9375$) and compare them with conventional SANE runs.
 
The sub-SANE disks show regular large-scale dynamo cycles with periods of about ten orbits. Decomposition of the induction equation shows that the turbulent dynamo term is stronger in 3D compared to 2.5D and balances advection in the saturated state, confirming sustained large-scale field generation. These dynamo-generated fields are advected inward with minimal time lag, producing correlated peaks in both poloidal and toroidal field strengths from $r_{\rm max}$ to the horizon. Early in the evolution, these peaks imprint directly onto the jet’s electromagnetic energy flux, indicating that the jet mirrors the dynamo wave.

Though jets form at early times, the sub-SANE runs eventually undergo jet shutdown. We show that this occurs when the magnetic field at the horizon loses coherence, as quantified by a decline in the signed-to-unsigned flux ratio $\mathcal{C}_{\rm BH}$ below $\approx 0.6$. In contrast, the SANE reference case with similar accretion rate and horizon magnetic flux maintains high magnetic coherence because its initial large-scale field persists, allowing its jet to survive. Our results show that both dynamo-driven field evolution and horizon magnetic-field coherence critically regulate jet longevity, establishing a direct dynamo–jet connection in GRMHD disks.
\end{abstract}


\section{Introduction}
Magnetic fields have a central role to play in accretion disks in high energy astrophysical systems like active galactic nuclei (AGNs), X-ray binaries (XRBs) and gamma ray bursts (GRBs). In particular, recent polarimetric observations from Event Horizon Telescope have indicated the presence of coherent large-scale fields in Sgr~A$^*$ and M\,87$^*$ \citep{collaboration_first_2021, collaboration_first_2024, collaboration_first_2024-1}. 
In general, direct observations of magnetic fields in accretion disks are limited but they can be indirectly inferred for different components of these systems. In X-ray binaries, spectral modelling and X-ray emission calculations infer field strengths of a few Gauss \citep{bosch-ramon_magnetic_2008} likely in the disk, while using QPO analysis and jet modelling, inferred fields can be as strong as   $10^5$ G in the corona or at the base of the jet \citep{del_santo_magnetic_2013, Giannis2005, huang2025revisitinginfraredxraycorrelationgx}. The estimates for the coronal magnetic field can also be on the lower side of tens of Gauss \citep{dallilar_precise_2017}.  

Most of these systems also show evidence for the presence of jets. Jets are characterized by high collimation and relativistic plasma outflows. Very Long Baseline Interferometry (VLBI) of AGNs and quasars reveals relativistic jets extending from parsec to kiloparsec scales, with superluminal motion and polarized synchrotron emission indicating ordered magnetic fields \citep{Lister2013_MOJAVEX, Wardle2013_MagneticFieldsPolarizationAGN, BoccardiKrichbaumRosZensus2017, ParkAlgaba2022}.
Faraday rotation gradients and multiwavelength variability constrain the jet launching region and field geometry near the BH \citep{Zavala2005_3C273_RMgradient, Hovatta2011_3C273_MOJAVE, Hovatta2019_3C273_mmRM}.
Radio and infrared observations of X-ray binaries detect transient, relativistic jets with Lorentz factors $\sim 2–10$, often coupled to accretion state changes where transitions between hard and soft states regulate the appearance, disappearance, and power of relativistic jets \citep{Mirabel1999_GalacticJetsReview, Dhawan2000_GRS1915_Jets, Fender2004_JetsXRB, 2009MNRAS.396.1370F,  Mirabel2007_Microquasars, Nemmen2012, Ghisellini2014, Saikia2019_LorentzXRB, 2020NatAs...4..697B}.
Observations from protostars to AGNs consistently reveal strong magnetic fields to be associated with jet activity, pointing towards the crucial role of magnetic processes originating within the accretion disk itself \citep{pudritz_magnetic_2012, MotterGabuzda2017_FaradayAGNJets, Lee2018_MagnetizedProtostellarJet, Ray2021_JetsYoungStars}. 

Jet formation in astrophysical systems around compact objects, particularly BHs, remains a major enigma in astrophysics. Some of the leading theoretical models: Blandford–Znajek (BZ) and Blandford–Payne mechanisms, 
propose that these jets are powered by tapping the rotational energy of the BH or disk via large-scale magnetic fields \citep{BlandfordZnajek1977, BlandfordPayne1982, Livio1999_JetsReview}.
General Relativistic Magnetohydrodynamic (GRMHD) simulations have been indispensable in shedding light on the physical mechanisms governing jet formation and evolution, bridging the gap between theoretical models and observational data \citep{tchekhovskoy_general_2012, McKinney2012_MCAFs, Davis2020_MHDAGN}. These studies have confirmed the dependence of jet efficiency on BH spin, broadly consistent with the BZ mechanism, wherein rotational energy is extracted from the BH ergosphere via magnetic fields threading the event horizon \citep{narayan_jets_2022}. 

In particular, these simulations have clarified the critical role of magnetic fields, demonstrating that jet launching often relies on magnetic flux accumulation near the BH horizon   
\citep{mizuno_grmhd_2022, Janiuk2022_MADJets}. 
In these simulations, the accreting plasma drags in magnetic field lines, which are then twisted by the rotating BH, driving collimated relativistic jets carrying energy and angular momentum outward. Depending on the amount of magnetic flux accumulated near the BH, two distinct accretion regimes are realized, known as Magnetically Arrested Disk (MAD) and Standard And Normal Evolution (SANE) \citep{Narayan2003_MAD, Narayan2012, 2019Univ....5..146B}.  
In SANE regime, the horizon carries only modest magnetic flux; consequently, any jets are weaker whereas in the MAD regime, the horizon magnetic fluxes are much larger leading to powerful jets. 
GRMHD simulations have systematically probed these accretion regimes such as SANE and MAD configurations revealing that the strength and structure of magnetic fields critically determine jet collimation and energetics \citep{Dhruv2024_GRMHDsurvey, PathakMukhopadhyay2025_ULXBlazar}. By incorporating radiative transfer calculations, these simulations now produce synthetic observational signatures that can be directly compared with data from instruments such as the Event Horizon Telescope, thus offering robust observational constraints on jet-launching theories \citep{collaboration_first_2021, collaboration_first_2024, Roder_2025}. 

Recent simulation efforts have further explored how accretion disk properties influence jet launching and variability, highlighting the  relationship between disk turbulence, magnetic flux dynamics, and resulting jet characteristics \citep{White_2020,Chashkina2021,Dihingia2023_TwoTemperatureGRMHD,Galishnikova_low_angumomentum}.
Simulations have also elucidated how magnetorotational instability (MRI)-driven turbulence and magnetic dynamos in accretion disks regulate magnetic flux transport, directly impacting jet structure, power, and variability \citep{hogg_influence_2018, Dhang_2018, Dhang_2023}. 
Others have explored the jets due to mean-field dynamos implemented directly in GRMHD \citep{DelZanna2007ECHO, fendt_bipolar_2018, tomei_are_2021, vourellis_relativistic_2021, MattiaFendt2020_I, MattiaFendt2020_II, MattiaFendt2022_Quenching}.
These developments underscore the importance of magnetic field evolution in shaping jet dynamics.  
This naturally motivates the study of magnetic dynamos and their role in maintaining the field structures necessary for powerful jet formation. 

GRMHD simulations typically start from an assumed magnetic field structure (e.g. a poloidal loop threading the disk) to seed MRI and allow jet launching. The chosen configuration (strength and topology) often forces evolution into either SANE or MAD states. While useful for exploring jet physics, this insertion of large-scale initial magnetic field in the disk by hand may not reflect realistic astrophysical conditions, where infalling plasma might carry only weak or tangled fields, not large-scale vertical ones.
In this respect, \cite{jacquemin-ide_magnetorotational_2024} report a quantitative identification of a magnetorotational dynamo that can produce large-scale vertical magnetic fields in global 3D GRMHD simulations, starting with initial fields that were fully toroidal. They find nonlinear interactions lead to poloidal field loops in the disk, which then are advected inward. This advective transport is key to building up a large-scale vertical field near the BH, as loops merge and accumulate. These new results are an important step toward understanding the interconnections between dynamos and jets: they indicate that a turbulent accretion disk can self-generate the magnetic fields required to launch relativistic jets. 

In the existing literature, relatively little attention has generally been given to how the jet itself is dynamically influenced by the dynamo operating within the accretion disk. 
Current global accretion disk simulations indicate that both the strength and topology of the initial magnetic field strongly shape disk/jet evolution. Low plasma beta $\beta$ (which is the ratio of plasma thermal pressure to magnetic pressure) suppresses MRI dynamo activity, and spatial variations of $\beta$ lead to uneven dynamo efficacy. Even for weaker initial field strengths (e.g., commonly used $\beta \simeq 100$), a coherent topology that resists reconnection can preserve magnetic memory for long. 
In this work, we address this gap by studying a disk initialized with a magnetic configuration that leads to an evolution far less coherent than that in the standard setups typically used to produce SANE or MAD states. 
With our specific initial condition, the system rapidly loses memory of its starting configuration due to magnetic reconnections, allowing dynamo action to arise  speedily. 

While jets are launched early in the evolution, they eventually shut down. Our analysis investigates the cause of this shutdown, revealing that the coherence of the magnetic field threading the BH horizon plays a critical role—an aspect often overlooked in previous studies. Furthermore, we find intriguing correlations between the dynamo-generated fields in the disk and the magnetic flux at the horizon, particularly in their cyclic variability.

The remainder of this paper is organized as follows. In Section 2, we describe the numerical setup and the diagnostics employed in our analysis. Section 3 presents the results, which are structured into three subsections focusing on jets, dynamos, and their interconnections. While we do not attempt a detailed exploration of the dynamo mechanism itself, our emphasis is on uncovering the links between disk dynamo activity and jet evolution.

\section{Numerical set-up}

We use the Black Hole Accretion Code (\texttt{BHAC}) for our GRMHD simulations \citep{Porth_2017,Olivares_AMR}. The code employs a finite-volume method to solve for the standard ideal GRMHD equations as mentioned in \citet{Porth_2017}. Unless otherwise stated, we use geometrized units with  $G=c=1$.

\subsection{Domain and the disk set-up}
We use the standard Fishbone--Moncrief torus, which is in hydrostatic equilibrium around a rotating BH of mass $M$ and spin $a$ \citep{Fishbone1976RelativisticFD, Uniyal:2024sdv}. The computational domain is expressed in horizon-penetrating Kerr--Schild coordinates $(r_{\mathrm{KS}}, \theta_{\mathrm{KS}}, \phi)$, spans $r_{\mathrm{KS}} \in [1.21M,\,800M]$, $\theta_{\mathrm{KS}} \in [0,\pi]$, and $\phi \in [0,2\pi)$. For simplicity, we will address $(r_{\mathrm{KS}}, \theta_{\mathrm{KS}}, \phi)$ as $(r,\theta,\phi)$ from now on. We use a grid that provides enhanced resolution near the BH and around the equatorial plane, employing a coordinate transformation from Kerr--Schild $(r_{\mathrm{KS}}, \theta_{\mathrm{KS}})$ to the modified Kerr--Schild coordinates  $(s, \vartheta)$. The original Kerr-Schild coordinates are related to the modified Kerr-Scild coordinates as follows:
\begin{subequations}
\begin{align}
    r_{\mathrm{KS}} &= R_{0} + \exp(s),\\
    \theta_{\mathrm{KS}}(\vartheta) &= \vartheta + \frac{2h\vartheta}{\pi^{2}}\,(\pi - 2\vartheta)(\pi - \vartheta),
\end{align}
\end{subequations}
where $R_{0}$ sets the inner radial offset and $h$ is a concentration parameter controlling the degree of grid compression toward the equatorial plane \citep{McKinney_2004}.
$R_0$ is set to be zero and $h$ is set to be 0.5 in both the 2D and 3D simulations. This value of $h$ sets the ratio between $\Delta\theta_{\rm{polar}}$ and $\Delta\theta_{\rm{equator}}$ to be $\sim 4$. We apply outflow boundary conditions at the radial edges of the domain, ensuring that no scalar or vector fields enter through these boundaries. At the $\theta$ boundaries, a small polar cone is excised and reflective conditions are imposed. The domain is periodic over $\phi$-direction. We set floor values for the density, $\rho_{\rm{floor}} = 10^{-5}r^{-3/2}$ and the pressure, $p_{\rm{floor}} = 1/3 \times 10^{-7} r^{-5/2}$.
The torus has an inner radius $r_{\rm{in}}$ at $6M$ and a density maximum $r_{\rm{max}}$ at $13.8M$. This sets the scale height of the disk which is defined as follows:

\begin{equation}
    \frac{H}{R} = 
   \left[
    \frac{
      \displaystyle \oint \rho \sqrt{-g}\, \left|\tfrac{\pi}{2} - \theta \right|^2 \, \mathrm{d}\theta\,\mathrm{d}\phi
   }{
       \displaystyle \oint \rho \sqrt{-g}\, \mathrm{d}\theta\,\mathrm{d}\phi
    }
    \right]^{1/2},
   \label{eq:H_over_R}
\end{equation}
where $g$ is the determinant of the Kerr spacetime metric~\citep{PhysRevLett.11.237}, and $\rho$ is the density field of matter around the rotating BH. This definition of disk scale height is similar to that of \citet{Porth_2019}, except for the time-averaging.
At time $t=0 M$, $H/R \sim 0.3$ around the density maximum.

\begin{figure}[!htbp]
    \includegraphics[width=\linewidth]{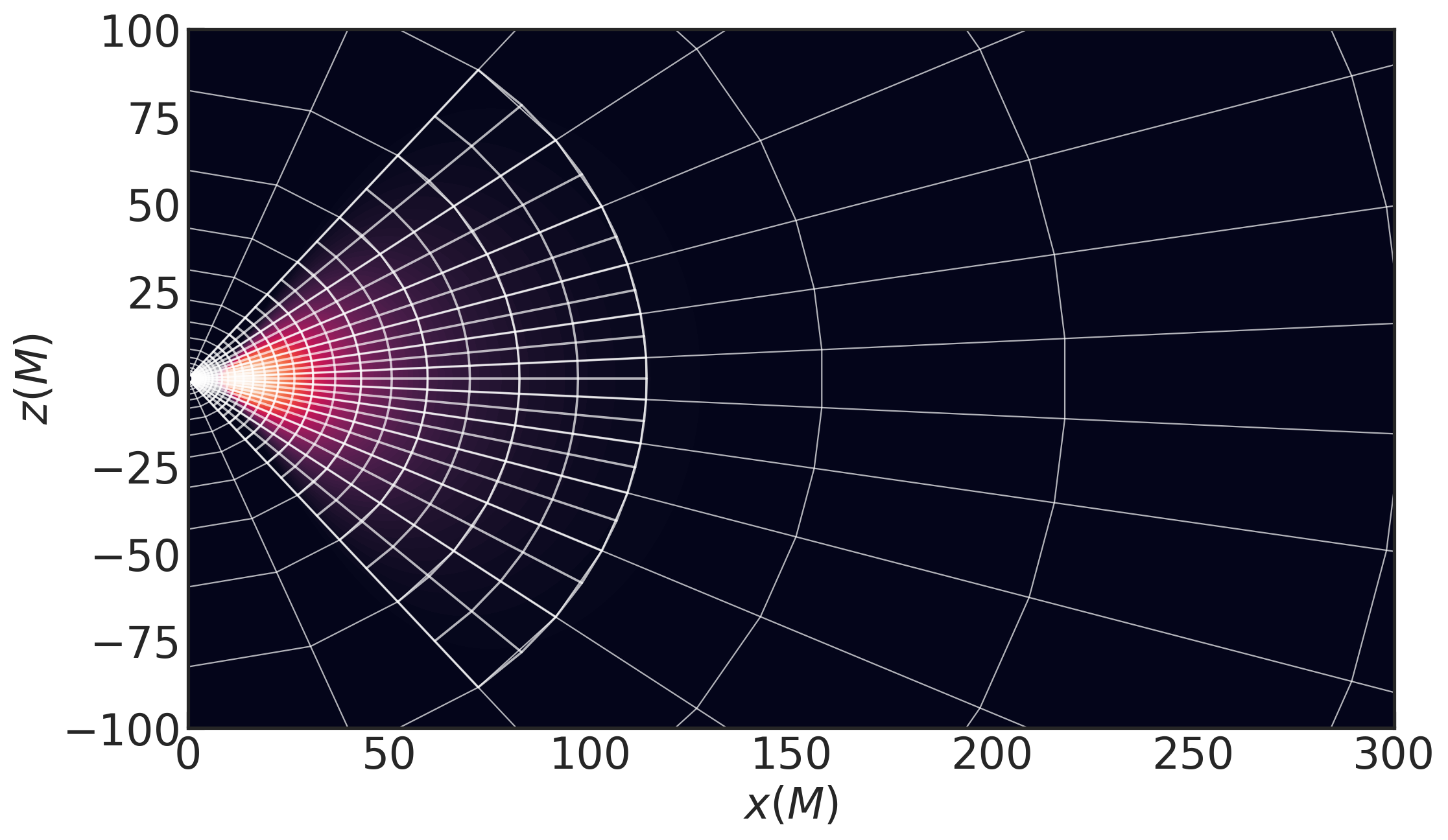}
        \caption{Block structure for the 3D simulations in the $\phi=0$ plane ($z$-axis contains the spin-axis of the BH). Static mesh refinement of level two is applied in all directions in the disk region, identified by the density contour in the background. The mesh is axisymmetric. However, there is no symmetry assumption involved in the evolution of 3D simulations.   
        }
    \label{fig:init_grid}
\end{figure}

\begin{figure}[!htbp]
    \centering
    \includegraphics[width=\linewidth]{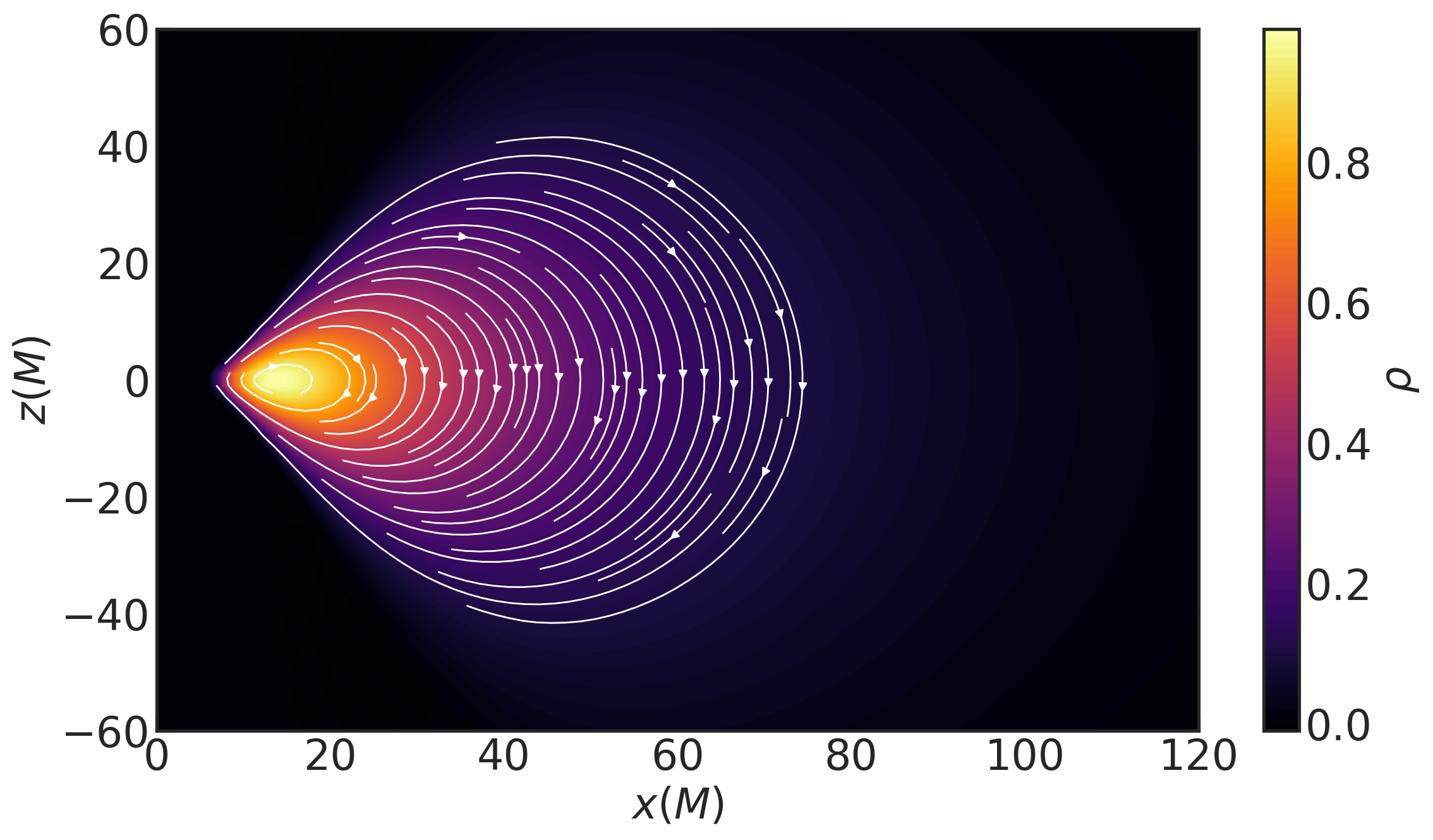}
    \caption{Initial magnetic field configuration in the $\phi=0$ plane (shown in solid while lines) for both the 2D and 3D SANE simulations set through the vector potential (see \Eq{saneic}), with the colormap showing density, $\rho$.}
    \label{fig:sane_init_condition}
\end{figure}

\begin{figure}[!htbp]
    \centering
    \includegraphics[width=\linewidth]{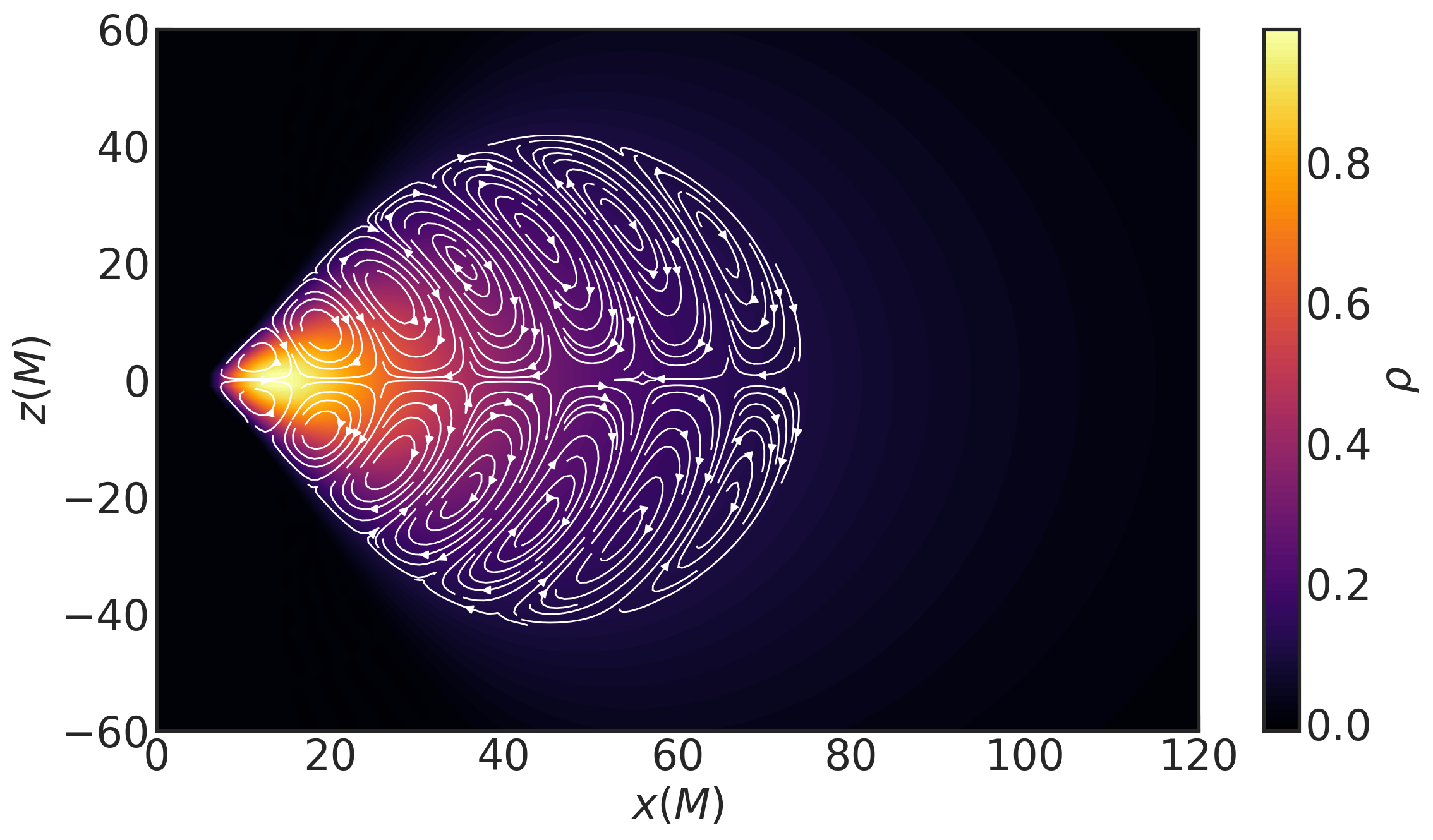}
    \caption{As \Fig{fig:sane_init_condition}  but for 3D sub-SANE simulations (see \Eq{subsaneic}).}
   \label{fig:ss_init_condition}
\end{figure}

\subsection{Description of simulations}
\label{simpars}

In the literature on GRMHD simulations, the categorization of the thick disk simulations is done based on a quantity often referred to as the MAD parameter $\mphi$ \citep{narayan_jets_2022}. This parameter quantifies the strength of magnetic field at the horizon and is given by, 
\begin{equation}\label{mad_BH}
\mphi = \frac{\sqrt{4\pi }\Phi_{\rm BH}}{\sqrt{\dot{M}_\mathrm{BH}}}, 
\end{equation}
where $\Phi_\mathrm{BH}$ is the magnetic flux at BH horizon,
\begin{equation}\label{phi_BH}
    \Phi_\mathrm{BH} = \frac{1}{2}\oint_{r=r_{\mathrm{BH}}} \sqrt{-g}|{B}^r|\text{d}\theta\text{d}\phi,
\end{equation}
and $\dot{M}_\mathrm{BH}$ is the mass accretion rate across the horizon, given by
\begin{equation}\label{acc_rate}
    \dot{M}_\mathrm{BH} = -\oint_{r=r_{\mathrm{BH}}} \sqrt{-g}\rho u^r\text{d}\theta\text{d}\phi,
\end{equation}
were $u^r$ is the radial component of the four velocity of the fluid. 
When $\mphi$ is larger than $\sim 40$, we find a MAD state. Simulations that achieve $\mphi \sim 5-15$ are classified as the SANE cases \citep{Tchekhovskoy2011_MADJets, Narayan2012, Sadowski2013, Sadowski2014}. 
This parameter $\mphi$ can be heavily influenced by the initial condition employed given that the field strength achieved at the horizon largely results from advection of the fields from the disk. We propose the term sub-SANE for disk evolution runs with $\boldsymbol{\Phi} \sim 1-3$ initialized with small-scale, multi-loop (of same polarity) magnetic fields. In our work, we only study sub-SANE and SANE runs and we specify below the initial conditions that lead to these runs. 

The magnetic field configuration for the SANE setup is initialized through the following vector potential \citep{Porth_2019},
\begin{equation}
    {A}_{\phi,\,\text{SANE}}= \text{max}\left({\rho}/{\rho_{max}}\ -\ 0.1, 0\right),
    \label{saneic}
\end{equation}
as shown in \Fig{fig:sane_init_condition}.
For the sub-SANE set up, the vector potential is derived from
\begin{equation}
    {A}_{\phi,\,\mathrm{sub\text{-}SANE}}
    =  \left|\, 
        {A}_{\phi,\,\mathrm{SANE}}
        \,\cos\!\big[(N_{l} - 1)\theta\big] 
        \sin\!\big(R_{l}\theta\big)
      \,\right|,
    \label{subsaneic}
\end{equation}
where 
$R_{l} = 2\pi\,\dfrac{r - r_{\mathrm{in}}}{\lambda_{r}}$, 
$N_{l} = 2$, and 
$\lambda_{r} = 20$ 
represent the radial wavenumber, the number of magnetic field loops across the mid-plane, 
and the radial wavelength of each loop, respectively (visually illustrated in \Fig{fig:ss_init_condition}). The plasma beta is defined as $\beta = 2p_\mathrm{max}/b^2_\mathrm{max}$ where $p_\mathrm{max}$ is the maximum pressure in the disk, and $b^2_\mathrm{max}/2$ is the maximum of fluid-frame magnetic energy in the disk and these locations may not necessarily coincide.  
We set $\beta$ to 100 at $t=0 M$. The adiabatic index is $5/3$. 

Our main dynamo simulations are 3D sub-SANE (3DSS) runs with two different values of the dimensionless spin on the central BH, $a=\{0.5,0.9375\}$, while rest of the system's parameters remain identical. We denote these runs as 3DSS-5 and 3DSS-9375.
The base resolution for these simulations is $320 \times 120 \times 64$ in $r$, $\theta$ and $\phi$ dimensions. 
We employ static mesh refinement in the disk region (see \Fig{fig:init_grid}) such that the effective resolution for the simulations is $640 \times 240 \times 128$. In these sub-SANE runs, we employ the initial magnetic field configuration as specified in \Eq{subsaneic} and as shown in \Fig{fig:ss_init_condition}.
To compare, we also perform two SANE runs in 2.5D\footnote{A 2.5D simulation solves for 3-vectors on a 2D grid. In our simulations, this pertains to axisymmetric disks.} and 3D, with the initial magnetic field configuration as given in \Eq{saneic} and shown in \Fig{fig:sane_init_condition}, both with BH spin $a=0.9375$. We will denote these as 2DS-9375 and 3DS-9375. 
Except for the magnetic field configuration, the SANE runs have the same disk set-up as the sub-SANE runs above. The 3D SANE run has the same resolution as 3D sub-SANE runs while the 2.5D run has an effective resolution of $1024 \times 512$.
All the simulations are evolved upto $t=10^4 M$ except for 3DS-9375. A summary of the runs is given in the Table~\ref{tab:sim_params}.

\subsection {Diagnostics}

\begin{deluxetable}{lccc}
\tablecaption{Simulation parameters and configurations.\label{tab:sim_params}}
\tablehead{
\colhead{Name} & \colhead{Dim} & \colhead{Magnetic field configuration} & \colhead{BH spin ($a$)}
}
\startdata
3DSS-9375 & 3 & Sub-SANE, SS(\Fig{fig:ss_init_condition}) & 0.9375, 1.348$M$ \\
3DSS-5 & 3 & Sub-SANE, SS(\Fig{fig:ss_init_condition}) & 0.5, 1.866$M$ \\
2DS-9375 & 2 & SANE, S(\Fig{fig:sane_init_condition}) & 0.9375, 1.348$M$ \\
3DS-9375 & 3 & SANE, S(\Fig{fig:sane_init_condition}) & 0.9375, 1.348$M$ \\
\enddata
\end{deluxetable}

The azimuthal average of any quantity $q$ is defined as
\begin{equation}
  \overline{q} = \frac{\int ^{2\pi} _0\sqrt{-g}~q~\text{d}{\phi}}{\int^{2\pi} _0\sqrt{-g}~\text{d}{\phi}}.
\end{equation}
In particular, we refer to the azimuthally averaged magnetic field as the large-scale field. 
The spatial average along a given coordinate direction ($x^i = r,\theta$) is computed as
 \begin{equation}
     \langle q \rangle_{i}=\frac{\int\sqrt{-g}~q~\text{d}x^{i}}{\int\sqrt{-g}~\text{d}x^{i}},
 \end{equation} 
where $i=r,\theta$.
The root mean square for a quantity is correspondingly computed as 
 \begin{equation}
     q_{rms} = \sqrt{\frac{\int \int\sqrt{-g}\, \overline{q^2} \,\text{d}{r} \text{d}{\theta}}{\int \int\sqrt{-g}\text{d}{r} \text{d}{\theta}}}.
 \end{equation}
If the quantities are averaged only in the disk, then they are integrated over the extent of $r\in[6M,50M],\ \theta\in[\pi/4,3\pi/4]$.

The magnitude of the poloidal component of magnetic field is calculated as follows:
$\overline{B}_{{pol}}=\sqrt{\overline{B^r}^2+\overline{B^{\theta}}^2}$. The magnetic field components $B^r$, $B^{\theta}$ and $B^{\phi}$ are in the locally non-rotating frame, unless explicitly mentioned otherwise.

\begin{figure}
    \centering
    \includegraphics[width=1.04\linewidth]{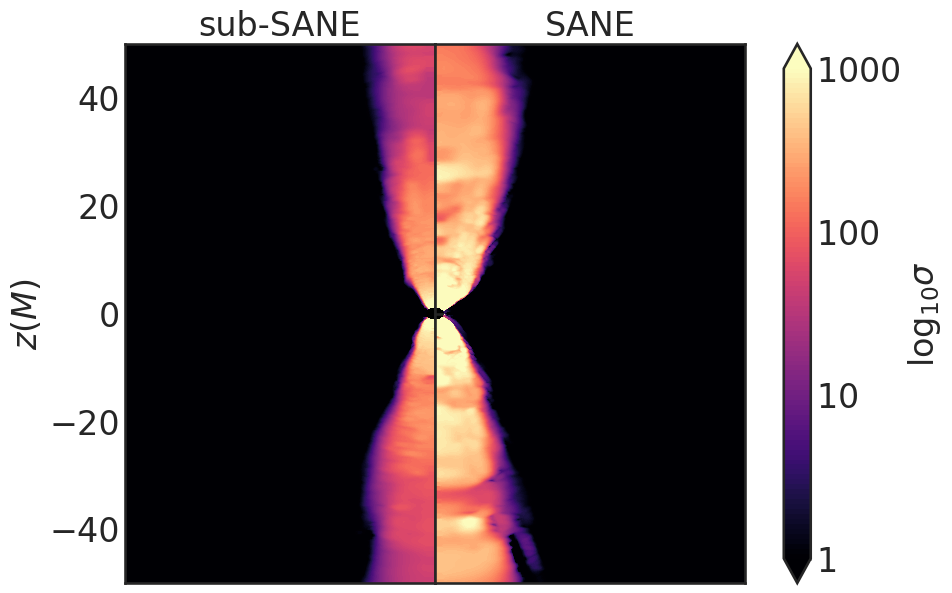}
    \includegraphics[width=1\linewidth]{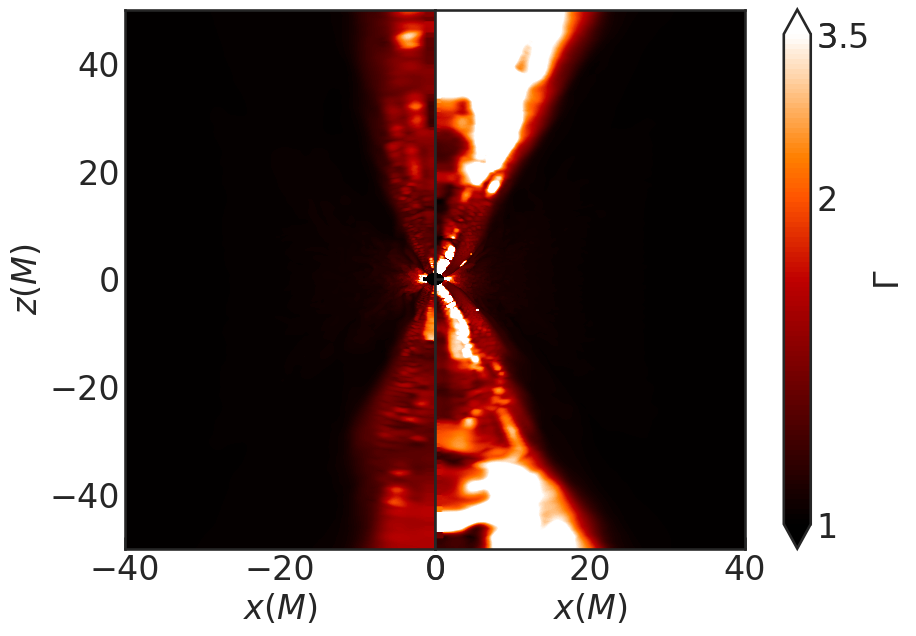}
    \caption{The top panel shows magnetisation, $\sigma=b^2/ \rho$ in the $x-z$ plane at $t=1000 M$ for 3DSS-9375 and 2DS-9375 in the left and right halves respectively.  The bottom panel shows Lorentz factor $\Gamma=(1-v^iv_i)^{-1/2}$  at $t=1000 M$ for 3DSS-9375 (left) and $t=1350 M$ for 2DS-9375 (right). The parabolic structure of the jet is more prominent in the SANE case.}
    \label{fig:jetmorph}
\end{figure}
\begin{figure*}
    \centering
    \includegraphics[width=0.85\textwidth]{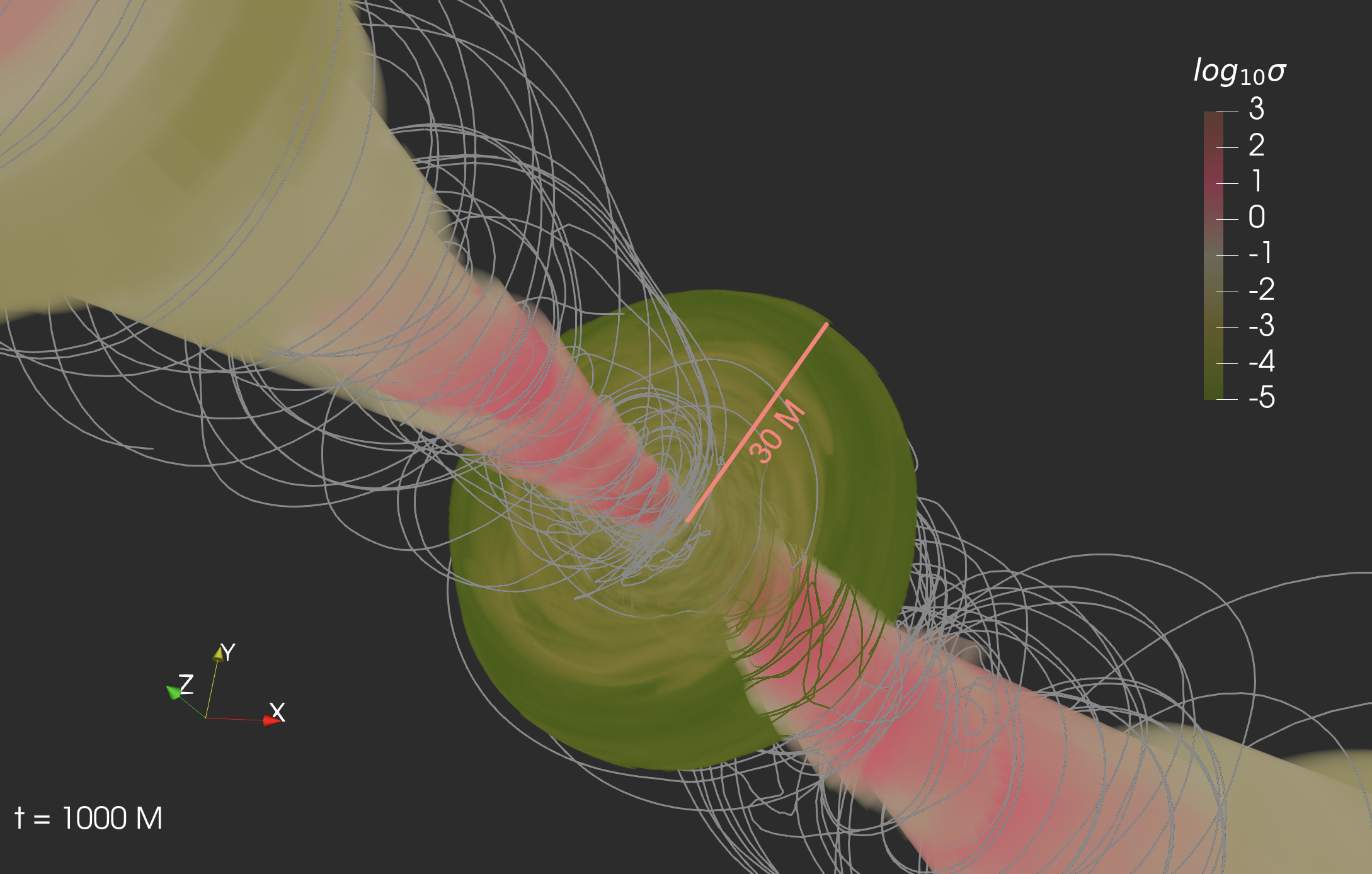}
    \includegraphics[width=0.85\textwidth]{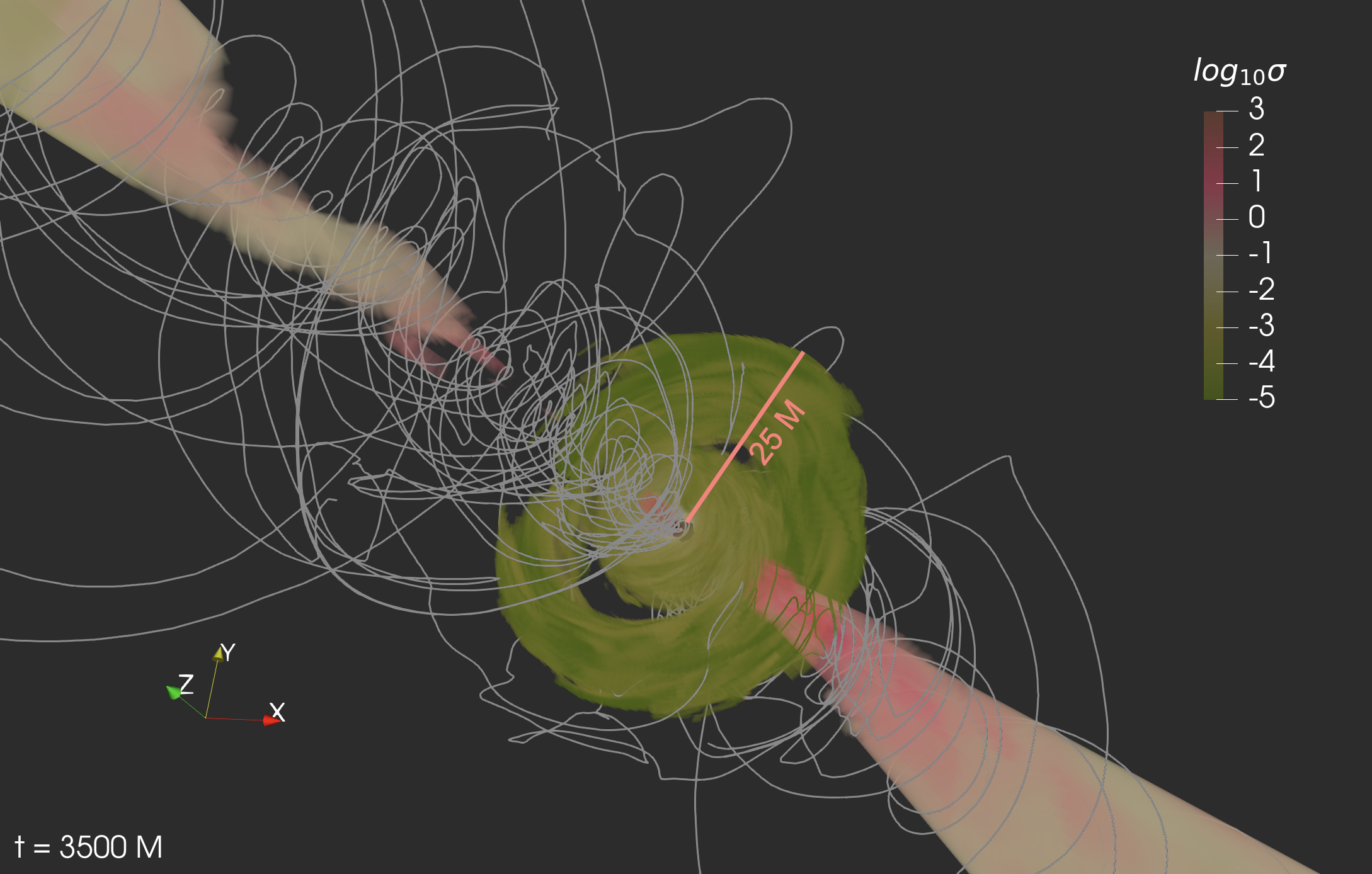}
    \caption{The top and bottom panels show magnetisation $\rm{log}_{10}\sigma$ in the disk ($\rho_\text{cutoff}=[0.6,1])$ and jet ($\Gamma_\text{cutoff}=[1.1,2.5]$) region for the  3DSS-9375 run at $t=1000M$ and $t=3500M$ respectively. The magnetic field lines are shown by the grey lines. We see that the field lines are more coherent in the funnel region at $t=1000M$ and they eventually become tangled at the base of the jet at $t=3500M$. This leads to the shut down of the jet, as seen by the diminished $\sigma$ in the jet.}
    \label{fig:3d_ss_sigma}
\end{figure*}

\section{Results}

We present results here from two main 3D sub-SANE runs termed 3DSS-9375 and 3DSS-5, which exhibit dynamo behaviour, corresponding to spin values of $a = 0.9375$ and $a = 0.5$. We have introduced all runs already in Section~\ref{simpars} and the list of runs can be found in Table~\ref{tab:sim_params} for ease of reference. We compare the sub-SANE runs with a 2.5D SANE simulation, 2DS-9375, with $a = 0.9375$. 

Note that sub-SANE runs were initialized with multiple magnetic field loops (of same polarity) distributed across the equatorial plane. This topology is susceptible to  magnetic reconnections, leading to a rapid loss of memory and thus early arisal of the dynamo. 
We aim to study both jets and dynamos and their interconnections. We defer the study of dynamo aspect to Section~\ref{subsec:dynamos}. We begin by examining the jets. 

\subsection{Jets}
\label{jets}

GRMHD simulations of jets have routinely reported the presence of a spine–sheath structure, along with parabolic collimation near the BH that transitions to a conical shape farther out \citep{HardeeMizunoNishikawa2007, Chatterjee2019, liska_large-scale_2020}. While both SANE and sub-SANE runs exhibit this characteristic spine–sheath morphology, the parabolic geometry is more accurately captured in the SANE case than in the sub-SANE case, as evident from the upper panel of \Fig{fig:jetmorph}. In this panel, we show contours of the magnetisation parameter $\sigma$ to illustrate the geometry of the relativistic jet, following the standard definition. The lower panel shows the Lorentz factor $\Gamma$ within the jet, reaching values of up to $\sim 3.5$ in the SANE case and up to $\sim 2$ in the sub-SANE case near the BH, at distances of about $50M$, and remaining comparable out to $\sim 500M$. The Lorentz factor contour map reveals higher values closer to the sheath in the SANE case, reminiscent of the arm-like structures observed in astrophysical jets: features typically attributed to limb brightening due to Doppler boosting \citep{Ghisellini2005StructuredJets, Asada2012ParabolicM87}. This characteristic is conspicuously absent in the sub-SANE case. We conjecture that this difference may arise from the jet related magnetic field being more coherent in the SANE disk, whereas in the sub-SANE case, increased field randomness could promote mixing; we revisit this possibility in detail later in the text. Over time, the jet in the SANE run becomes stronger and more stable, while in the sub-SANE case, the jet appears thinner, wispy and wobbly, eventually shutting down. This can be seen in \Fig{fig:3d_ss_sigma} where we show the 3D structure of the disk and jet parametrized by the magnetisation $\sigma$, for the 3D sub-SANE higher spin case. At early times (see top panel in\Fig{fig:3d_ss_sigma}), the jet is robust and we observe helical structure of the magnetic field lines, as expected and they remain coherent within the funnel region. As the system evolves, however, the field lines become increasingly tangled (see bottom panel in \Fig{fig:3d_ss_sigma}), and the characteristic dipolar jet structure gradually diminishes—consistent with the eventual shutdown of the jet.

In the following paragraphs we discuss more quantitative aspects of these jets.

\begin{figure}
\includegraphics[width=\linewidth]{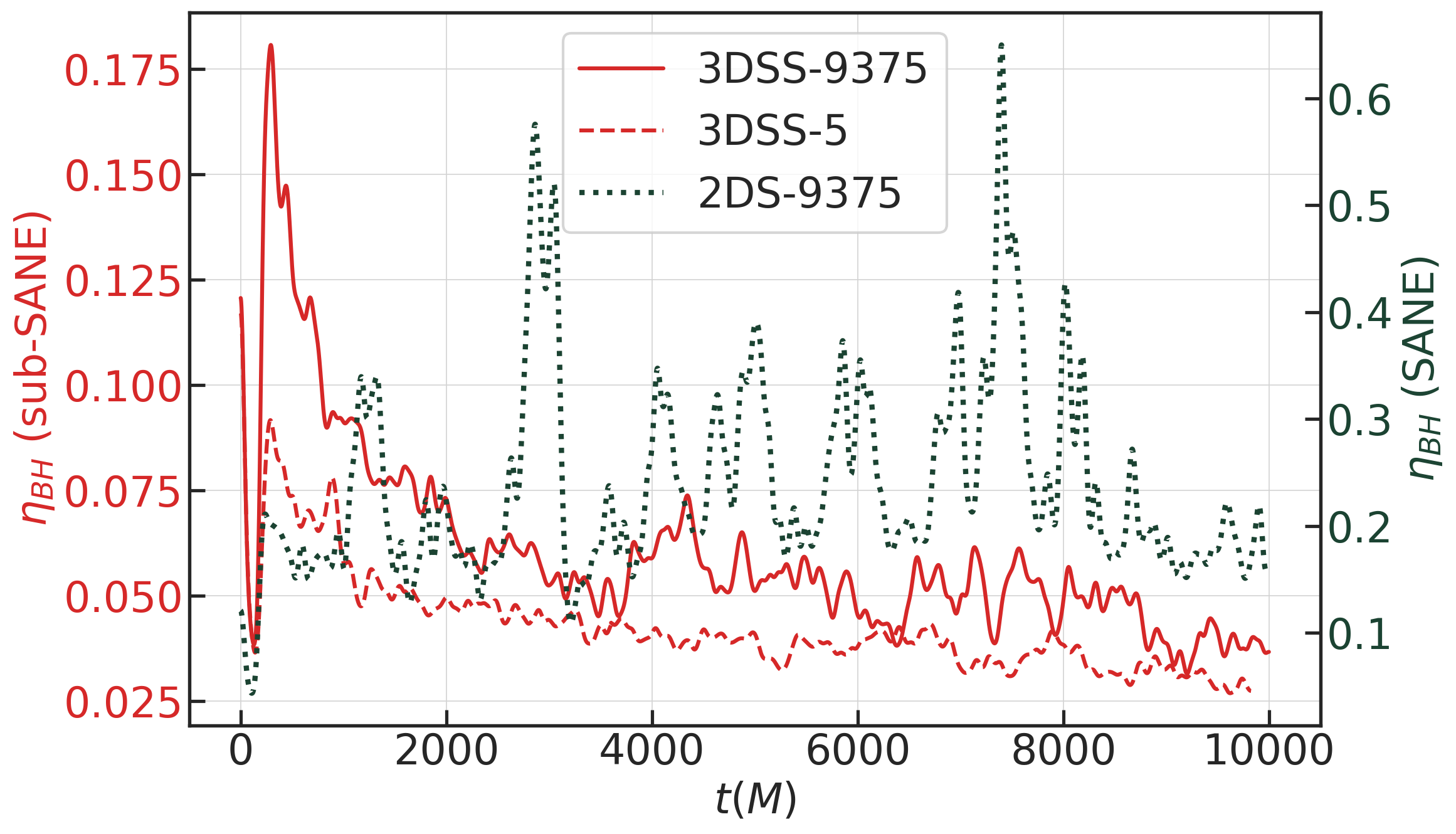}
\caption{Efficiency of the outflow, $\eta_{\mathrm{{BH}}}$ 
at the horizon for 2DS-9375 (SANE), 3DSS-9375 (sub-SANE) and 3DSS-5 (sub-SANE) simulations as denoted by the grey dotted, red solid and red dashed lines respectively. We see  $\eta_{\mathrm{{BH}}}$ drops significantly at later times for both the sub-SANE simulations 
while it fluctuates around $\sim 0.25 $ in the SANE run.} 

\label{fig:eta}
\end{figure}

\begin{figure}
\includegraphics[width=\linewidth]{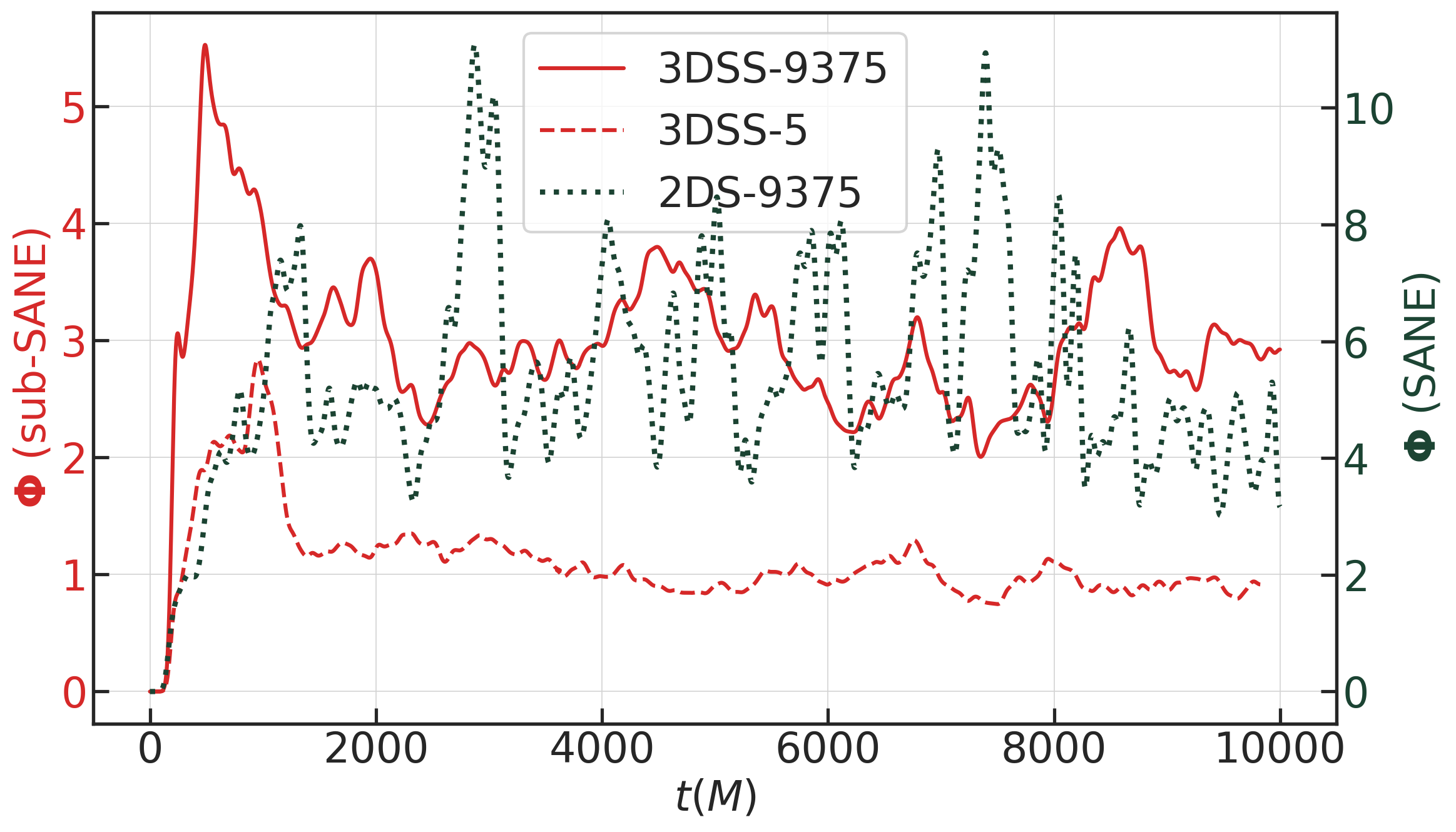}
\caption{The evolution of MAD parameter $\boldsymbol{\Phi}$ at the horizon for 2DS-9375, 3DSS-9375 and 3DSS-5 simulations. $\boldsymbol{\Phi}$ for the sub-SANE runs are 2-6 times smaller when compared to the SANE run.}
\label{fig:mad} 
\end{figure}

\begin{figure}
\includegraphics[width=\linewidth]{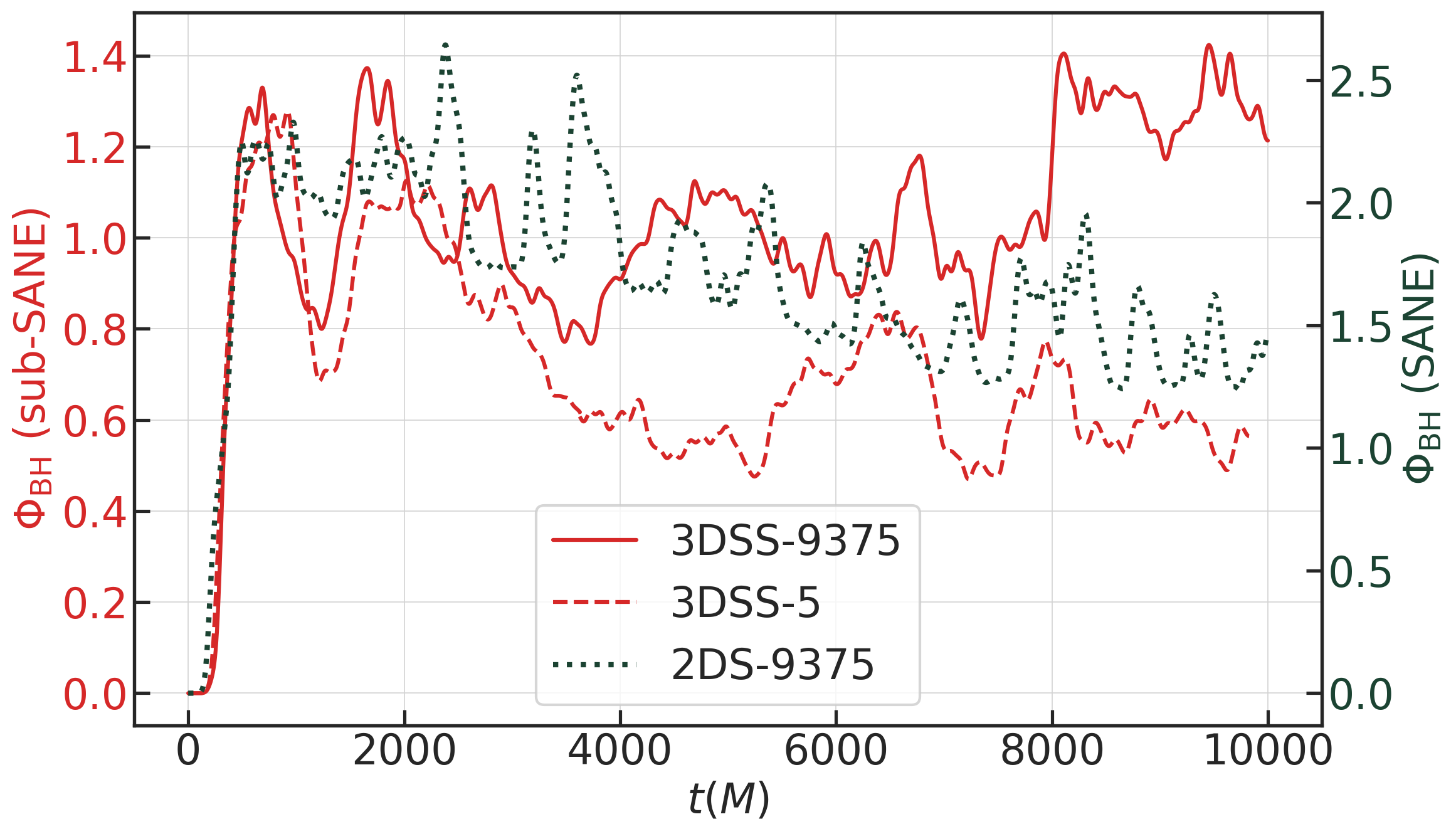}
\caption{The evolution of magnetic flux $\Phi_\mathrm{{BH}}$ at the horizon for 2DS-9375, 3DSS-9375 and 3DSS-5 simulations. Eventually, $\Phi_\mathrm{{BH}}$ becomes very similar between both SANE and sub-SANE runs with $a=0.9375$. 
}
\label{fig:phiBH}
\end{figure}
\begin{figure}[!htbp]
\includegraphics[width=\linewidth]{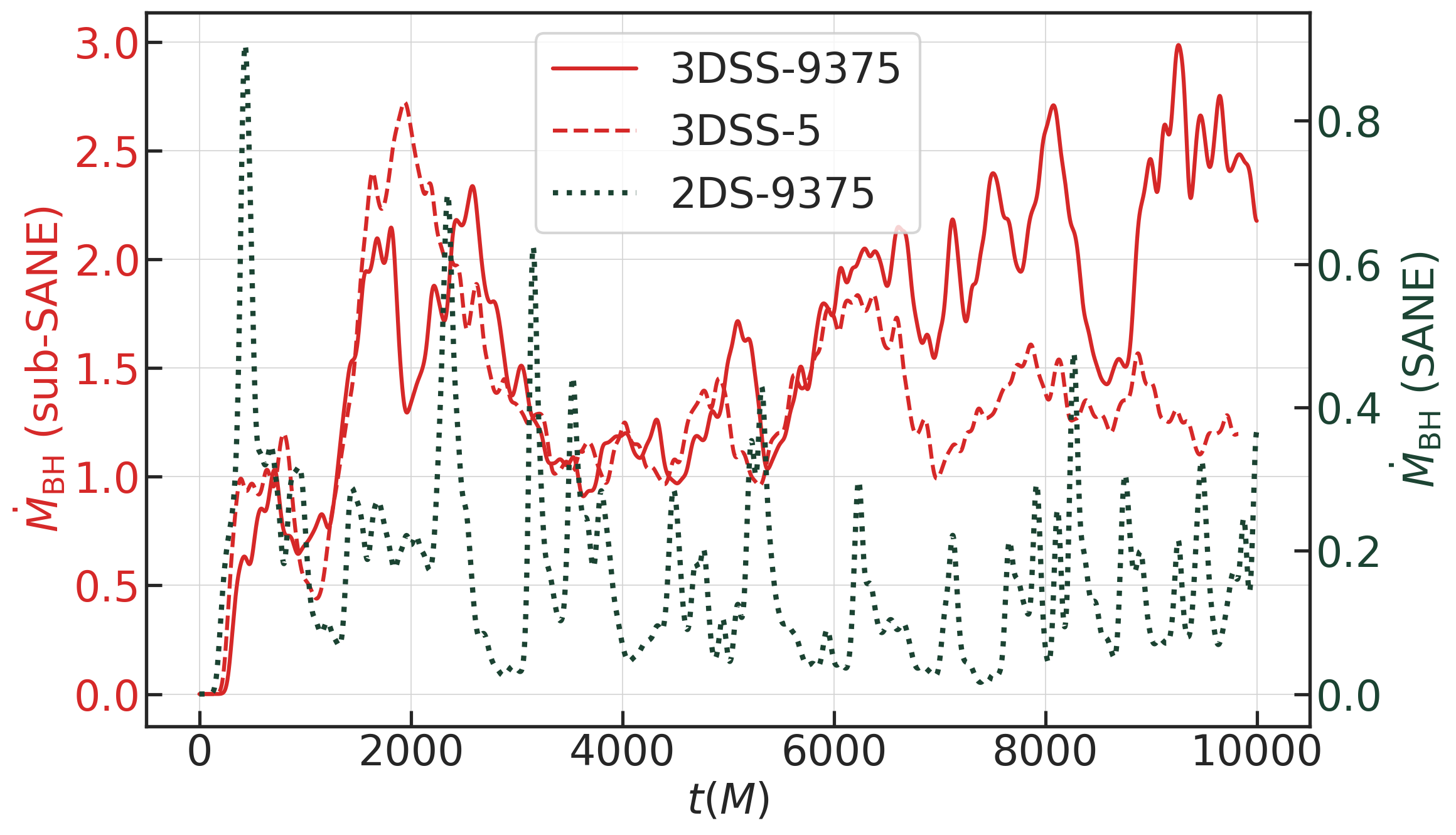}
\caption{The evolution of accretion rate $\dot{M}_\mathrm{BH}$ at the horizon for 2DS-9375, 3DSS-9375 and 3DSS-5 simulations. $\dot{M}_\mathrm{BH}$ is an order of magnitude lower for the 2DS-9375 run when compared to 3DSS-9375.}
\label{fig:mdot}
\end{figure}

The primary quantity used to assess jet strength is the jet efficiency, defined as 
\begin{equation}\label{eta_BH}
\eta_{\mathrm{{BH}}}= \frac{\dot{E} - \dot{M}_\mathrm{BH}}{\dot{M}_\mathrm{BH}} = \frac{{\dot{E}_\mathrm{MAKE}+\dot{E}_\mathrm{EM}}}{\dot{M}_{\mathrm{BH}}} 
\end{equation}
Here, $\dot{E}_\mathrm{MAKE}$ is the free particle and enthalpy component of the energy outflow rate, $\dot{E}_\mathrm{EM}$ is the electromagnetic component of the same \citep{McKinney2012_MCAFs} and $\dot{M}_{\mathrm{BH}}$ is the mass accretion rate at the horizon as defined in \Eq{acc_rate}. This efficiency serves as a figure of merit for how effectively accretion onto the BH is converted into outflow power. In \Fig{fig:eta}, we show the evolution of jet efficiency for the three runs: 3DSS-9375, 3DSS-5, and 2DS-9375. In the SANE case, $\eta_{\mathrm{BH}}$ fluctuates around 0.25  throughout the simulation. In contrast, the 3DSS-9375 run shows $\eta_{\mathrm{BH}}$ initially increasing to $\sim 0.18$, followed by a rapid decline to 0.06 by $t \sim 3000M$, and then a slow decrease to 0.04 by $t = 10,000M$. This temporal behavior is consistent with visual inspection of movies of the magnetisation parameter $\sigma$ in the $x$–$z$ plane, which reveal a robust jet in the SANE case, intensifying around $t \sim 4000M$, whereas in the sub-SANE cases, the jet gradually weakens and eventually dies out after $t \sim 3000M$. We also find correlation of  the jet shutdown with $\eta_{\mathrm{BH}}$ becoming smaller than around 0.05. We now turn to investigating the physical causes of this jet shutdown.

\begin{figure}
\includegraphics[width=\linewidth]{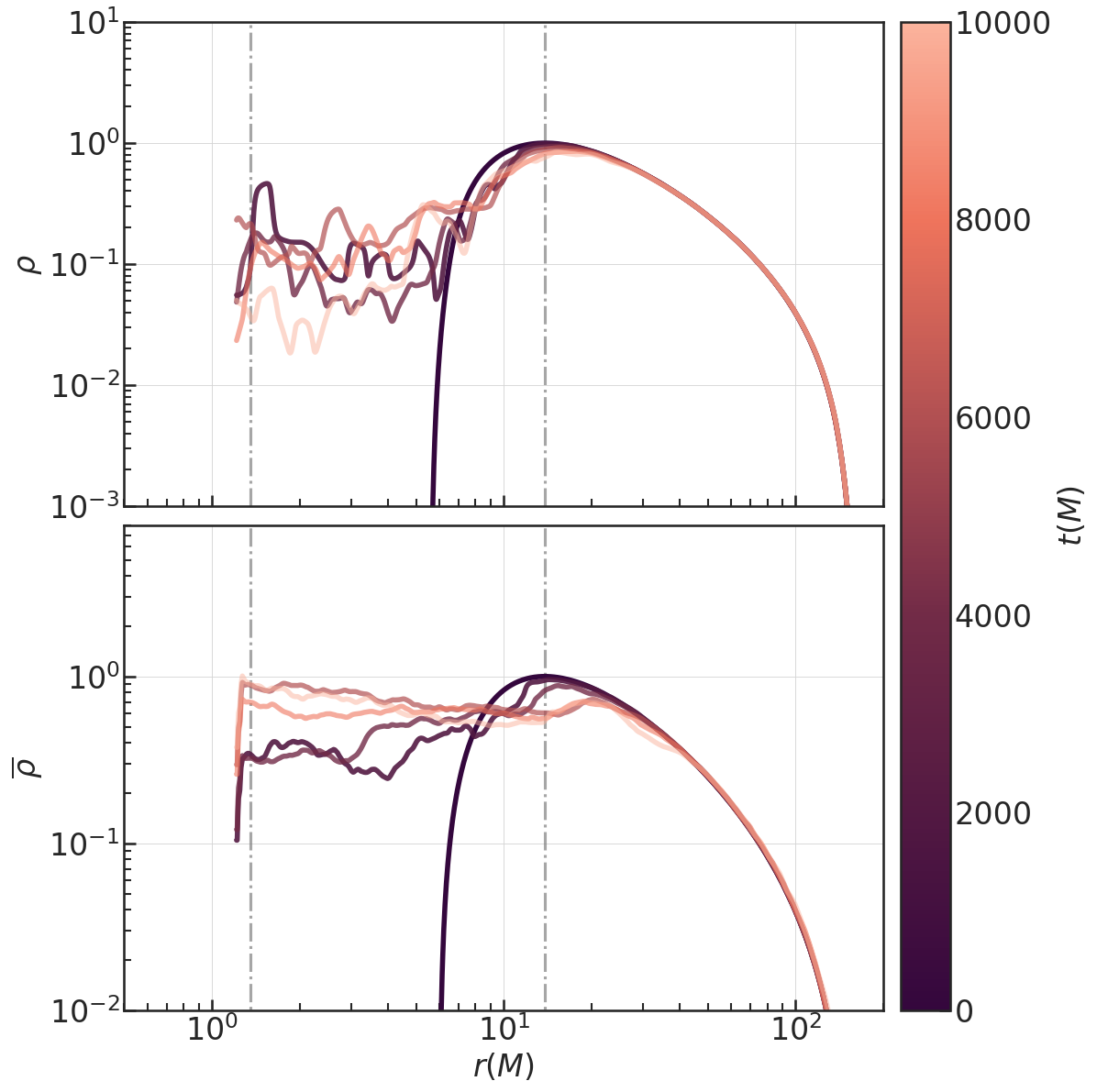}
\caption{The top panel shows the radial profile of density $\rho$ in the mid-plane ($\theta=\pi/2$) of the disk for 2DS-9375 run.  
A density depression is maintained at later times in the inflow region - in sync with the lower $\dot{M}_\mathrm{BH}$. The bottom panel shows axisymmetrized density $\overline{\rho}$ in the mid-plane for the 3DSS-9375 simulation, which develops a power law bridging the gap between $r_{\rm{max}}$ and the horizon at later times. }
\label{fig:rhoprof}
\end{figure}

According to the BZ theory of jet production, $\eta_{\mathrm{BH}}$ depends primarily on three factors: the magnetic flux at the horizon $\Phi_{\mathrm{BH}}$, the mass accretion rate $\dot{M}_\mathrm{BH}$, and the BH spin $a$, 
\begin{equation}
    \eta_{\mathrm{BZ}} \sim 
    \boldsymbol{\Phi}^{2}\Omega_{\mathrm{H}}^{2},
\label{bzeq}
\end{equation}
where $\Omega_{\rm{H}} = a/r_\mathrm{BH}$ is the angular velocity of the horizon.
In \Fig{fig:mad}, we show the time evolution of $\boldsymbol{\Phi}$, which exhibits a qualitative correspondence with the $\eta_{\mathrm{BH}}$ curve. In the SANE case, $\boldsymbol{\Phi}$ falls within the range $\sim 4–8$, consistent with the typical definition of a SANE state, whereas in the sub-SANE cases it remains much lower, in the range of $1-3$. To understand this difference, we examine the individual contributions from $\Phi_{\mathrm{BH}}$ and $\dot{M}_\mathrm{BH}$. As shown in \Fig{fig:phiBH}, the SANE case initially ($t<2000M$) exhibits a $\Phi_{\mathrm{BH}}$ higher than that of the sub-SANE cases by a factor of $\sim 2$, but this value gradually decreases with time. At late times ($t \gtrsim 8000M$), the runs 3DSS-9375 and 2DS-9375 converge to similar values, around $\Phi_{\mathrm{BH}} \simeq 1.5$. Thus we surmise, that the contrasting jet behaviour between the two cases at late times perhaps  arises from differences in $\dot{M}_\mathrm{BH}$ rather than in $\Phi_{\mathrm{BH}}$.

In \Fig{fig:mdot}, we show the evolution of the mass accretion rate, $\dot{M}_\mathrm{BH}$, for the two runs. It is evident that $\dot{M}_\mathrm{BH}$ in the SANE case remains lower by a factor of $\sim 2$–$8$ compared to the sub-SANE cases. In the sub-SANE run of 3DSS-9375, $\dot{M}_\mathrm{BH}$ increases from around unity to nearly twice that value fairly early in the simulation, after $t \sim 1000M$. This enhanced accretion rate leads to increased mass loading of the jet, which in turn weakens its magnetisation and is perhaps ultimately one of the main causes for the jet to shut down.

We next ask what causes the SANE run to attain a lower $\dot{M}_\mathrm{BH}$ compared to the sub-SANE cases. In \Fig{fig:rhoprof}, we show the radial density profiles from 3DSS-9375 and 2DS-9375 in the lower and upper panels, respectively. Although both simulations start with nearly identical radial density distributions, their evolution differs significantly. As the simulation progresses, the SANE case preserves a pronounced density depression from $r = r_{\mathrm{in}}$ down to the horizon, whereas this depression is gradually filled in the sub-SANE case, resulting in a density that actually increases toward smaller radii. This difference naturally explains the lower $\dot{M}_\mathrm{BH}$ seen in \Fig{fig:mdot}.

The contrasting behaviour of $\dot{M}_\mathrm{BH}$ and the density profiles in the SANE and sub-SANE cases can be attributed in a large-part to their initial magnetic field configurations. In the SANE case, the magnetic field is relatively coherent and large-scale, which gets advected towards the horizon, where it accumulates and compresses, amplifying the magnetic flux. The resulting magnetic pressure near the BH becomes comparable to, or even exceeds, the gas pressure, opposing the ram pressure of the inflowing material. This magnetic “throttling” of the accretion flow—similar in spirit to what occurs in MAD systems—reduces the inflow rate and maintains the density depression near the horizon. However, there is an important difference between the 2D and 3D SANE cases. In 3D, with time, the turbulence develops fully and can disrupt the magnetic-pressure barrier, allowing more gas to penetrate and this is reflected in a secular increase in the accretion rate at the horizon (see \Fig{3dsane-accr} in Appendix~\ref{3dsaneplots}).

In contrast, the sub-SANE case begins with a multi-loop, small-scale magnetic field configuration. Such a field is highly susceptible to reconnection, which rapidly destroys large-scale coherence and produces a more tangled magnetic structure. The fields advected toward the horizon are therefore more disordered, leading to weaker and spatially fluctuating magnetic pressure. With less magnetic support, the gas flows more freely toward the BH, resulting in a higher mass accretion rate and a density profile that follows a smooth negative power-law scaling with radius \citep{MitraMaityDihingiaDas2022, Chatterjee_2022, MitraDas2024LowAngMomGRMHD}. 
\begin{figure}[!htbp]
\includegraphics[width=\linewidth]{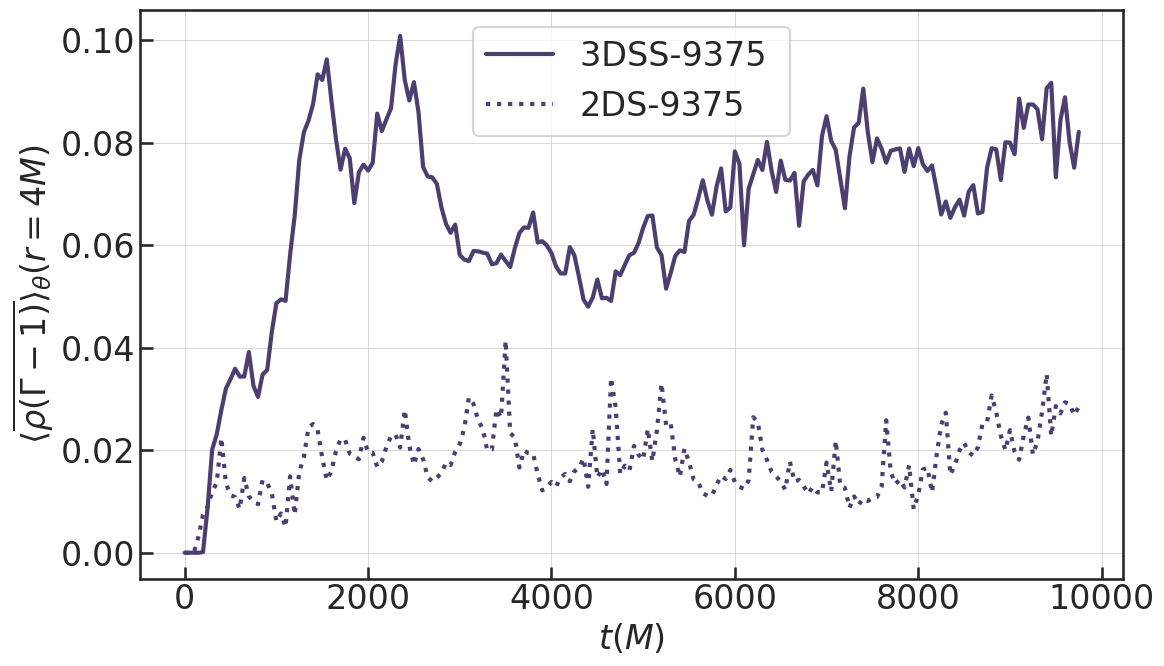}
\caption{The evolution of axisymmetrized kinetic energy density averaged over $\theta$, at $r=4 M$  for 3DSS-9375 and 2DS-9375. The lower kinetic energy density in 2DS-9375 is persistent in the inner flow region and is consistent with resulting lower $\dot{M}_\mathrm{BH}$.}
\label{fig:kin}
\end{figure}

 \begin{figure}
\includegraphics[width=\linewidth]{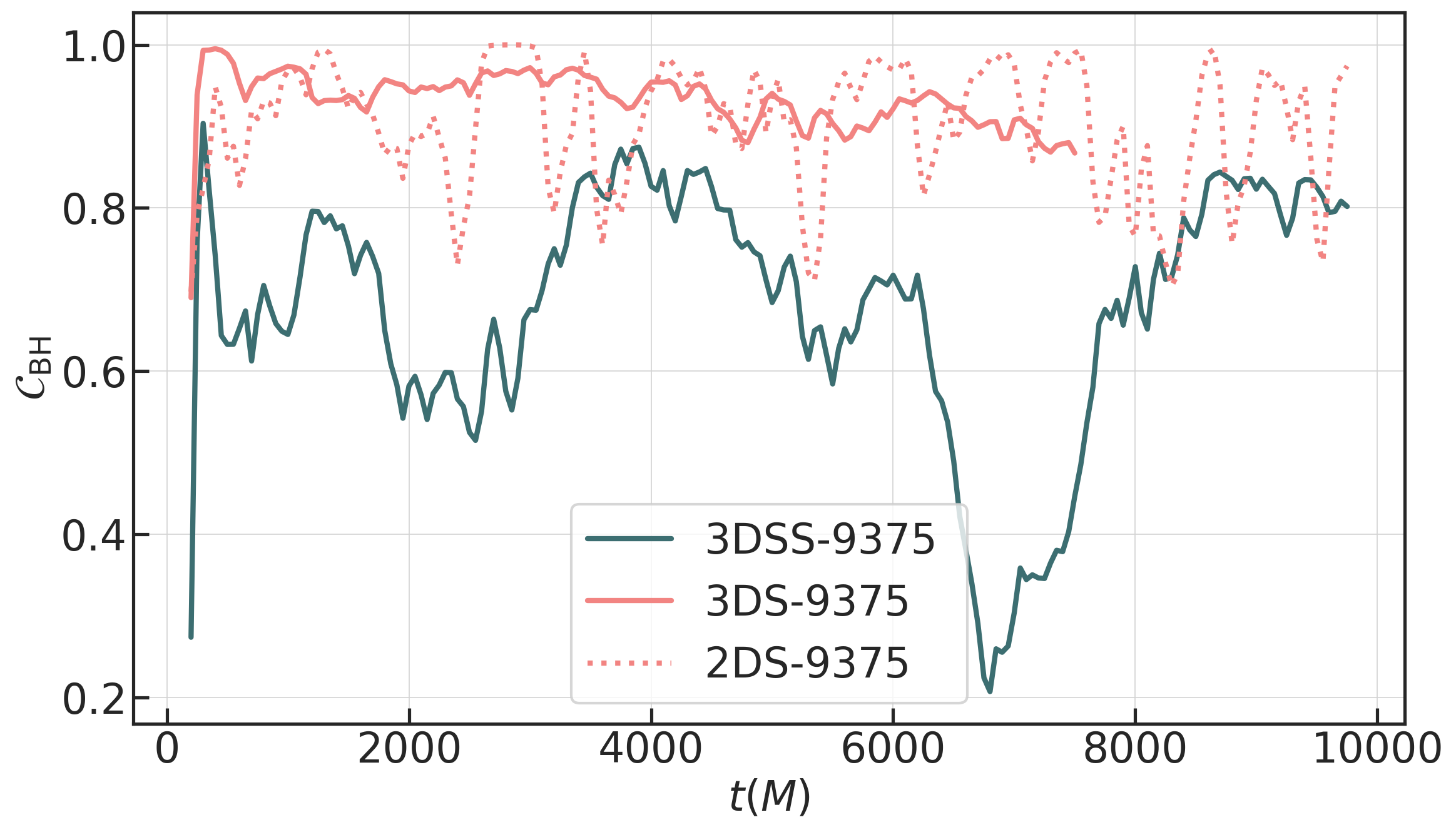}
\caption{The evolution of coherence parameter for the magnetic flux at the event horizon, $\mathcal{C}_\mathrm{BH}$ for the 3DSS-9375, 3DS-9375 and 2DS-9375 as defined in \Eq{ratio_BH}. $\mathcal{C}_\mathrm{BH}$ is $\sim0.8$ or more for the SANE runs, consistent with the observation of jets. For 3DSS-9375, $\mathcal{C}_\mathrm{BH}$ fluctuates widely, particularly drops below 0.6 after $t=2000 M$, indicating that the poloidal fields are more random at the horizon, leading to a jet shutdown. The magnetisation $\sigma$ and the Lorentz factor $\Gamma$ in the jet also eventually vanish around this time for 3DSS-9375.}
\label{fig:ratio-signed}
\end{figure}

The density depression in the SANE run also reduces the local turbulent kinetic energy, further supporting the observed reduction in the mass accretion rate. In \Fig{fig:kin}, we show the evolution of the kinetic energy density at $r = 4M$.
The amplitude of the curve in the SANE case is significantly smaller—by a factor of $\sim 4$—compared to the sub-SANE case. 

Intriguingly, we find that $\eta_{\mathrm{BH}}$ in the run 3DS-9375, at late times, is similar to that in sub-SANE run 3DSS-9375. We find $\eta_{\mathrm{BH}}$ in 3DS-9375 run goes down to a value of $0.05$ around $t\sim 6500M$ (which is similar to the sub-SANE case where it happens fairly early; see left panel of \Fig{fig:poynting_3Dsane} in the Appendix~\ref{3dsaneplots}), nonetheless there is no corresponding jet shutdown. Thus we find that quantification of the jet by $\eta_{\mathrm{BH}}$ alone is not satisfactory\footnote{We find $\eta_{\mathrm{BH}}$ itself maybe satisfactorily described by the BZ formula in \Eq{bzeq} even in sub-SANE cases as shown in \Fig{fig:eta_bz_3DSS} in Appendix~\ref{bzmatch}.}.
We surmise that another possible cause for the shutdown could arise from the structure of the fields that support the jet.  
Towards this we plot the ratio of signed to unsigned flux at the horizon $\mathcal{C}_\mathrm{BH}$ \citep{Rodman_2024} as follows:
\begin{equation}
    \mathcal{C}_\mathrm{BH} = \frac{\int \sqrt{-g}B^r \text{d}\theta\text{d}{\phi}}{\int \sqrt{-g}|B^r| \text{d}\theta\text{d}{\phi}}
\label{ratio_BH}
\end{equation}
The integral goes from $\phi = [0, 2\pi]$ and $\theta=[0,\pi/2]$ for the 3D case, which is the northern hemisphere of the domain at $r=r_{\mathrm{BH}}$ and in 2D case, it comprises of $\theta=[0,\pi/2]$. We term $\mathcal{C}_\mathrm{BH}$ as the coherence parameter associated with the fields at the horizon. 
We find from \Fig{fig:ratio-signed} that in both 2DS-9375 and 3DS-9375 runs, the value of the coherence parameter $\mathcal{C}_\mathrm{BH}$ is fairly close to unity indicating that the magnetic fields at the horizon are fairly coherent and less random. But in the sub-SANE run, the field becomes {\it significantly} more random by $t\gtrsim 2000 - 3000M $ as inferred from decay in $\mathcal{C}_\mathrm{BH}$ to a value below 0.6, which is also co-incident with the jet shutdown as seen in \Fig{fig:3d_ss_sigma}. Thus, we find that the coherence parameter is also an important factor to accounted for besides the $\eta_\mathrm{BH}$ as defined in \Eq{bzeq}. The evolution of  $\mathcal{C}_\mathrm{BH}$ in 3DSS-9375 run beyond the jet shutdown point, shows increase back to a value closer to unity and another major dip later.  However,  the jet remains turned off because by $t=3500M$, $\eta_{\mathrm{BH}}$ is also fairly low and remains at that level for rest of the simulation. 
Overall, we find that while $\eta_{\rm BH}$ is more of an indicator of the possibility of jet shutdown, the coherence parameter $\mathcal{C}_\mathrm{BH}$ has a direct causal link to it.
\begin{figure}
 \centering
    \includegraphics[width=1\linewidth]{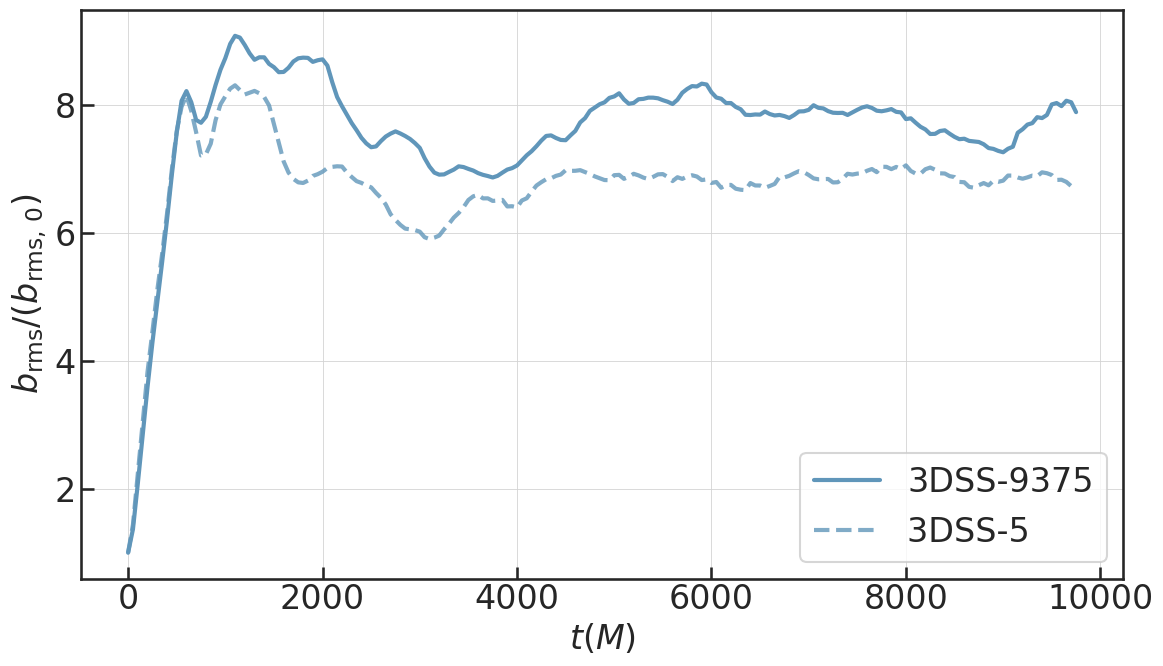}
    \caption{\small The evolution of fluid frame $b_{rms}$ (density-averaged) for 3DSS-9375 and 3DSS-5 runs, normalized by its initial value. The sustained RMS field in the disk is due to turbulent dynamo action and not by any memory of the initial magnetic field configuration.
    }
    \label{fig:brms}
\end{figure}

\subsection{Dynamos}
\label{subsec:dynamos}
In conventional MAD or SANE setups, the initial magnetic field typically consists of a single large-scale poloidal loop across the equator. Such configurations retain their memory for a long time, with the magnetic flux near the horizon primarily determined by the advection of this pre-existing large-scale field from the disk rather than the advection of disk-dynamo generated fields. In contrast, in our sub-SANE setup, we initialize the disk with multiple small-scale magnetic loops distributed across the equator.
This configuration promotes rapid reconnection and loss of memory of the initial field, effectively randomizing the magnetic field on very short timescales. Under these conditions, robust dynamo activity emerges from the ensuing turbulence fairly early in the simulation. In the following, we present clear evidence for such dynamo behaviour in our simulations.

\begin{figure}
 \centering
    \includegraphics[width=\linewidth]{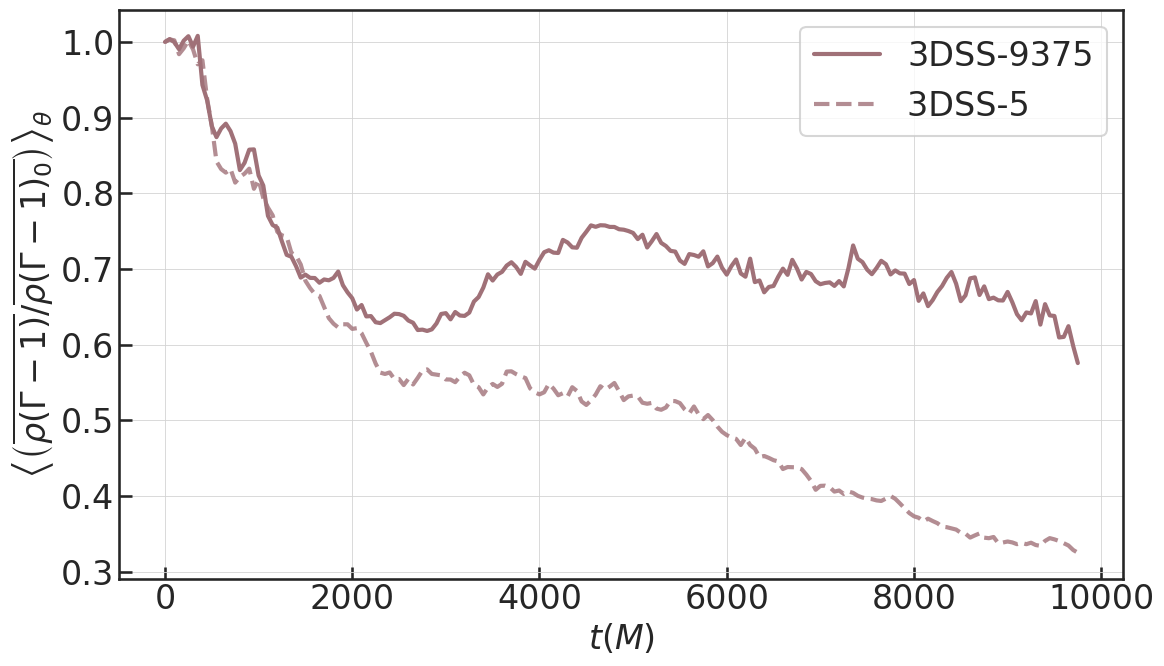}
    \caption{The evolution of axisymmetrized kinetic energy $\rho(\Gamma-1)$ at density maximum, $r_\mathrm{max}$, normalised with its initial value (at $t=0M$) and averaged over $\theta$ in the disk. The lowering of this kinetic energy in both the cases can be attributed to the slight density loss at $r_\mathrm{max}$ due to inflow of matter. 
    }
    \label{fig:ekin}
\end{figure}

In \Fig{fig:brms}, we show the time evolution of the RMS magnetic field strength (averaged over the relevant disk volume) for the two 3D sub-SANE runs. Interestingly, the lower-spin case saturates at a smaller field amplitude. In dynamo theory, saturation typically occurs when the Lorentz force associated with the magnetic field becomes strong enough to counteract further amplification by the turbulent flow. Although the MRI-driven dynamo departs from the classical picture—in which externally driven turbulence amplifies the magnetic field—here the magnetic field itself drives the instability and generates the turbulence. Nevertheless, one still expects an approximate equipartition between magnetic and kinetic energies.

In \Fig{fig:ekin}, we show the evolution of the 
 kinetic energy in the disk at $r = \rmax$. The 3DSS-9375 run exhibits a higher kinetic energy compared to the 3DSS-5 run at later times, with the differences first appearing at around $t\sim 1500 M$. The overall decreasing trend in kinetic energy is likely a consequence of the decline in local density at $\rmax$, as seen earlier in \Fig{fig:rhoprof} for the 3DSS-9375 case. 
\begin{figure}
    \includegraphics[width=\linewidth]
    {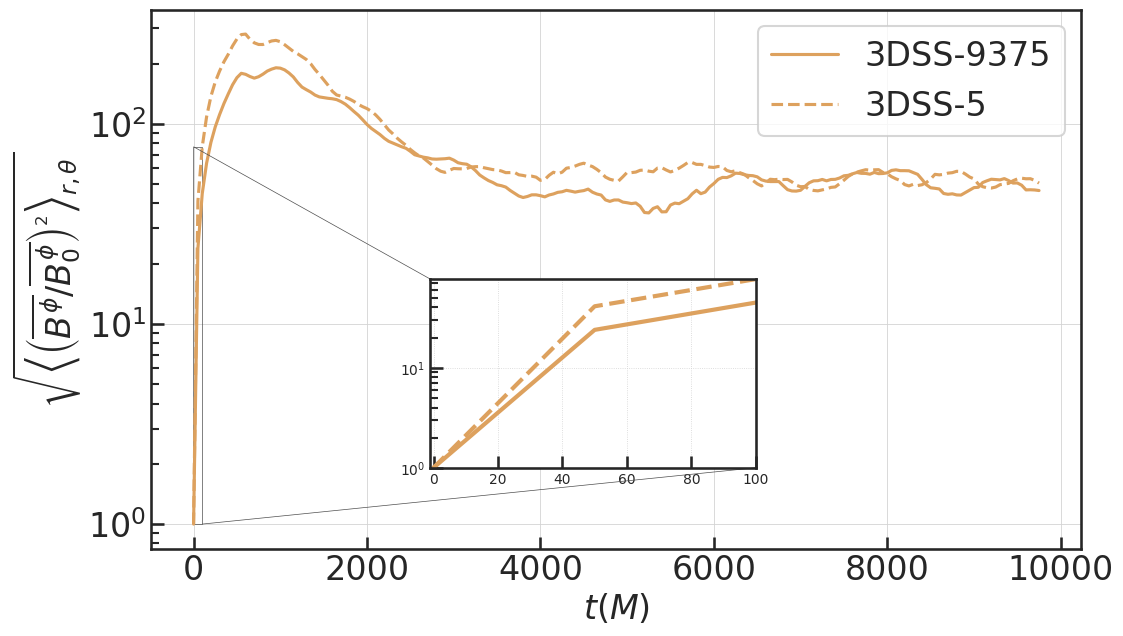}
    \caption{The evolution of large-scale azimuthal field $\overline{B^{\phi}}$, density-averaged over the disk region for 3DSS-9375 and 3DSS-5 and normalized by their corresponding initial values. Both simulations exhibit an initial exponential growth driven by the MRI-dynamo, after which the large-scale field saturates at a level nearly two orders of magnitude larger than its initial amplitude.
    }
    \label{fig:B-LS-field}
\end{figure}

\begin{figure}
    \centering
    \includegraphics[width=1\linewidth]{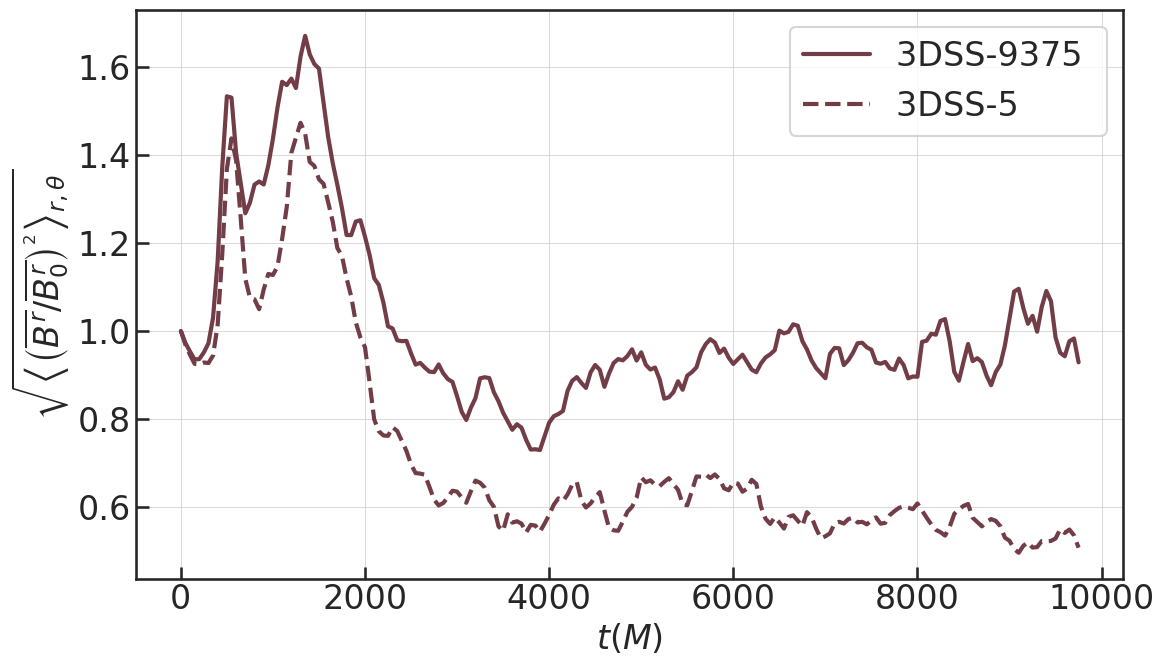}
    \caption{The evolution of the large-scale radial field $\overline{B^r}$, density-averaged over the disk region for 3DSS-9375 and 3DSS-5, normalised with the initial value. For 3DSS-9375, the amplitude of the field eventually saturates at the same level as the initial field while it saturates at a relatively lower value for 3DSS-5.}
    \label{Br_LS}
\end{figure}

\begin{figure*}
 \centering
     \includegraphics[width=0.85\textwidth]{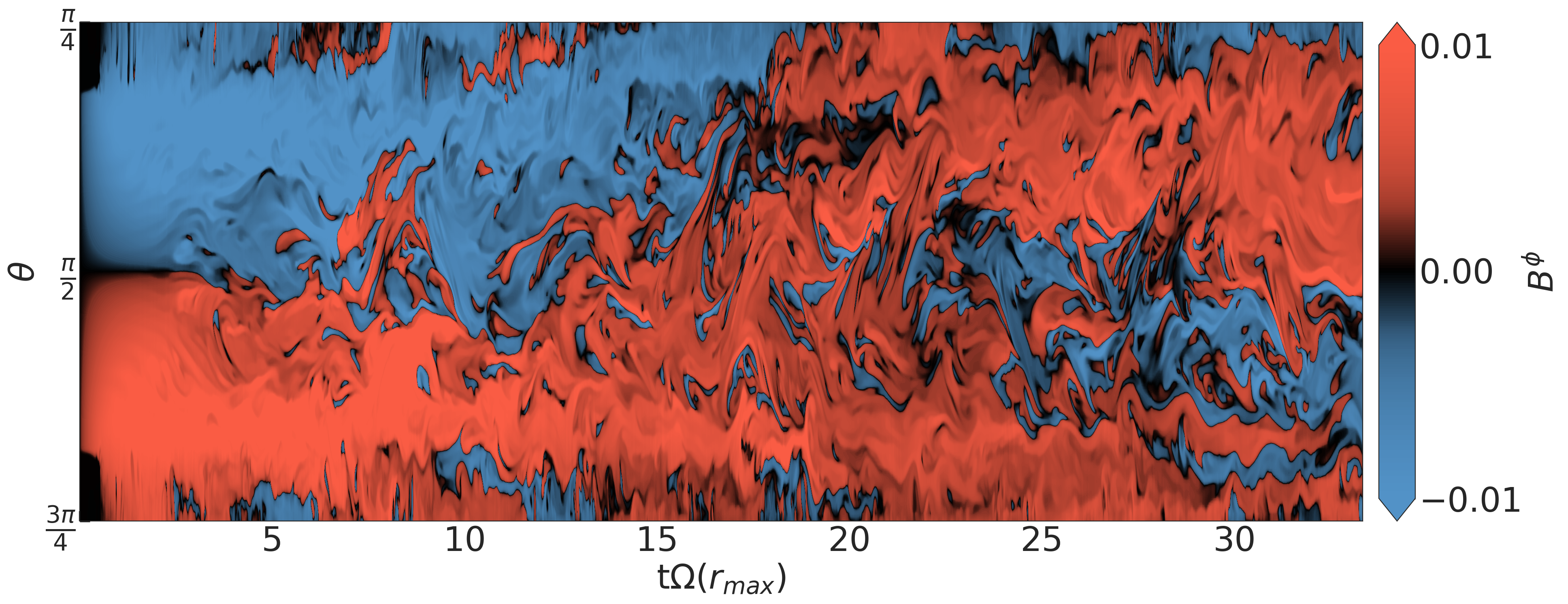}
    \includegraphics[width=0.85\textwidth]{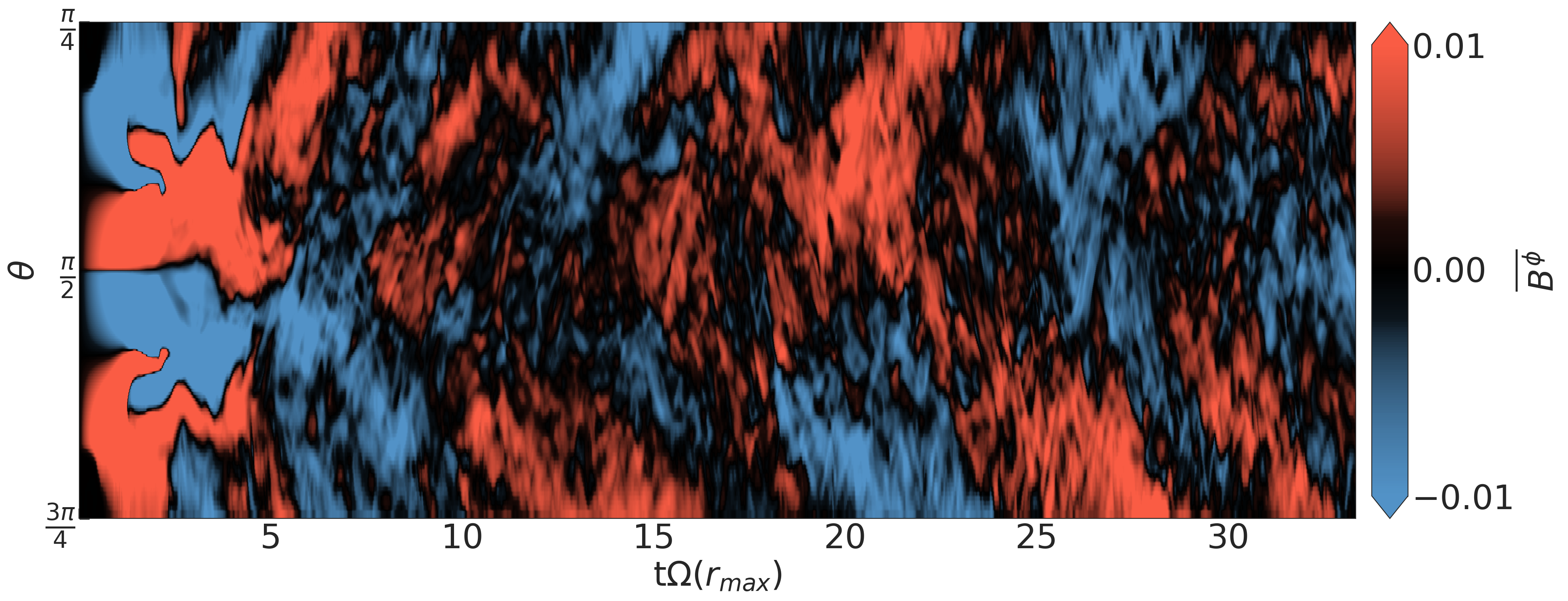}
    \caption{Large-scale azimuthal field  $\overline{B^{\phi}}$ at r=$r_{\rm{max}}$ as a function of $(\theta,t)$. The time $t$ is normalized with the orbital time period at $r_{\rm{max}}$. The top panel is for 2DS-9375 where there is no butterfly pattern and the initial memory of the magnetic field configuration is retained for a longer period of time. The bottom panel is for 3DSS-9375, where a regular butterfly pattern is clearly seen indicating robust large-scale dynamo action.}
    \label{fig:butterfly}
\end{figure*}

We think this difference in both magnetic and kinetic energy in the two 3D sub-SANE runs with different spins can arise due to the overall energetics, where a significant component can be attributed to the initial jet and in general, to outflows. 
The higher-spin case attains a larger $b_{\rm rms}$ possibly via the following chain of processes. The black-hole spin strengthens the magnetized wind, which extracts angular momentum vertically. This enhanced angular-momentum loss increases the inflow speed, thereby affecting how much magnetic flux is advected inward and compressed at each radius through flux freezing. 
The lower-spin case experiences the same processes, but weaker vertical torque keeps the advection smaller and thus $b_{\rm rms}$ saturates at a lower level. Note that the argument focuses on the effect of advection/compression rather than the dynamo action, which is not expected to be affected by the spin of the blackhole. 

Since we started out with an initial condition that consisted fully of only poloidal fields, the differential rotation in the disk should naturally lead to the production of toroidal or azimuthal fields, also known as the $\Omega$-effect in the dynamo literature. In \Fig{fig:B-LS-field}, we show the evolution of rms strength of the large-scale (defined as the field component averaged over the azimuthal direction) toroidal field. It can be seen that large-scale toroidal field grows exponentially from very small values, as expected, due to MRI large-scale dynamo action. 

The large-scale poloidal fields, as shown in \Fig{Br_LS}, themselves grow a bit early on due to the advection and compression of the fields at the inner radii. They then suffer from turbulent dissipation after $t=1000M$ before reaching a saturated stage where there is a balance between growth and decay.

\begin{figure*}
    \centering
    \includegraphics[width=0.8\textwidth]{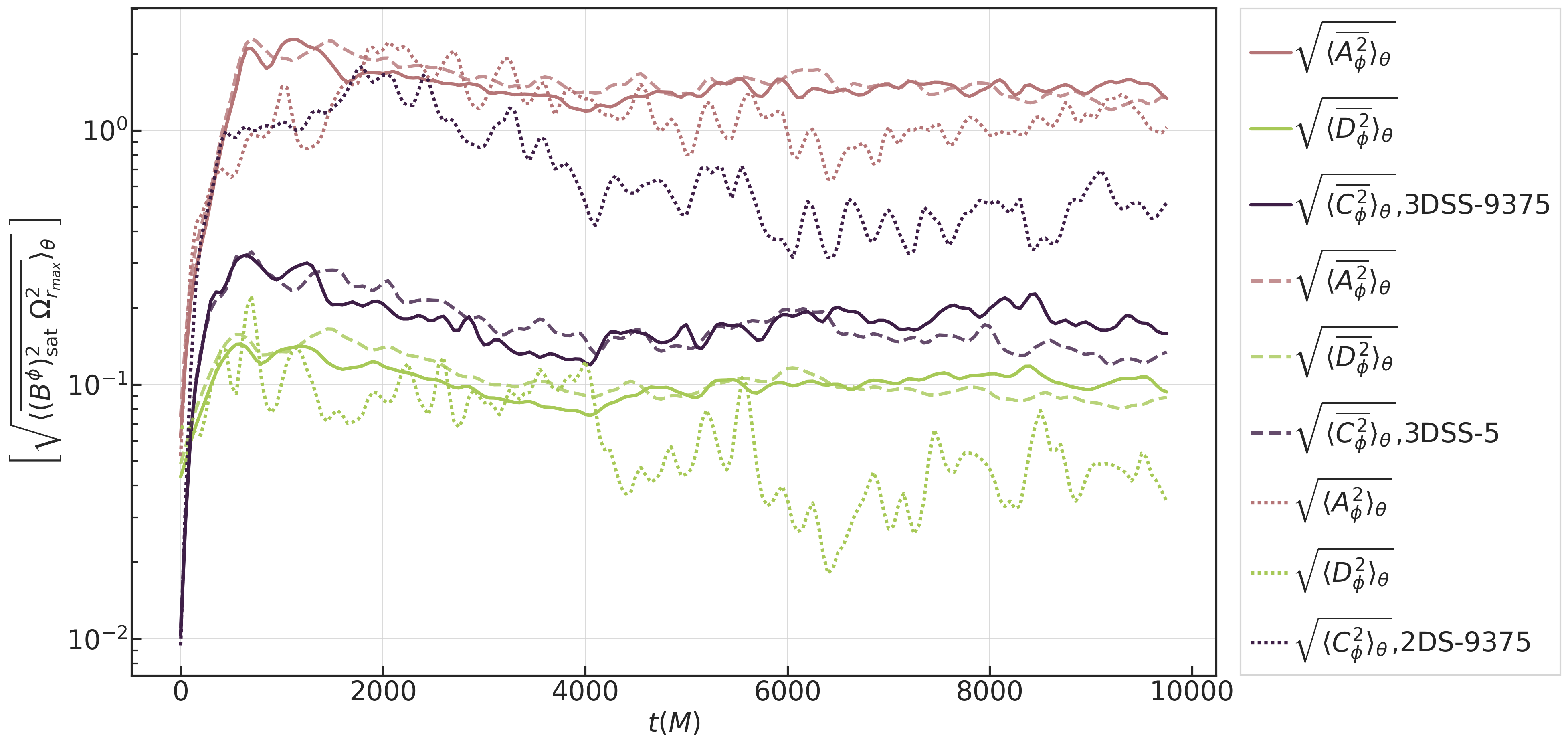}
    \caption{The RMS of compression $\text{C}_{\phi}$, advection $\text{A}_{\phi}$ and dynamo $\text{D}_{\phi}$ terms, normalised by  $\sqrt{\langle\overline{{({B^{\phi}})^2_{\rm sat}\ \Omega_{r_{max}}^2}}\rangle _{\theta}}$ calculated at $r=r_\mathrm{max}$ as in \Eq{ind_eqn_Bphi} for the simulations - 3DSS-9375, 3DSS-5 and 2DS-9375 as shown in solid, dashed and dotted lines respectively. While the advection term dominates overall across these simulations, the dynamo term (and the other two terms) in 2DS-9375 drops after $t\sim4000M$ coinciding with emergence of stronger turbulence. Correspondingly, we see the butterfly pattern in $B^{\phi}$ becoming more stochastic in the top panel of \Fig{fig:butterfly}. }
    \label{fig:TermCompare}
\end{figure*}

\begin{figure*}
    \centering
    \includegraphics[width=0.7\textwidth]{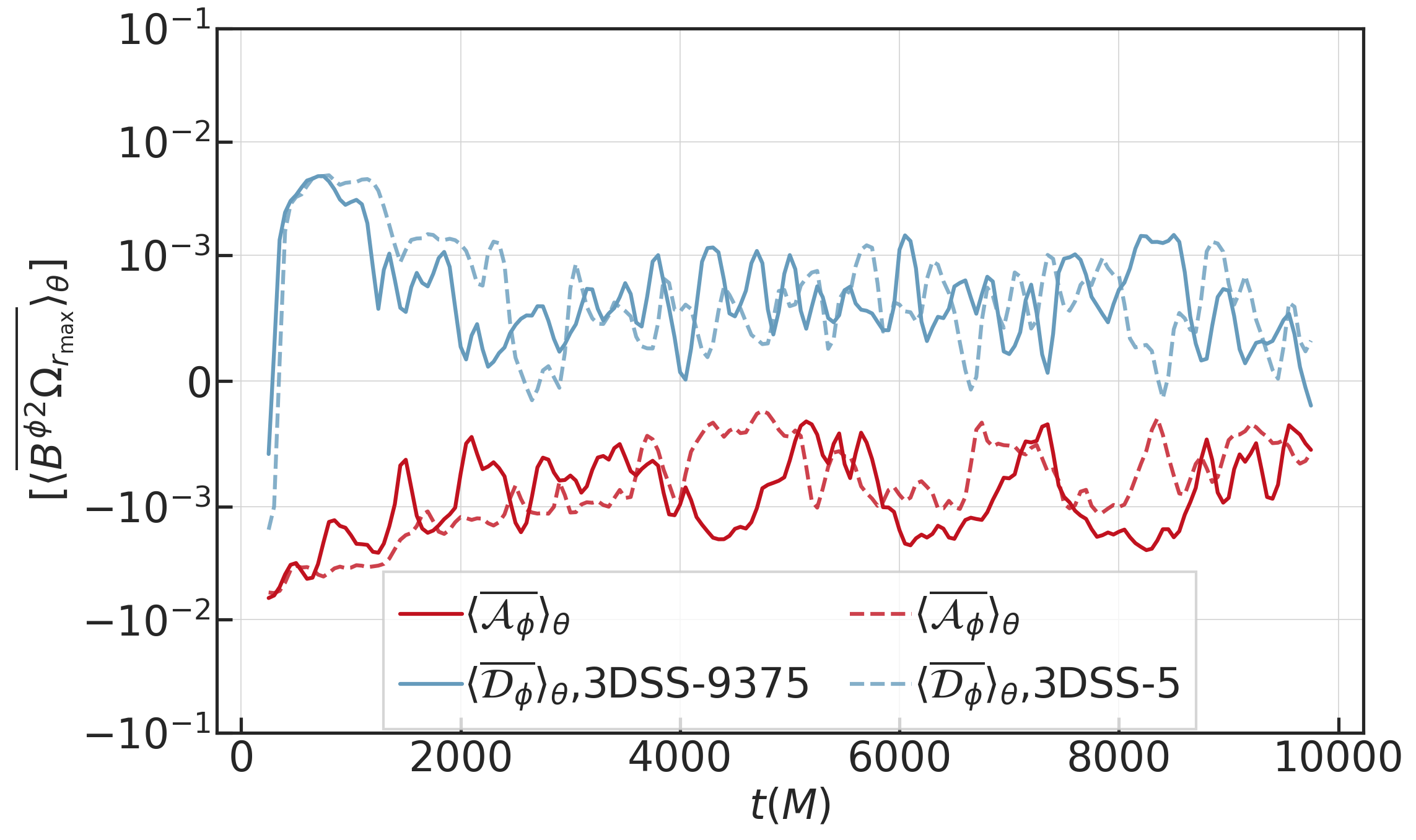}
    \caption{The dynamo term $\mathcal{D}^{\phi}$ and the advection term $\mathcal{A}^{\phi}$, each normalised by $\langle\overline{{{B^{\phi}}^2\Omega_{r_\mathrm{max}}}}\rangle _{\theta}$, represent contributions to the toroidal-field energy budget averaged over $\theta$ in the disk region at $r = r_{\rm max}$, and are computed following \Eq{energy_budget}. Results for the 3DSS-9375 and 3DSS-5 runs are shown by the solid and dashed lines, respectively. The two terms behave similarly in the high- and low-spin cases, with the dynamo and advection contributions largely balancing one another.}
    \label{fig:energy_budget}
\end{figure*}

The large-scale component of the flow i.e. the differential rotation can lead to the possibility of \textit {dynamo waves}, an oscillatory, spatially‐propagating pattern of large‐scale magnetic field.
In \Fig{fig:butterfly}, we show space-time plots of evolving large-scale magnetic field i.e. the contravariant component, $\overline{B^{\phi}}$ (or just $B^{\phi}$ in 2.5D case) as a function of the polar angle $\theta$ and time $t$ in the two runs: 3DSS-9375 (lower panel) and 2DS-9375 (upper panel) 
Unlike in 2.5D, we see that $\overline{B^{\phi}}$ component flips polarity and exhibits a quasi-cyclic pattern in time in 3D simulations. These cycle-patterns are a robust indicator of  the presence of large-scale 
dynamo in the disk \citep{Axel1995}.
The butterfly pattern in \Fig{fig:butterfly} are from the dynamo generated magnetic field at $r=r_\mathrm{max}$. 
In the 2.5D SANE run, we do observe some dynamo activity, despite the absence of any dynamo waves. Yet, anti-dynamo theorems imply that large-scale field amplification in an effectively two-dimensional flow can only be temporary and cannot lead to a long-lived field \citep{shukurovkandubook}. The conspicuous lack of a dynamo cycle is therefore a clear signature of the missing third dimension, notwithstanding the presence of strong large-scale shear. The dynamo wave has a period that lasts about ten orbits (at $\rmax$), similar to what has been previously reported \citep{ONeill2010_global_dynamo_cycles, Dhang2024_MRI_dynamo}. 
We find that the dynamo cycles in our simulations are more robust as compared to \citet{jacquemin-ide_magnetorotational_2024}, where they likely had interference from the strong initial field. In \citet{hogg_influence_2018}, they reported increasingly stochastic butterfly patterns with increasing $H/R$. This may, again, be a result of the interference from the remnant initial magnetic field (which becomes larger with increasing $H/R$ and thus more difficult to erase) in thicker disks.

Next, we compare the dynamo action between our simulations. 
To do so, we examine the terms in the induction equation. 
We consider the induction equation in the 3+1 spacetime formulation,
\begin{equation}
    \frac{\partial\bm{B}}{\partial t} = \boldsymbol{\nabla} \times (\alpha \bm{v} \times \bm{B} -\bm{\beta} \times \bm{B}).
\label{eq:ind}
\end{equation}
Here, $\alpha$ is the lapse function and $\bm{\beta}$ is the shift vector as in \citet{Porth_2017}. $\boldsymbol{\nabla}$ denotes the proper covariant derivative, with the determinant of the spatial metric ${\gamma}$ included for appropiate weightage.
Upon expanding and writing this equation for the toroidal component of the magnetic field, we have,
\begin{align}
\frac{\partial B^{\phi}}{\partial t} 
&= V^{\phi} \nabla_{i} B^i
   - B^{\phi} \nabla_{i}V^i 
   - V^i \nabla_i B^{\phi}
   + B^i \nabla_i V^{\phi} \nonumber \\
&= \text{Div}(B) + \text{C}_{\phi} + \text{A}_{\phi} + \text{D}_{\phi}.
\label{ind_eqn_Bphi}
\end{align}
Here $V^i = \alpha v^i - \beta^i$.
The first term in the expansion contains the divergence of magnetic field, which is zero. The other three terms are the compression $\text{C}_{\phi}$, the advection $\text{A}_{\phi}$ and the dynamo $\text{D}_{\phi}$ terms respectively. 
We verify that the LHS of \Eq{ind_eqn_Bphi} indeed matches with the RHS, both being computed from simulation data, and discuss it in Appendix~\ref{sec:ind_eqn_match}.

We inspect the individual terms on the RHS of \Eq{ind_eqn_Bphi}, and show their rms values calculated at $r=r_\mathrm{max}$ for all three runs: 3DSS-9375, 3DSS-9375 and 2DS-9375 in \Fig{fig:TermCompare}. Across all three runs, the advection term consistently dominates, maintaining nearly the same amplitude in each case. As expected, the dynamo term is larger in the 3D simulations—by roughly a factor of two at late times—and remains relatively steady, indicative of a robust, self-sustaining turbulent dynamo that continuously regenerates the field. In the 2.5D run, the dynamo term is initially comparable to that in the sub-SANE 3D cases up to 
$t\sim4000M$, owing to the presence of a strong $\omega$-effect even in axisymmetry.  However, after $t\sim4000M$, turbulent diffusion outweighs coherent amplification, causing the dynamo term to decline.
Interestingly, in the 2.5D run the compression term is actually much larger and nearly comparable to the advection term and thus, can mimic dynamo-like growth of fields. The compression term in the 3D cases is only slightly larger than the dynamo term, underlining its role in such global simulations. 
Also, the dynamo terms for lower spin are similar to the higher spin case, consistent with the expectation that it should be independent of the BH spin.

Overall, it is clear that advection is an important process governing the evolution of the fields in such disks. However, the dynamo and compression terms convert kinetic energy to magnetic energy and are important for the long term evolution of fields in, both, the disk and at the horizon. We have been unable to run the simulations for too long but it is expected that the magnetic flux at the horizon will increase due to the advection of disk dynamo generated fields. This was already seen in previous GRMHD dynamo simulations \citep{liska_large-scale_2020,jacquemin-ide_magnetorotational_2024}. 

Next we study, in particular, the evolution of large-scale fields.
We provide the magnetic energy equation for the  large-scale component of the toroidal field at $r=r_\mathrm{max}$,
\begin{equation}\label{energy_budget_full}
   \frac{1}{2} \frac{\partial\langle{{\overline{B^{\phi}}}^2}\rangle_{\theta}}{\partial t} =\langle\overline{\mathcal{D}^{\phi}}\rangle_{\theta} + \langle\overline{\mathcal{A}^{\phi}}\rangle_{\theta},
\end{equation}
where 
\begin{equation} \label{energy_budget}
   \begin{split}
        \langle{\overline{\mathcal{D}^{\phi}}}\rangle_{\theta}
        &= \langle\,\overline{\boldsymbol{{B}^{\phi}}}\cdot 
        \overline{\boldsymbol{\nabla} \times \,(\boldsymbol{V^{\phi}}\times \boldsymbol{B^{\mathrm{pol}})}}\,\rangle_{\theta}, \\
        \langle{\overline{\mathcal{A}^{\phi}}}\rangle_{\theta}
        &= \langle\,\overline{\boldsymbol{B^{\phi}} }\cdot 
        \overline{\boldsymbol{\nabla} \times \,(\boldsymbol{V^{\mathrm{pol}}} \times \boldsymbol{B^{\phi})}}\,\rangle_{\theta},
   \end{split}
\end{equation}
are the large-scale dynamo and advection terms respectively. Here, $\boldsymbol{ B^{\rm{pol}}}=(B^r,B^{\theta})$. Note that all the components of both magnetic and velocity fields involved in \Eqs{ind_eqn_Bphi}{energy_budget_full} are the contravariant components.

In \Fig{fig:energy_budget}, we have shown the terms specified in \Eq{energy_budget}, for the two main sub-SANE  runs. Note that these terms contribute to the evolution of the mean field or the large-scale field but we have not separated the contributions from mean quantities and from mean of the correlations in fluctuating quantities i.e. the electromotive force (EMF), $\overline{\bm{V'}\times \bm{B'}}$ (here the total field $\bm{Q}=\bar{\bm{Q}}+\bm{Q'}$). Thus, $\overline{\mathcal{D^{\phi}}}$ contains both the stretching due to large--scale shear ($\Omega$ effect) and the EMF. 
In \Fig{fig:energy_budget}, we find that the dynamo term, $\langle{\overline{\mathcal{D^{\phi}}}}\rangle_{\theta}$, is responsible for growing the large-scale field at that radius whereas the advection causes its depletion. 
Here, since the LHS of \Eq{energy_budget_full} oscillates around zero in steady state, the two terms approximately balance out, as we see in \Fig{fig:energy_budget}. 
Finally, the amplitude of these terms are independent of the spin of the BH. This calculation establishes beyond doubt the existence of the large-scale dynamos in the 3D sub-SANE runs.

\subsection{Dynamo-Jet connections}

\begin{figure}[!htbp]
    \includegraphics[width=\linewidth]{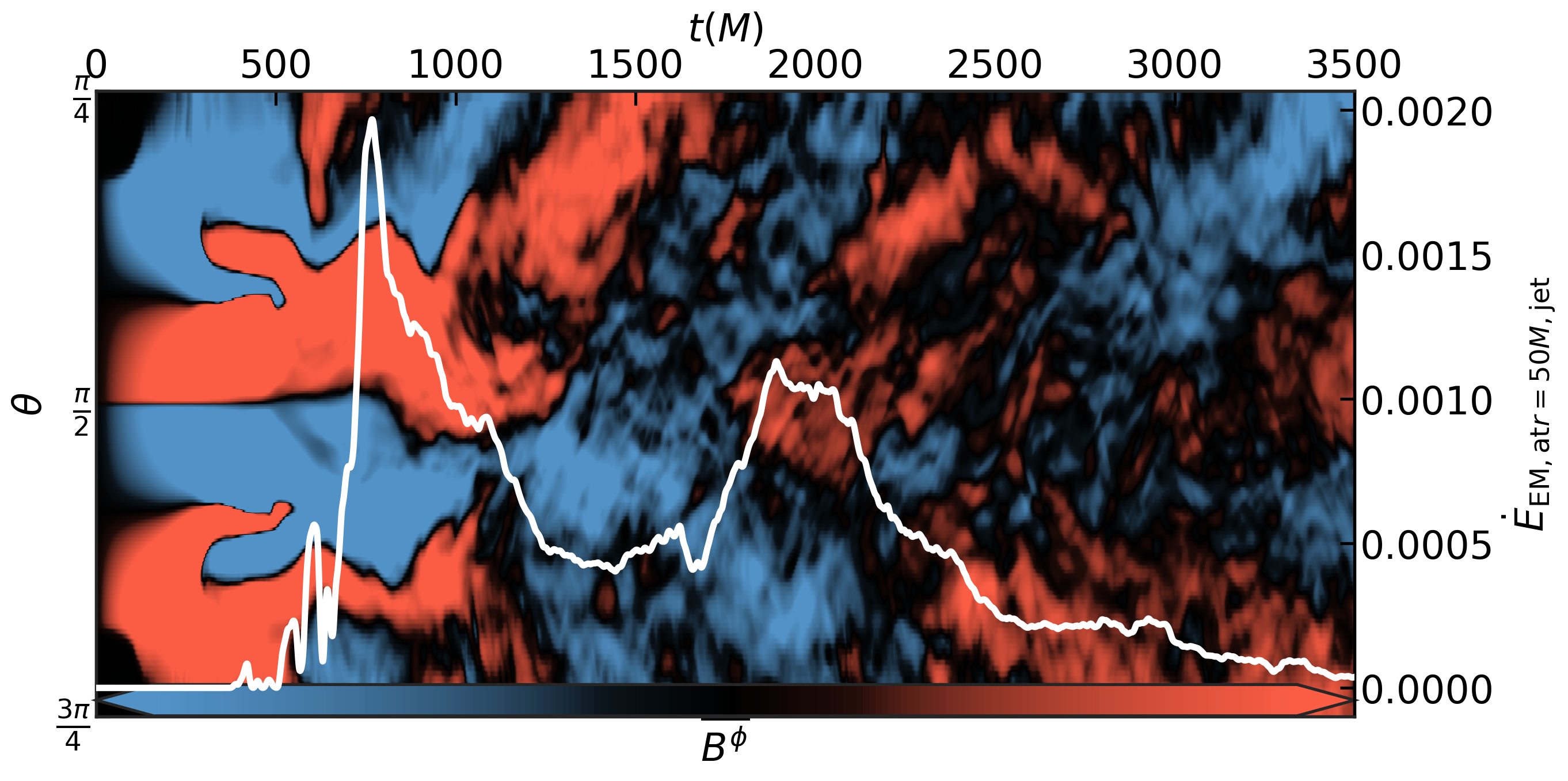}
\caption{The contour shows the butterfly pattern for the large-scale azimuthal field $\overline{B^{\phi}}$ as a function of $(\theta,t)$ for 3DSS-9375. The white solid line shows the Poynting flux in the jet region at $r=50 M$. The peaks in this flux match in time with the first two patterns of wings in the butterfly diagram indicating that the jet is reflecting the dynamo wave.}
\label{fig:overlap} 
\end{figure}

\begin{figure}[!htbp]
\includegraphics[width=0.95\linewidth]{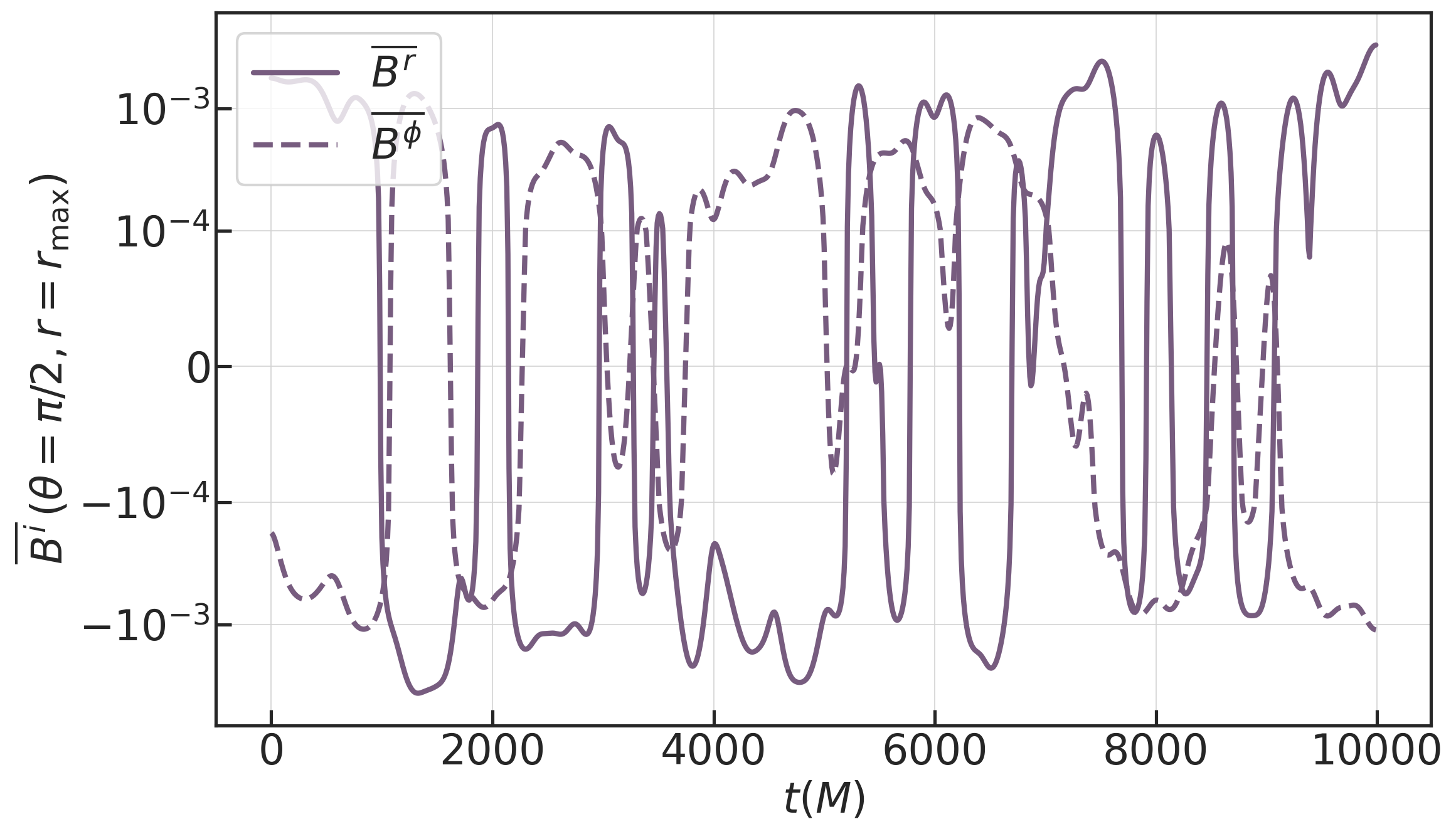}
\includegraphics[width=0.95\linewidth]{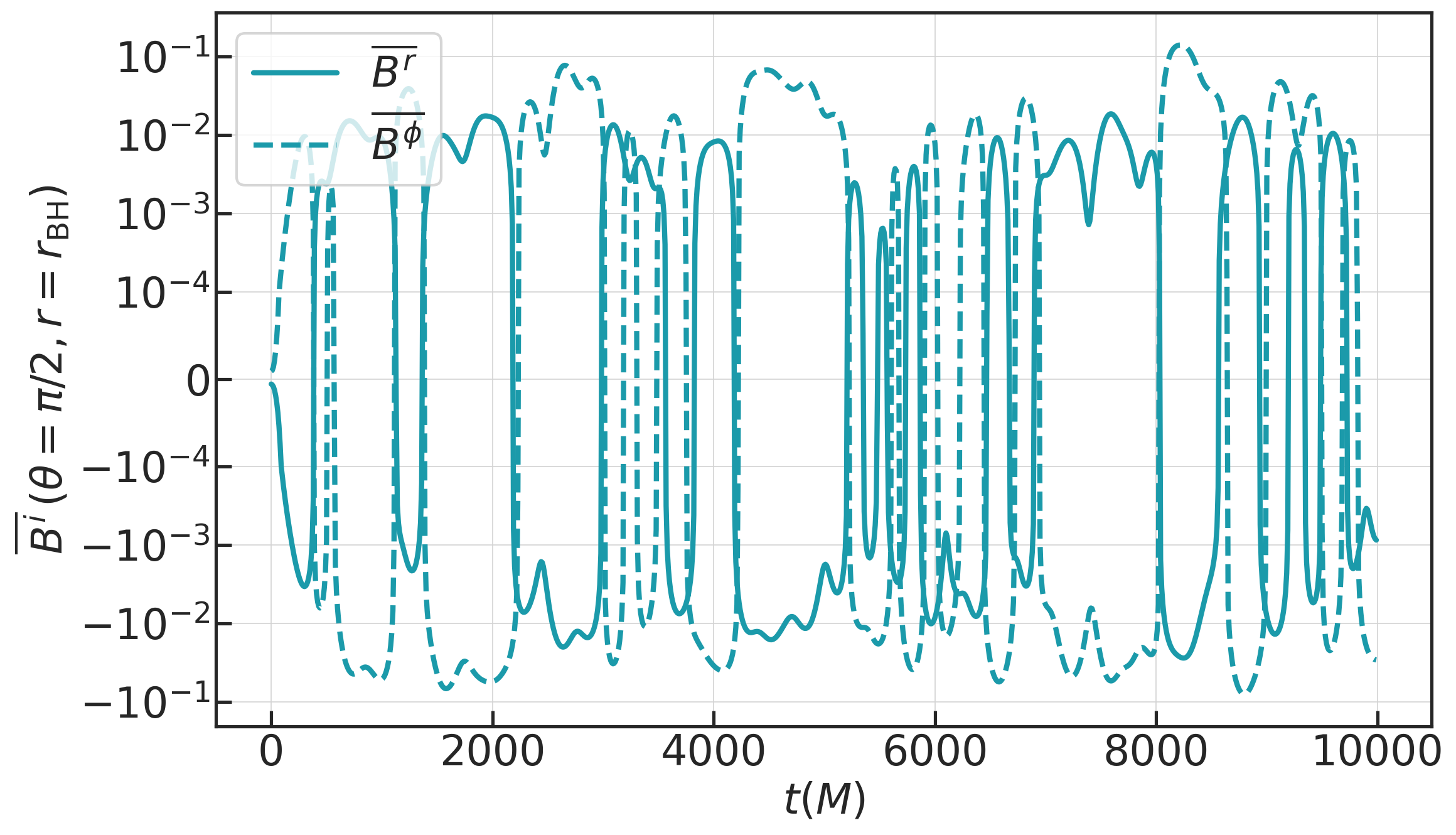}
\caption{Evolution of large-scale field components $\overline{B^r}$ and $\overline{B^{\phi}}$ in the mid-plane $\theta=\pi/2$ for 3DSS-9375 indicated by solid and dashed lines respectively. The top panel shows these components at r = $r_{max}$ which are anti-correlated -- a typical signature of large--scale dynamo. The bottom panel shows these components at the horizon where the anti-correlated behaviour is still retained.}
\label{fig:Bdiffr}
\end{figure}

\begin{figure*}
    \centering
    \includegraphics[width=0.95\textwidth]{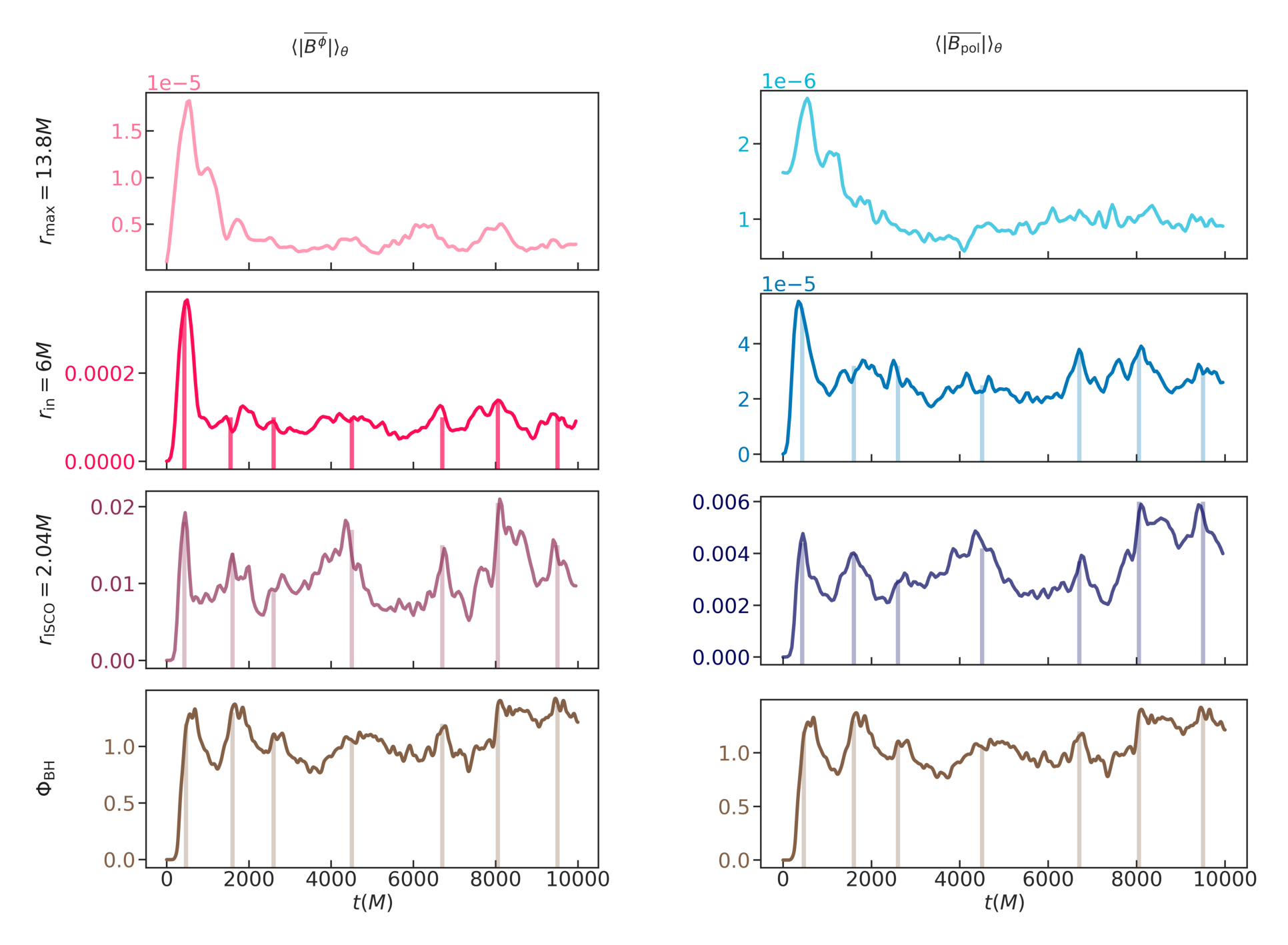}
    \caption{Large-scale fields $|\overline{B}^{\phi}|$ (left) and $|\overline{B}_{\rm{pol}}|$ (right) plotted as a function of time at different radii, at $r$= $r_{\rm{max}}$, $r_{\rm{in}}$, $r_{\rm{isco}}$ as we move downward. The bottom-most panel shows $\Phi_{\rm{BH}}$. Vertical lines mark the peaks observed in $\Phi_{\rm{BH}}$. These times are shown in every panel to enable tracking of the corresponding features across radii. These features in the curves don't change as we move inwards from $r_{\rm{max}}$ to the horizon, showing how as large-scale fields are advected, their structure is preserved. Also, the rightward shift in the peak is quite small showing the time delay in the inward transport of the magnetic field is negligible.}
    \label{fig:compPhiBH}
\end{figure*}

Here, we examine the link between the dynamo in the disk and the jet emanating from the BH. In \Fig{fig:overlap}, we show the butterfly diagram at $r=\rmax$ against the electromagnetic Poynting flux $\dot E _ {\rm{EM}}$ from the BH at the distance of $r=50M$ in the vertical direction. We find that the peaks in this flux emanation match in time with  the first two patterns of wings in the butterfly diagram. Note that at the time of occurrence of the peak, the pattern has blue in the wings above equator representing negative field. This corresponds to positive field in $\overline{B^r}$ aligning with the jet polarity (not shown here). However, the flux dwindles thereafter indicating near shutdown of the relativistic jet. Nonetheless, it is interesting that the jet emanation reflected the dynamo waves for the first two cycles. This suggests a direct connection between the magnetic fields at the horizon and those generated due to the dynamo. 

In the upper panel of \Fig{fig:Bdiffr}, 
we show the two large-scale magnetic field components, 
$\overline{B^r}$ and $\overline{B^{\phi}}$, from the mid-plane at $r=\rmax$. We see that the both $\overline{B^r}$ and $\overline{B^{\phi}}$ are out of phase with each other, and that $\overline{B^{\phi}}$ follows $-\overline{B^r}$ with a small lag. This is the classic dynamo signature of large-scale toroidal field $\overline{B^{\phi}}$ being generated from the large-scale poloidal field via the large-scale shear due to the disk differential rotation~\citep{Axel1995}. 
These quasi-periodic peaks are due to the dynamo cycle pattern. 
Remarkably, these peaks remain evident in the large-scale field components even at the horizon, as shown in the lower panel of \Fig{fig:Bdiffr}. Note, however, that these curves represent large-scale fields evaluated at the mid-plane, whereas the magnetic flux at the horizon, $\phibh$, which powers the jets, is obtained from an additional average over $\theta$. Thus, next, we study  quantities with $\theta$ averaging as well.

\begin{figure}[!htbp]
    \centering
    \includegraphics[width=\linewidth]{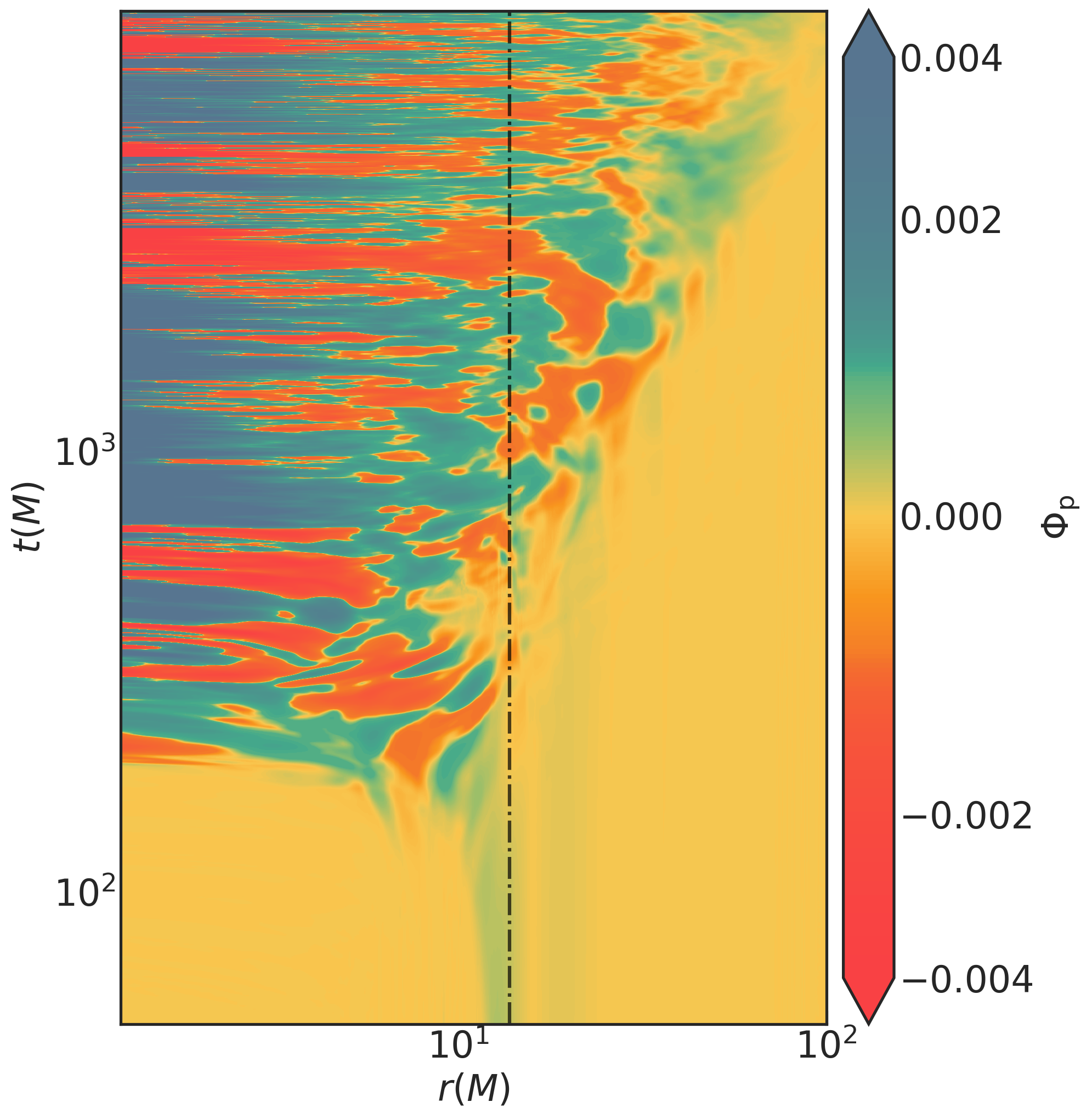}
    \caption{$\Phi_{\rm{p}}$ as in \Eq{phi_p} plotted as a function of $(r,t)$. Structures in the large-scale $B^r$ generated around $r_{\max}$ denoted by the dash-dot line, are quickly advected to the horizon, while also growing stronger due to compression. The slanted line indicates the expansion of turbulence to larger radii and thus enabling dynamo action progressively farther out.}
    \label{fig:adv-map}
\end{figure}

\begin{figure}[!htbp]
\includegraphics[width=\linewidth]{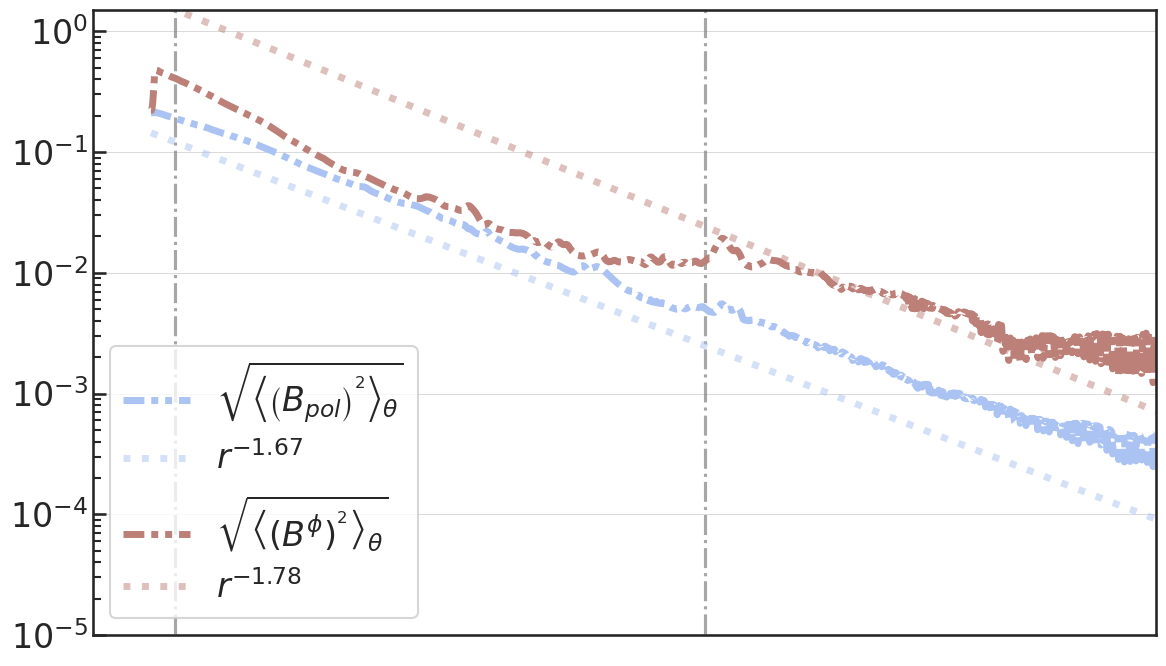}
\includegraphics[width=\linewidth]{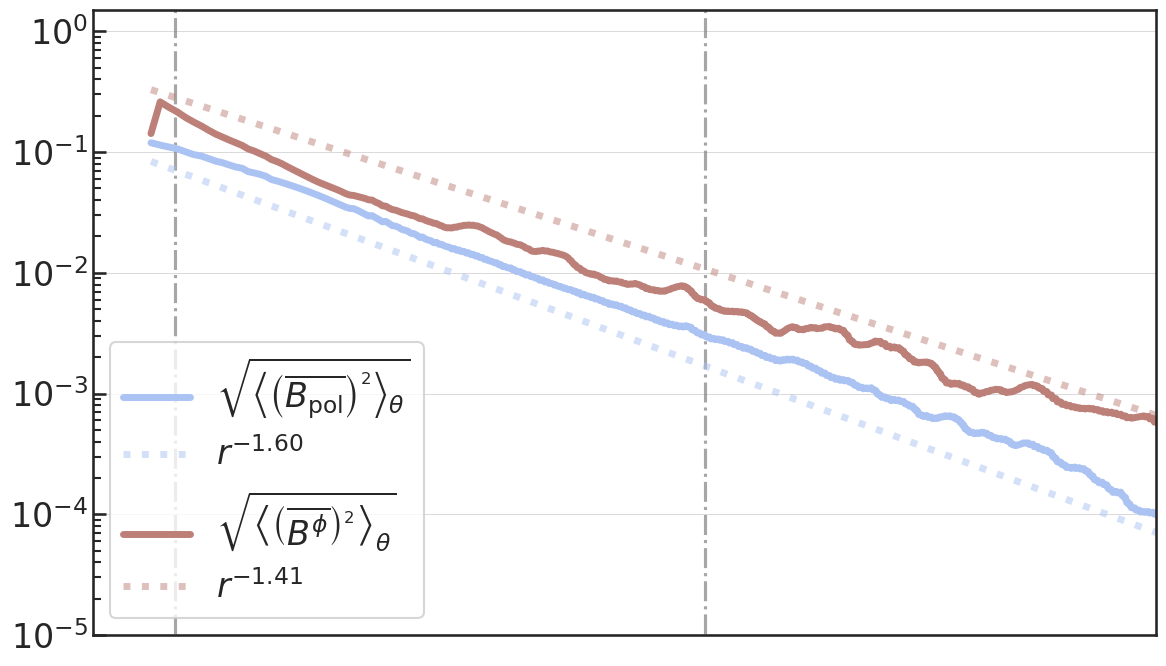}
\includegraphics[width=1.03\linewidth]{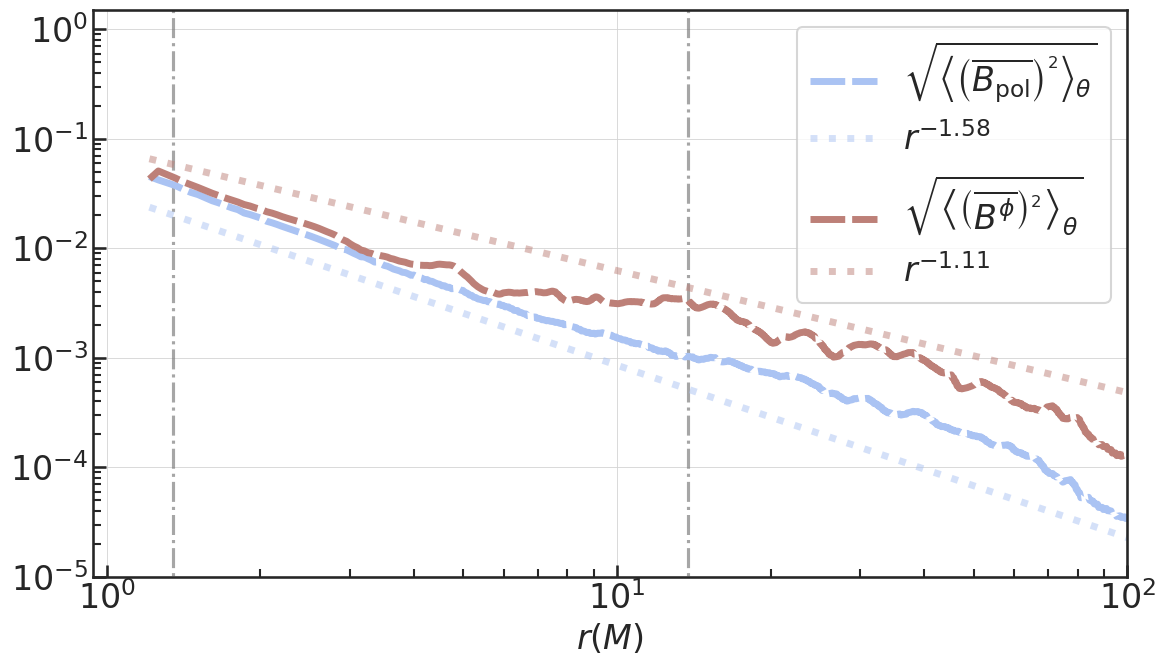}
\caption{Large-scale fields, $\overline{B}_{\rm{pol}}$ and $\overline{B}^{\phi}$ (averaged over $\theta$) vs radius, time averaged between t=[9990,9999]M. The upper, middle and lower panels are for 2DS-9375, 3DSS-9375 and 3DSS-5 respectively. The fits are made only by considering the slopes of the curves in the extent $r=[2,50]M$. While the 2DS-9375 case shows $\sim r^{-2}$ scaling, the other two cases show shallower scalings, likely owing to the dynamo action. }
\label{fig:Brad-prof}
\end{figure}

In \Fig{fig:compPhiBH}, we show the $\theta$-averaged, large--scale field components in the two columns, with each row representing that from different radii, decreasing as we go from top to bottom. It is clearer in the case of $\langle|\overline{B^{\phi}}|\rangle_{\theta}$ that as we go from $\rmax$ closer to the BH, the peaks in time clearly correlate at these different radii. At $\rmax$, the first peak at just less than $t\sim 1000M$ is much larger than the subsequent ones by a factor of two. This difference reduces as we go to smaller radii. And thus the peaks show up more dramatically and match with that in $\phibh$. The peaks in $\langle|\overline{\Bpol}|\rangle_{\theta}$ are much less visible at $\rmax$ again due to the dynamic range being larger. However, the peaks in $\langle|\overline{\Bpol}|\rangle_{\theta}$ as well become increasingly clearer and more dramatic closer to the BH. We interpret that these fields are essentially being advected from $\rmax$ to the horizon, 
which is why the peaks in time are preserved along the radius. 
Note that the lag in peaks appearing at horizon is not much. This is because the advection timescale involved is quite small amounting to about $20M - 40M$ or so.  
Further, the field strength has increased significantly at the horizon as compared to that at $\rmax$ due to the accumulation and compression of the fields close to the horizon. 

This picture that we have developed by the analysis in \Fig{fig:compPhiBH}, is further supported by the \Fig{fig:adv-map}
where we show the contours of the following integrated quantity, the magnetic potential, 
\begin{equation}
    \Phi_{\rm{p}}= \int_{\pi/2 -\delta}^{\pi/2 +\delta} \sqrt{-g}\ \overline{B^r} d\theta,
\label{phi_p}
\end{equation}
where $\delta = \pi/12$. 
Thus, from \Fig{fig:adv-map}, we can visually gauge that the fields generated at $r=\rmax$ are advected to smaller radii, which results in emergence of stronger fields around $t\sim 300M$. This space-time map for our high spin sub-SANE run in \Fig{fig:adv-map} is very similar to that in \citet{jacquemin-ide_magnetorotational_2024} (see Fig.~2). Note that our definition of the magnetic potential in \Eq{phi_p} evaluates the flux around the equatorial region, whereas their formulation computes it over a hemisphere. 
Thus fields that are generated and advected to the BH here appear turbulent and thus show frequent reversals in polarity at any given radius along time. A conspicuous similarity is that our space-time map also shows a slanted line suggestive of ``ejection" which we interpret as the turbulence spreading to larger radii over time and  extending the dynamo action (initially mainly at $\rmax$) to larger radii. However, unlike their results, we do not observe a corresponding dynamical separation of field polarities between the inner and outer disk\footnote{Our preliminary calculations do not show a long-lived feature of reversal in the flow velocity at some radius in the disk.}. Further, we find that the fields generated at these larger radii are then advected inward—which can be gauged by how the field structures curve inward. 

Next, we examine the effects of compression on the field components.
The evolution of magnetic-field components under advection and compression follows directly from flux freezing. For the poloidal field, conservation of vertical magnetic flux, $\Bpol~r^2=const.$ implies that the field strength scales as $\Bpol \propto r^{-2}$. In contrast, the toroidal field, constrained by flux conservation across a meridional surface of area $\sim rH$, where $H$ is the disk scale height. For thick disks, $H \propto r$. This yields $B^{\phi} \propto r^{-2}$ as well.   

In \Fig{fig:Brad-prof}, we show the radial profile of the large-scale magnetic field components from the 2D SANE and 3D sub-SANE runs, shown at late times closer to $t=10,000M$.
The uppermost panel of \Fig{fig:Brad-prof} is for the SANE run. We see a nice power law scaling of the field with radius. Even though the power law indices aren't exactly 2 (instead 1.67 and 1.78 for poloidal and toroidal cases respectively),  they are close to the theoretically expected $r^{-2}$ scaling. 
 
Next, we show the sub-SANE cases in the middle and lower panels of \Fig{fig:Brad-prof} for the higher and lower spin cases respectively. The field components do exhibit a power law fall off with radius, here as well. However, the  power law exponent for the poloidal field is closer to the expectation while the toroidal field deviates from it. The slope for toroidal case is shallower indicating that the robust dynamo in the sub-SANE cases is actually boosting the field a bit over and above that predicted by flux-freezing compression.  This becomes more glaring when we note that at lower radii, the slope is somewhat steeper and similar to that in the poloidal case but becomes shallower at larger radii, coinciding with regions of magnetic field generation. 

A particularly curious feature in the upper and middle panels of \Fig{fig:Brad-prof}, is the small but distinctive increment in $B^{\phi}$, towards the horizon in the large spin runs, where the radial profile departs from the underlying power-law trend. We believe this is due to the frame-dragging which is giving rise to a dynamo-like $\omega$-effect which shears out the poloidal into a toroidal component. Although the effect is modest in amplitude, it is nonetheless quite conspicuous, particularly because an analogous feature is conspicuously absent in the lower–spin case shown in the lower panel of \Fig{fig:Brad-prof}.

\section{Discussions}
\subsection{Initial magnetic fields in GRMHD simulations of accretion disks}

Earlier GRMHD simulations of jet launching have typically employed initially coherent vertical magnetic fields that drive the system toward either SANE or MAD states. In such setups, the large-scale flux is externally supplied, so the magnetic structure at the horizon primarily reflects advection rather than in-situ field generation, which then determines whether the system becomes MAD or SANE and sets the jet efficiency.

\citet{Rodman_2024} showed that even a relatively weak toroidal field ($\beta=200$) can leave a long-lasting imprint on the disk, with the system retaining memory of its initial magnetisation to late times.  This persistence places some tension on the interpretation of \citet{liska_large-scale_2020}, whose coherent poloidal-field growth relied on particularly strong initial magnetisation, even though the dynamo that arises later is robust. 

\citet{White_2020} employed small-scale poloidal loops similar in size to ours; however, their alternating-polarity configuration (unlike our non-alternating one) inhibits reconnection and thereby preserves the memory of the initial magnetic topology. Their long-duration GRMHD simulations of radiatively inefficient disks showed that the flow structure remained sensitive to initial conditions, with polar inflow and outflow patterns dictated by the seed field geometry. Despite extended evolution, no self-similar steady state emerged, suggesting that advection-dominated flows can retain magnetic memory over tens of thousands of dynamical times. Similarly, in the simulations of \citet{Dhang_2023}, the MRI-driven dynamo remained largely perfunctory: their oppositely oriented large-scale loops were again unfavorable for reconnections, delaying the onset of the large-scale dynamo. Even in the pseudo-Newtonian study of \citet{hogg_influence_2018}, a comparable effect likely persists. Their initial fields, not conducive to reconnections, appear to influence the late-time behavior, and the loss of coherent butterfly patterns in disks with larger H/R ratio, could reflect an inability to fully erase this initial magnetic memory.  

In contrast to all of the above studies, in our simulations the initial multiple  small-scale loops of magnetic fields of, importantly, non-aletrnating configuration are conducive to reconnection and thus rapidly lose memory of the initial topology enabling the dynamo to establish itself early. We find robust dynamo cycles or butterfly patterns as well. 

\subsection{Dynamo-Jet missing link}

This work bridges a missing link between MRI-driven turbulence, dynamo-generated fields, and jet evolution — a connection that previous GRMHD studies (e.g., \citep{liska_large-scale_2020,Dhang_2023, jacquemin-ide_magnetorotational_2024}) only  suggested but didn't explore in any detail. 
In this work, we have shown large-scale dynamo waves can modulate the jet power and there is sustained migration of quasi-periodic butterfly peaks towards the horizon and eventually are reflected in the $\Phi_{\rm{BH}}$, establishing a dynamo-jet connection.  

\citet{liska_large-scale_2020} presented an important demonstration that a disk dynamo can, in principle, lead to robust jet formation. In their simulations, however, the dynamo operates only after a long evolution, given the strong initial toroidal field, and relies on a few “lucky” buoyant loops being trapped near the stagnation surface. In contrast, the dynamo in our simulations arises more promptly and self-consistently, without requiring such favorable conditions. Nevertheless, the large radial extent of their disk allows a sustained supply of dynamo-generated magnetic flux to the BH, supporting sustained jets once the MAD state is reached.

\citet{jacquemin-ide_magnetorotational_2024}, a follow-up of \citet{liska_large-scale_2020}, examined the dynamo action in great detail. However, they did not specifically analyze the correlation between the dynamo activity in the disk and the jet properties. Although they reported the presence of a butterfly pattern, somewhat less regular than in our simulations, and noted some bursty jet activity associated with the early merging of magnetic loops near the BH, they did not investigate the direct correspondence between the magnetic-field evolution within the disk and that at the horizon.

In \citet{Dhang_2023}, they reported that the dynamo in their simulation failed to produce sustained flux accumulation at the horizon. In general, they concluded that the dynamo in advection dominated disks are inefficient and thus cannot be relied on for generation of fields that allow for jet formation. However it is not clear that they had the right domain (a wedge instead of a full disk) to recover a robust dynamo. Moreover, the magnetic flux at the horizon might have continued to grow steadily, as seen in \citet{liska_large-scale_2020}, had they not interrupted the run by injecting additional magnetic loops midway through the simulation to mimic external field supply from the ISM. 

We find in our high-spin simulation, a steady buildup of horizon-threading magnetic flux (see \Fig{fig:phiBH}) that can be attributed to sustained dynamo action at $r_\mathrm{max}$ continually supplying the fields that are accreted onto the BH. We have been unable to run the simulation for very long timescales, but we expect this continuous supply of fields can, in principle, lead to the required fields for re-launching the jets, as seen in \citet{jacquemin-ide_magnetorotational_2024}.  

\subsection{Dynamo action and its relevance}

Multiple diagnostics have indicated the presence of a robust dynamo in our simulations. The cyclic reversals of $B^\phi$ with a period of roughly ten orbits at $r_{\rm max}$ (Fig.~\ref{fig:butterfly}) confirm sustained large-scale dynamo activity. We find a well-defined butterfly pattern even for a disk with $H/R \approx 0.3$ (in the run with $a=0.9375$), whereas the butterfly pattern in \citet{jacquemin-ide_magnetorotational_2024, Rodman_2024} appears fairly irregular even with a lower value of $H/R \approx 0.2$. This is contrasting with the results of \citet{hogg_influence_2018}, who reported that the dynamo becomes increasingly inefficient and stochastic as $H/R$ increases. As discussed earlier, that trend may reflect the lingering influence of un-erased initial magnetic topology on the dynamo rather than an intrinsic suppression in thicker disks. 

A complementary indication of dynamo action comes from the decomposition of the induction equation (Eq.~\ref{eq:ind}; Fig.~\ref{fig:TermCompare}), which explicitly shows the presence of a stronger dynamo term in the 3D sub-SANE runs, in contrast to the 2D SANE case. In all cases, however, advection remains the dominant contributor, leading to the following picture: the dynamo operates initially near $r_{\rm max}$, where it generates coherent toroidal and poloidal fields, while advection transports this flux inward, producing the observed growth of $\Phi_\mathrm{BH}$. With time, the region of active dynamo expands to larger radii as advection continues to feed the horizon.  

Another signature of active field generation appears in the radial magnetic profiles: we find $B^{\phi} \propto r^{-n}$ with $n < 2$, indicating a shallower-than-flux-freezing scaling. This behavior suggests that dynamo amplification at and beyond $r_{\rm max}$ contributes to maintaining the field strength at large radii. 

\citet{jacquemin-ide_magnetorotational_2024} suggested that the maximum magnetic flux generated in a disk is determined by its radial extent and geometric thickness. We differ in our opinion, that if one efficiently erases memory of its initial field configuration—the saturation value of $\Phi_\mathrm{BH}$ and the timescale to reach it should asymptotically become independent of disk size. In other words, once a self-sustained dynamo is fully developed, the magnetic flux accumulated at the horizon is regulated more by the intrinsic turbulence than by the global dimensions of the disk. 

A detailed quantitative analysis of the dynamo operation will be presented in a future paper.

\subsection{Jet weakening due to incoherence of horizon magnetic fields}

A similar theme has been explored in the work of \citet{Chashkina2021} previously. 
A key distinction from their work lies in the origin and evolution of the magnetic fields driving jet modulation. In their multiloop setup, oppositely magnetized loops are embedded in the torus from the outset and advected sequentially to the horizon, each reversal producing quasi-striped jets. 
The flux is therefore externally supplied and its polarity changes prescribed by the initial conditions. This kind of radial modulation in the initial condition can lead to intermittent suppression of jets due to reconnections in the jet boundaries. In contrast, as mentioned before, our initial condition is such that the disk loses its magnetic memory rapidly leading to very early onset of dynamo and thus horizon fields are those provided by the dynamo action in the disk. The resulting quasi-periodic nature of $\Phi_\mathrm{BH}$ emerges naturally from the dynamo cycle. The weakening of the jet between two dynamo cycles reflects inward advection of oppositely directed poloidal fields; however, the limited spatial extent of our system prevents sufficient flux accumulation for recovery (after the second dynamo wave), leading instead to a complete jet shutdown.

Importantly, we have demonstrated that the decay of jet power in sub-SANE disks is not simply a consequence of reduced flux magnitude (or reduced $\eta_\mathrm{BH}$) but rather of declining field coherence. The coherence parameter, $\mathcal{C}_{\rm {BH}}$ (\Eq{ratio_BH}; \Fig{fig:ratio-signed}), drops below $\sim 0.6$ once the poloidal fields at the horizon become topologically disordered, coincident with the collapse of magnetisation and Lorentz factor in the jet. 

\subsection{Observational implications}
A number of quasi-periodic behaviours observed across accreting BHs such as: the heartbeat-like Q-cycles in XRBs \citep{Belloni2000, Altamirano2011}, the optical/UV damping times in AGN \citep{MacLeod2010, Burke2021}, and the flaring and polarization-swing episodes in jetted quasars \citep{Marscher2008, Blinov2015}, often involve clear changes in jet or outflow activity. In XRBs, specific variability states are associated with discrete radio flares and transient jet ejections \citep{FenderBelloni2004, Neilsen2011}, while in AGN, large optical/$\gamma$-ray flares and EVPA rotations frequently coincide with structural changes in the relativistic jet \citep{Chatterjee2009, Hovatta2014}. These consistent disk–jet correlations suggest that jets may be regulated by a shared, quasi-periodic driver originating in the accretion flow.

Our simulations point to the large-scale disk dynamo as the underlying driver. The dynamo generates coherent magnetic cycles that modulate the poloidal flux reaching the BH, producing corresponding variations in jet power. In this view, the jet effectively traces the disk’s magnetic state and inherits its dynamo-driven variability \citep{ZhouLai2024}. Since large-scale dynamo periods are typically 10–50 orbital times, the associated physical timescales naturally overlap with those of above mentioned quasi-periodic phenomena, suggesting that a disk dynamo may offer a unifying explanation for them.

\section{Conclusions}

Our simulations demonstrate that when GRMHD accretion flows are initialized with small-scale magnetic fields with multiple loops of same polarity, they rapidly lose memory of their initial topology and speedily give rise to a robust large-scale dynamo. Our identification of a robust large-scale dynamo rests on analyzing the decomposition of the induction equation into advection, compression, and dynamo terms; measurement of their rms contributions at $r_{\rm max}$ ; and construction of butterfly diagrams that reveal clear, quasi-periodic field reversals. 

To understand the origin of the magnetic structures delivered to the BH, we studied flux transport across the disk using space-time maps of the poloidal flux, large-scale field amplitudes at multiple radii, and their radial profiles. These calculations show that the dynamo-generated fields at $r_{\rm max}$ are efficiently advected inward with minimal lag, imprinting their temporal cycles onto the horizon. The correspondence between cycle-driven peaks in disk magnetic fields and variations in the horizon-threading flux establishes a direct causal link between disk-scale dynamo activity and the magnetic environment of the BH. 

These simulations also highlight a new insight governing jet survival: the coherence of the poloidal field at the event horizon. While the BZ process requires sufficiently magnetic flux and low accretion rate, our results show that these are not predictive of jet persistence. Instead, jet shutdown in sub-SANE runs occurs when the poloidal field at the horizon becomes too disordered, as quantified by a decline in the coherence parameter, even though the total flux and accretion rate may remain comparable to that in SANE disks. Overall, our study establishes that self-consistent disk dynamos can imprint their cyclic signatures onto BH magnetospheres and jets, but sufficiently coherent magnetic structure is essential for maintaining relativistic outflows.

We thank Samik Mitra for useful comments on the manuscript. We thank Prasun Dhang and Oliver Porth for helpful discussions in the early stages. Santhiya thanks the ICTS AstroPlasma group, including Vinay Kumar, Chandranathan Anandavijayan, Subham Ghosh, and Muhammed Irshad P for insightful discussions. Santhiya thanks Rajarshi Chattopadhyay, Ankur Barsode, and Ritwick Kumar Ghosh for valuable discussions. We acknowledge support from Rohith V. S. at  Sankhya Sutra Labs for assistance with data transfer.
This research was supported by the Department of Atomic Energy, Government of India, under Project No. RTI4001. All simulations were performed on Rudra cluster from Sankhya-Sutra, Contra and Sonic computing clusters at the International Centre for Theoretical Sciences (ICTS), Tata Institute of Fundamental Research. Santhiya acknowledges the Pushkala and Ramani Travel Fellowship awarded by ICTS for enabling travel to Beijing to present this work in the `International Symposium on Cosmic Magnetic Fields'. 

\appendix
\section{Quality factor}
\begin{figure}[!htbp]
    \centering
    \includegraphics[width=0.5\linewidth]{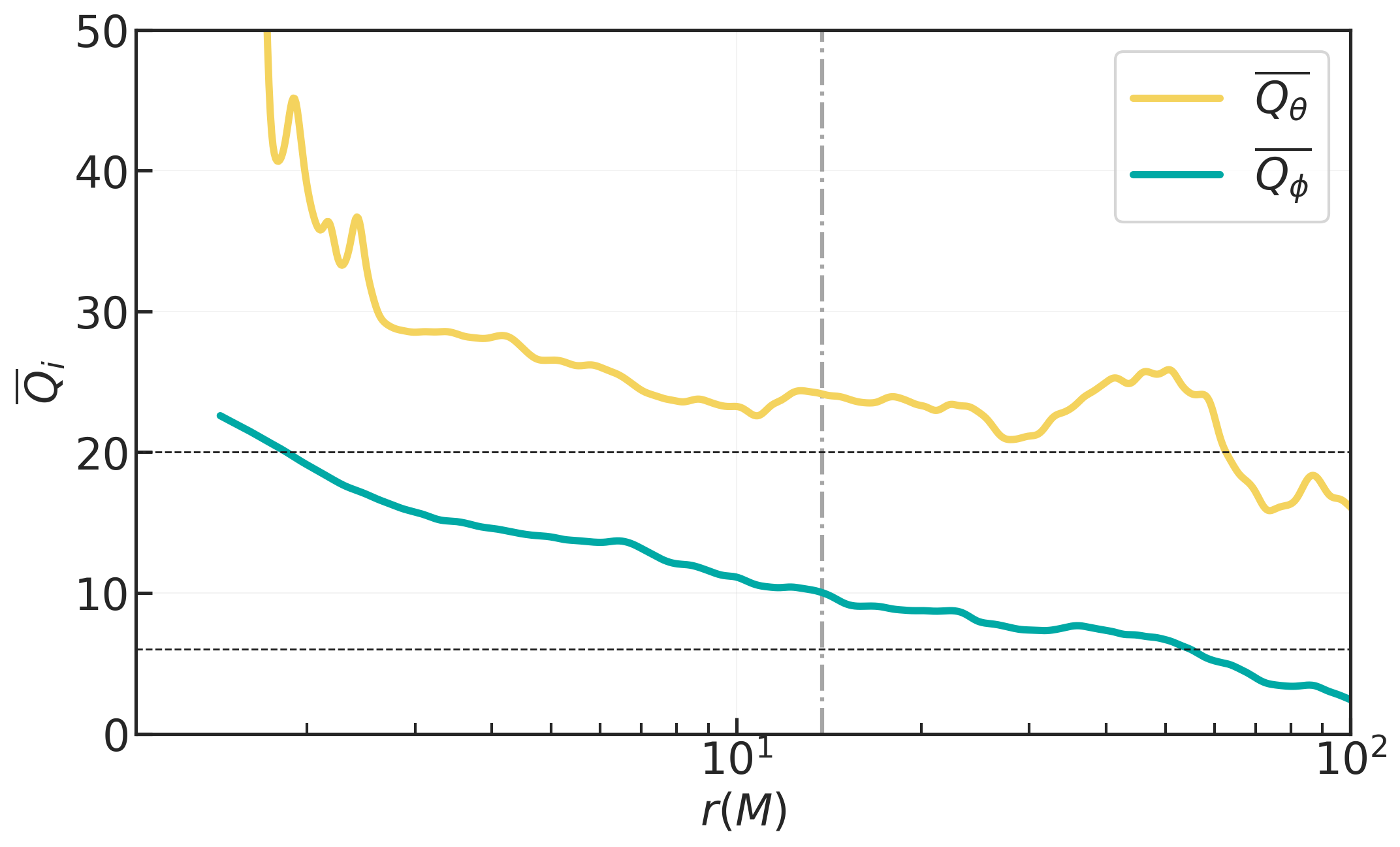}
    \caption{Axisymmetrized quality factors $Q_\theta$ and $Q_{\phi}$ in the mid-plane of the disk, time averaged over the window $t=[3000M,10,000M]$ for the 3DSS-9375 simulation. The horizontal dashed lines mark values 20 and 6. The dash-dotted vertical lines marks $r_{\rm{max}}$ and the product of $Q$ factors are above 200 around $r_{\rm{max}}$.}
    \label{fig:Qfac}
\end{figure}

We perform ideal GRMHD simulations and we ensure that our runs are `sufficiently' resolved by the following criteria. 
It is expected that we have 40-70 cells per disk scale height to capture the fastest growing MRI modes \citep{Hawley_2011, Dhang_2023}. It is also necessary to ensure that we resolve the fastest growing MRI modes in both the vertical ($\theta$) and azimuthal ($\phi$) direction in the disk. For this, the simulations must satisfy the standard quality-factor thresholds, $Q_\theta\ \geq 6$ and $Q_\phi \geq 20$ to achieve \textit{`numerical convergence'} for a given resolution.
The quality factors are defined as follows in the GRMHD simulations \citep{Porth_2019}:
\begin{equation}
\begin{split}
Q_\theta = \frac{2\pi}{\sqrt{(\rho h + b^2)}\,\Omega\,\Delta x_{\theta}}\, b^\mu e^{(\theta)}_{\mu},\\
Q_\phi = \frac{2\pi}{\sqrt{(\rho h + b^2)}\,\Omega\,\Delta x_{\phi}}\, b^\mu e^{(\phi)}_{\mu}
\end{split}
\end{equation}
We see that our simulations satisfy this criteria by examining the quality factors averaged over time $t=3000 M$ to $10000 M$. From \Fig{fig:Qfac}, we see that $\overline{Q_\theta} \times \overline{Q_{\phi}} \geq 200 $  upto $r=60M$ and the disk averaging in our analyses is upto $r=50M$ for all simulations. We also have $\sim 70$ cells per disc scale height, particularly around $r=r_{\rm{max}}$.

\section{Blandford-Znajek efficiency of Jets}
\label{bzmatch}
\begin{figure}[!tbp]
\centering
\gridline{
\includegraphics[width=0.5\linewidth]{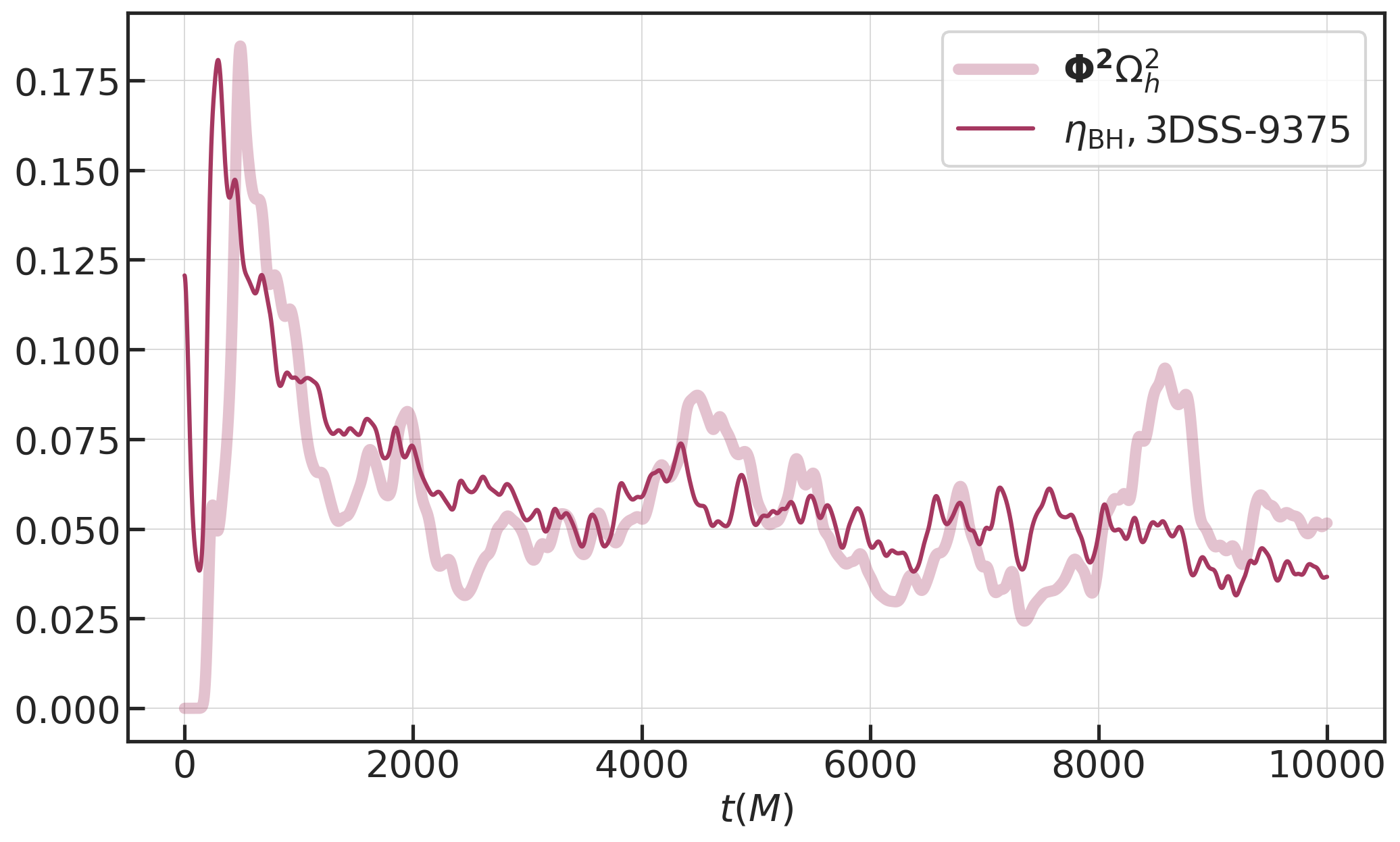}
\includegraphics[width=0.5\linewidth]
{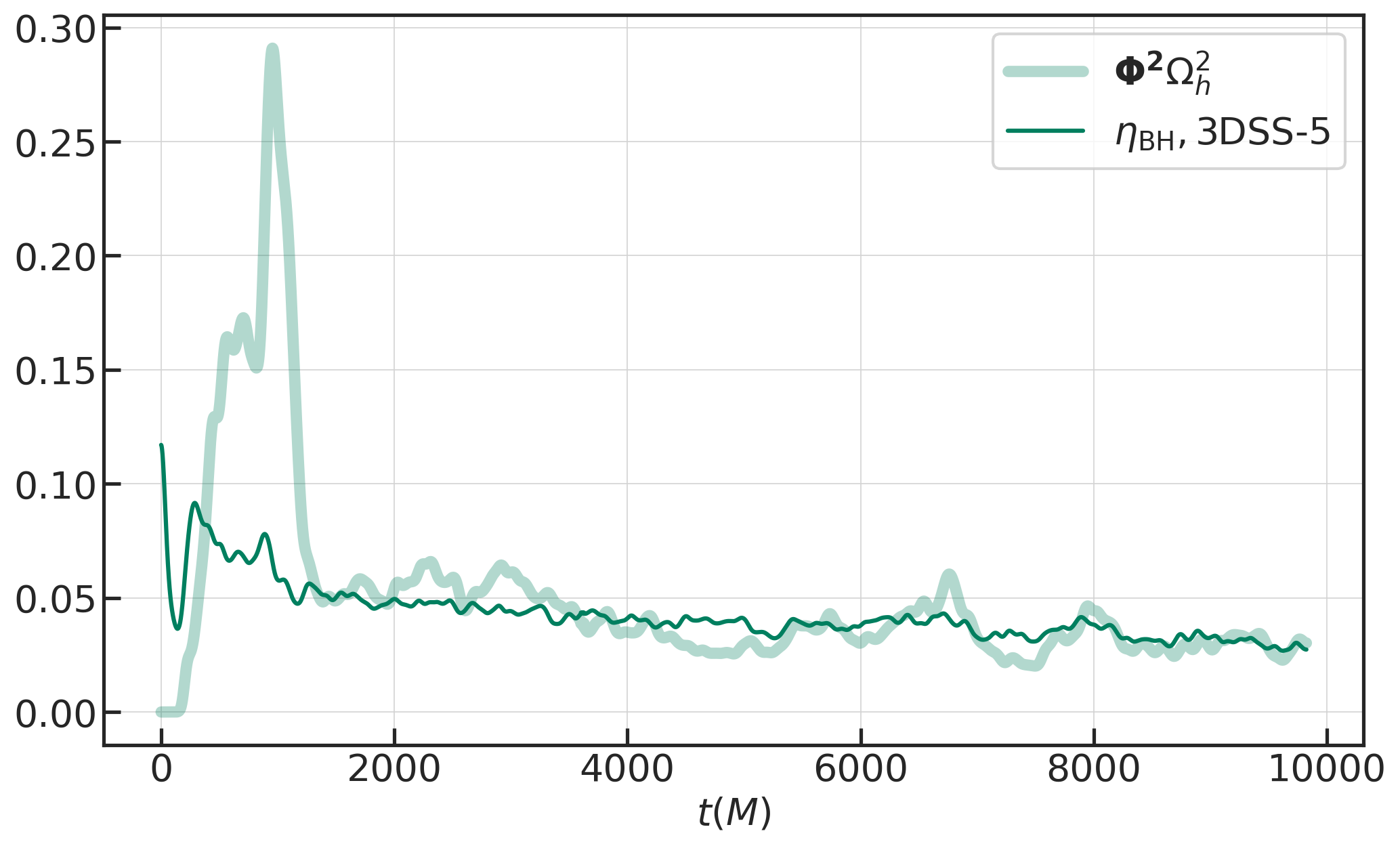}
}
\caption{Jet efficiency $\eta_\mathrm{BH}$ calculated as in \Eq{eta_BH} compared with $\boldsymbol{\Phi}^2\Omega_\mathrm{BH}^2$. For visual comparison with $\eta_\mathrm{BH}$, $\boldsymbol{\Phi}^2\Omega_\mathrm{BH}^2$ is rescaled by factors of $1/6$ and $2$ in 3DSS-9375 and 3DSS-5 respectively.}
\label{fig:eta_bz_3DSS}
\end{figure}

According to the BZ mechanism, the efficiency of the jet $\eta_{BZ}\propto\boldsymbol{\Phi}^2\Omega_\mathrm{BH}^2$ where $\Phi$ is the MAD-parameter and $\Omega_h = a/r_\mathrm{BH}$ is the angular velocity of the horizon. As we see from the previous studies \citep{Tchekhovskoy_2010,narayan_jets_2022}, this proportionality is well satisfied in simulations that reach the MAD state, leading to strong jets. Although our simulations do not produce sustained relativistic jets and remain in the sub-SANE regime throughout, it is still useful to examine to what extent the BZ scaling holds. As shown in Fig.~\ref{fig:eta_bz_3DSS}, the proportionality between $\eta_\mathrm{BH}$ and $\boldsymbol{\Phi}^2\Omega_\mathrm{BH}^2$ is very approximately maintained in both of our 3D sub-SANE runs, with the agreement being marginally better in the higher-spin case. The proportionality constant `$\kappa$' ($\eta_{BZ}=\kappa\boldsymbol{\Phi}^2\Omega_\mathrm{BH}^2$) is related to the magnetic field geometry\citep{McKinney_2005} in the MAD simulations and it is observed to be consistently the same value across BH spins\citep{narayan_jets_2022}. Irrespective of the same initial magnetic field configuration for the two sub-SANE runs, we couldn't recover that interpretation here since `$\kappa$' is significantly different for these two runs. 

\FloatBarrier

\section{Induction Equation}
\label{sec:ind_eqn_match}
The induction equation in the $3+1$ covariant formalism is given by
\begin{equation*}
\partial_t \mathbf{B}
= \nabla \times \big[(\alpha \mathbf{v} - \boldsymbol{\beta}) \times \mathbf{B}\big]
\end{equation*}
where $\alpha$ is the lapse function, $\boldsymbol{\beta}$ is the shift vector, and $\mathbf{v}$ is the fluid 3-velocity.
In section~\ref{subsec:dynamos}, we presented an analysis comparing the individual, averaged contributions of the various terms on the right-hand side of \Eq{eq:ind}. To further validate the robustness of those calculations, it is important to demonstrate that the full form of the induction equation is satisfied by the simulation data. In other words, the left-hand side should match the right-hand side when evaluated directly from the simulation output data.
\begin{figure}[!tbp]
\centering
\includegraphics[width=1\linewidth]{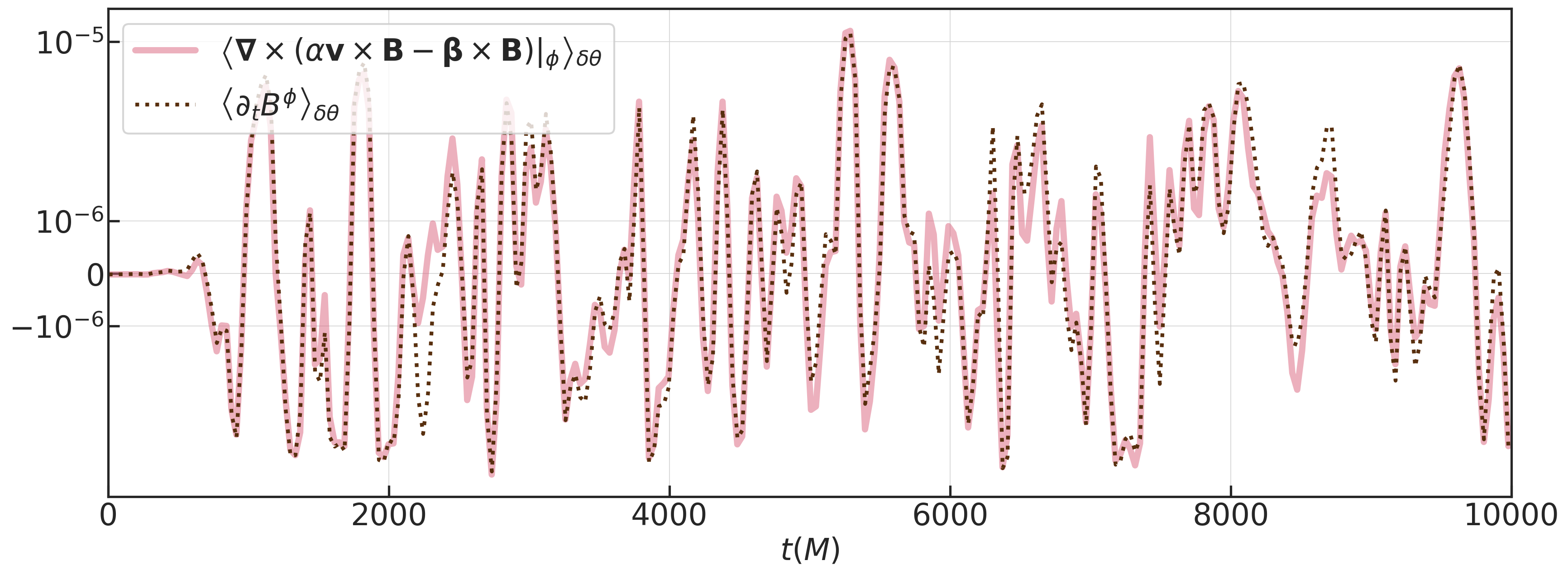}
\caption{Comparison of the left-hand side and right-hand side of the induction equation in the $3+1$ framework for the 3DSS-9375 simulation, evaluated at $r = r_{\max}$, $\theta = \pi/2 \pm \delta\theta$, and $\phi = 0$, as defined in \Eq{ind_eqn_Bphi}.
}
\label{fig:ind_eqn_match}
\end{figure}
\FloatBarrier

In \Fig{fig:ind_eqn_match}, we examine the $\phi$-component of \Eq{eq:ind} for the 3DSS-9375 run, evaluated at $r = r_{\max}$, averaged over a small opening angle $\delta \theta$ around the mid-plane, $\theta = \pi/2$, and $\phi = 0$. The close agreement between the left-hand side and right-hand side demonstrates that the induction equation is indeed well satisfied. This consistency gives us confidence that the diagnostic quantities calculated the in section~\ref{subsec:dynamos} can be trusted, even though the quantities themselves are very small in the steady-state of the disk (fluctuating around zero). Similar consistency checks were done for all the runs reported.

\begin{figure}[!tbp]
    \centering
    \includegraphics[width=\linewidth]{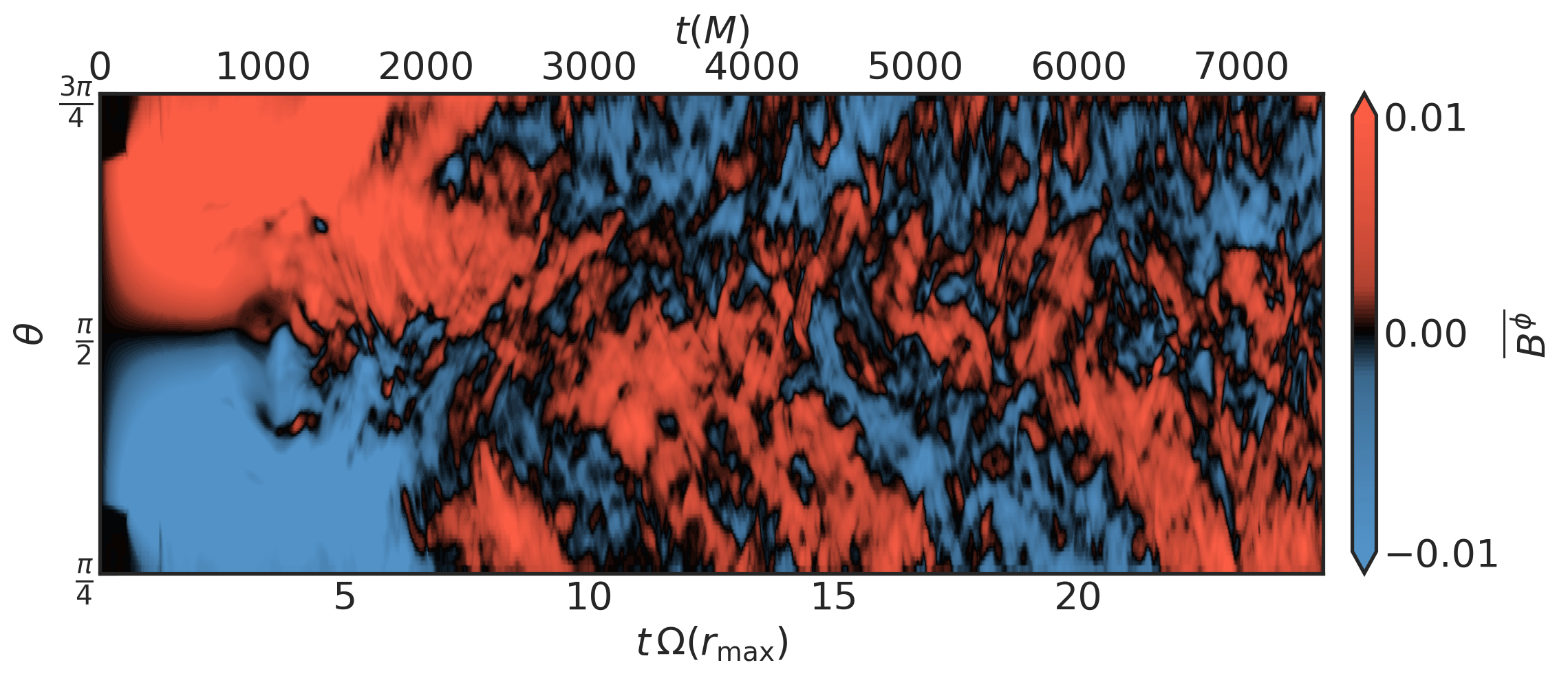}
    \caption{Butterfly diagram for 3DS-9375. The system retains a memory of its initial magnetic configuration for an extended period. As turbulence strengthens, a dynamo wave does emerge—though significantly less regular than in the 3DSS-9375 case in \Fig{fig:butterfly}. The jets show no response to these cycles as the horizon-threading fields remain dominated by the flux advected from the initial disk configuration and are only weakly influenced by these dynamo generated fields.}
    \label{fig:butterfly_3Dsane}
\end{figure}

\begin{figure}[!tbp]
    \centering
    \gridline{
\includegraphics[width=0.5\linewidth]{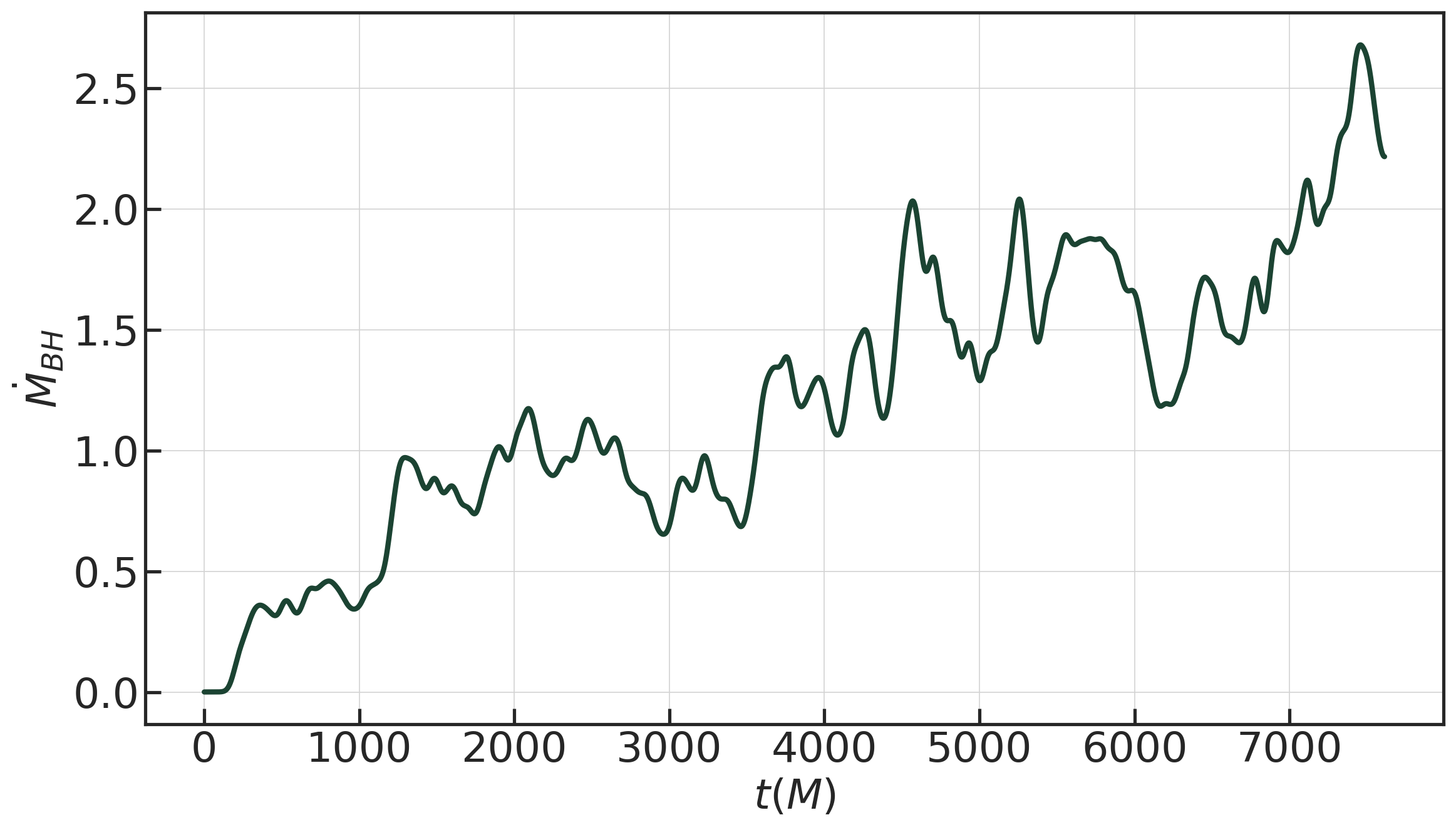}
\includegraphics[width=0.5\linewidth]{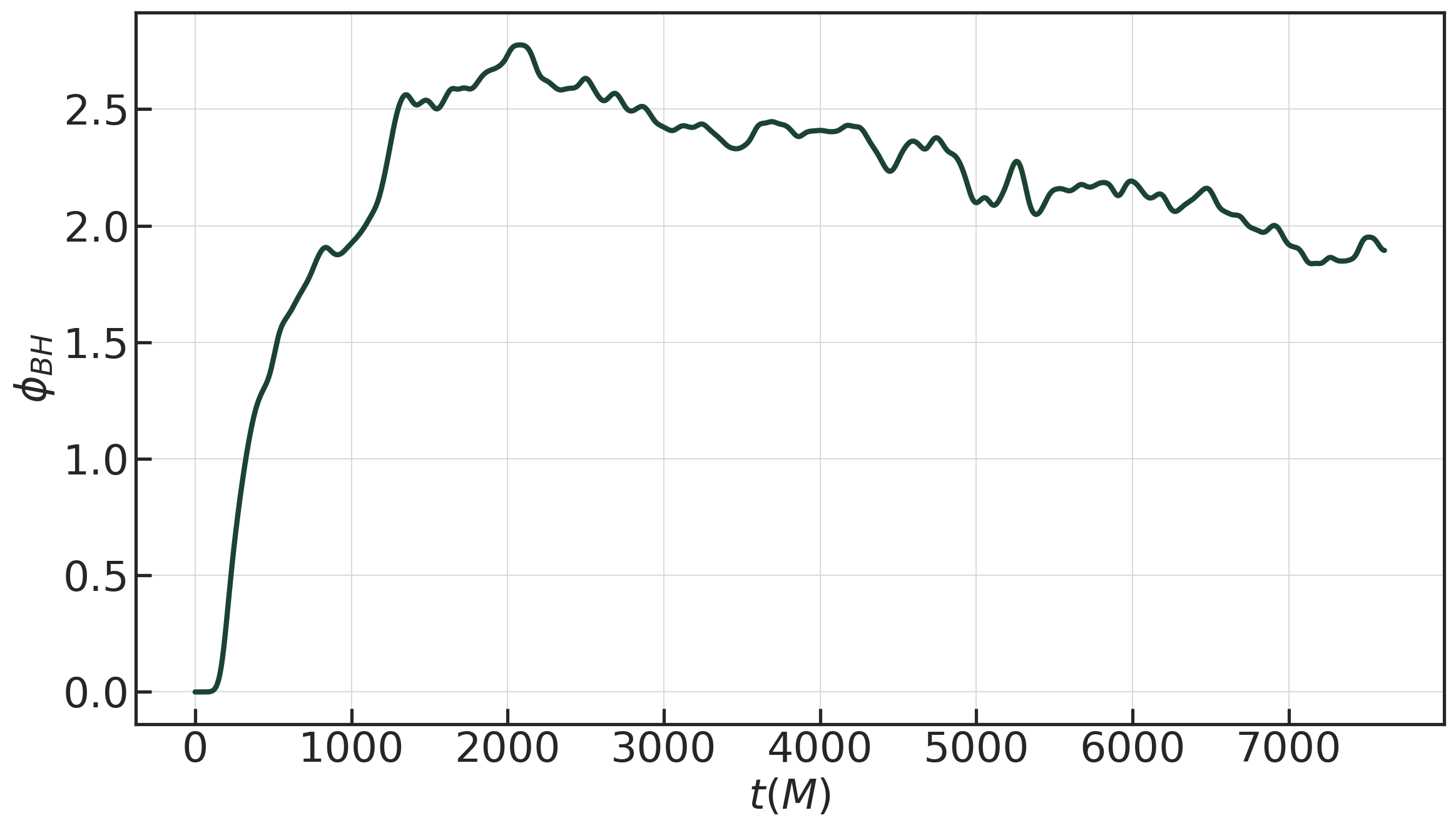}
\label{3dsane-accr}
}
\caption{The left panel shows the accretion rate, $\dot{M}\mathrm{BH}$, for the 3DS-9375 run. The accretion rate eventually rises, unlike in the 2DS-9375 case, and becomes similar to that of 3DSS-9375. The right panel shows the magnetic flux at the horizon, $\Phi_\mathrm{BH}$, for 3DS-9375. The initial value of $\Phi_\mathrm{BH}$ is $\sim 2.5$, comparable to that of 2DS-9375. Although it decreases slightly over time, it still remains marginally larger than the flux in 3DSS-9375.} 
\label{fig:mdot_phi_3Dsane}
\end{figure} 
\section{3D SANE simulation}
\label{3dsaneplots}
We perform a 3D SANE simulation using the same numerical resolution and physical parameters as the 3DSS-9375 run (evolved upto $t=7500\ M$). The initial magnetic field configuration follows the single loop configuration shown in \Eq{saneic} and is illustrated in \Fig{fig:sane_init_condition}. Although certain aspects of the evolution resemble the 2DS-9375 case, the loss of axisymmetry in full 3D introduces notable differences in the disk and jet dynamics. Below, we summarise the key aspects.

We first study the butterfly diagram showing $\overline{B^{\phi}}$ as a function of $(\theta,\ t)$ shown in \Fig{fig:butterfly_3Dsane}. We see that the pattern in the butterfly diagram is different from the 2DS-9375 case post $t=2000M$, with emergence of the large-scale dynamo wave. However, these large-scale field reversals begin noticeably later and exhibit a less organised pattern when compared to 3DSS-9375, perhaps due to the lingering effect of the initial field.

Next, we show in \Fig{fig:mdot_phi_3Dsane} the accretion rate evolution, which is initially low when compared to that of the 3DSS-9375 but gradually increases as turbulence develops in the disk, as expected. The magnetic flux at the horizon $\Phi_{\mathrm{BH}}$, is slightly larger than in 3DSS-9375 at early times. Even though we observe a quasi-cyclic butterfly pattern in 3DS-9375, there is no such pattern in the $\Phi_\mathrm{BH}$ as seen in 3DSS-9375 case. The increasing $\dot{M}_\mathrm{BH}$ and steadily declining $\Phi_\mathrm{BH}$ brings a gradual drop in the MAD parameter $\boldsymbol{\Phi}$ from $10$ to $\sim 5$ in this setup, unlike the 2DS-9375. This affects the jets, bringing down the efficiency of the outflows $\eta_\mathrm{BH}$ comparable to that of the 3DSS-9375. 
\begin{figure}[!htbp]
    \centering
    \gridline{
        \includegraphics[width=0.5\linewidth]{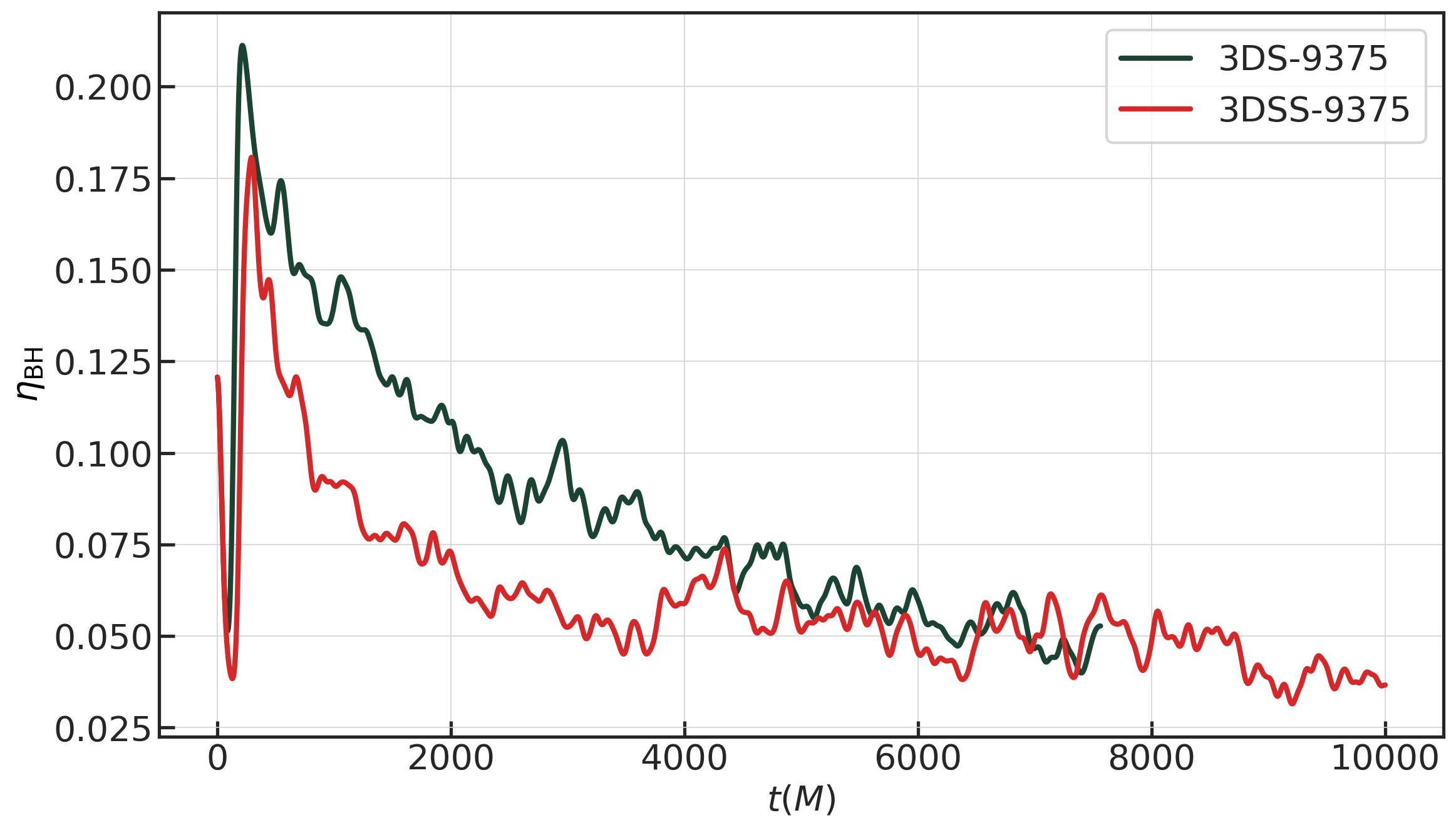}
    \includegraphics[width=0.5\linewidth]{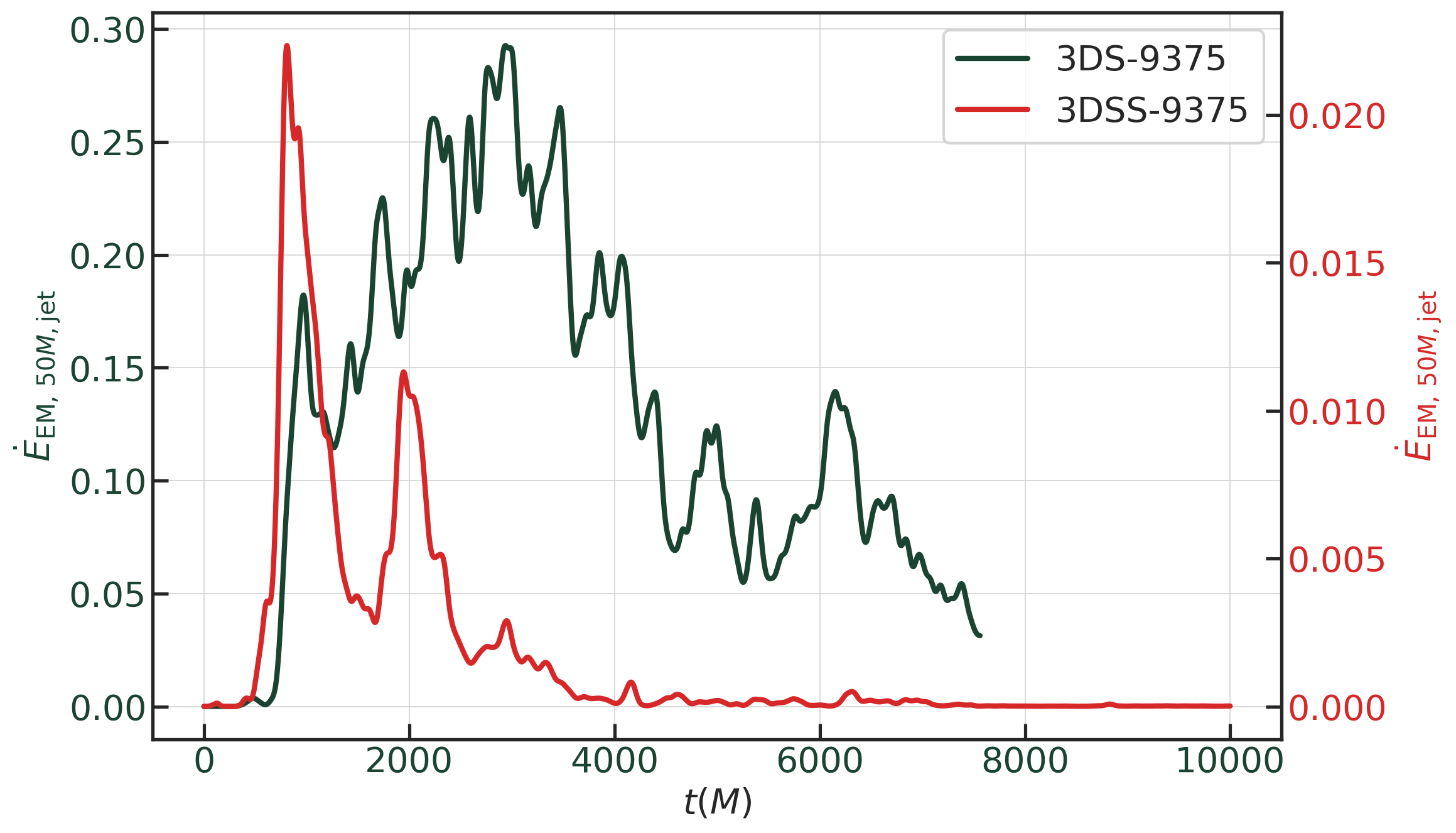}
    }
\caption{The left panel shows the outflow efficiency, $\eta_\mathrm{BH}$ (defined in \Eq{eta_BH}), for the 3DS-9375 and 3DSS-9375 runs. Although $\eta_\mathrm{BH}$ is initially slightly higher in 3DS-9375, it eventually becomes comparable to that of 3DSS-9375. The right panel shows the Poynting flux, $\dot{E}_\mathrm{EM}$, measured at $r = 50M$ in the jet and normalized by the total energy flux $\dot{E}$ at the horizon for both runs. While the efficiencies converge, the normalized Poynting flux remains non-negligible in 3DS-9375, indicating the presence of a weak jet. In contrast, it drops to nearly zero in 3DSS-9375, signaling a complete shutdown of the jet.}
\label{fig:poynting_3Dsane}
\end{figure}

\begin{figure}[!htbp]
    \centering
    \gridline{
    \includegraphics[width=0.5\linewidth]{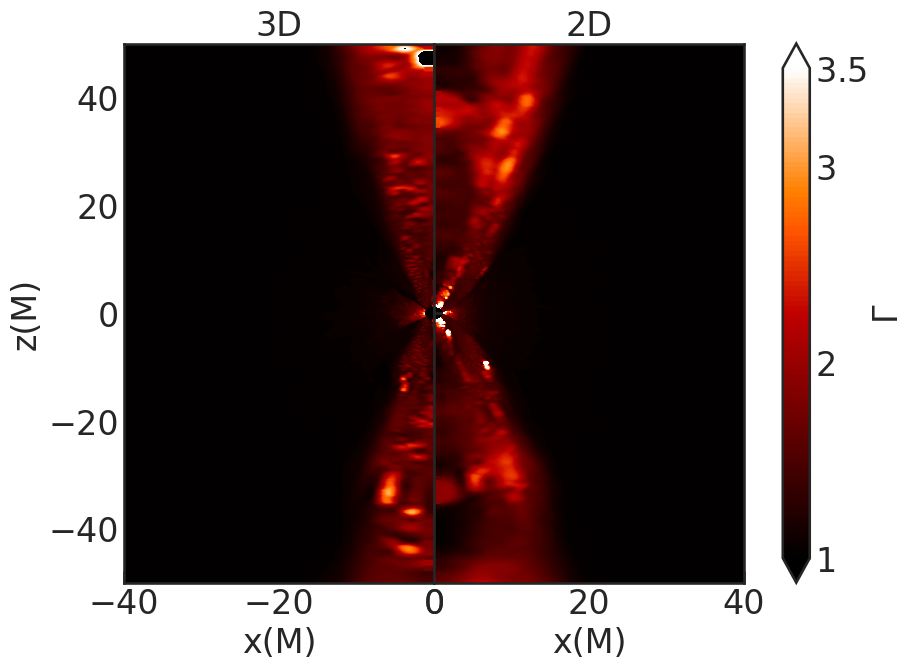}
    \includegraphics[width=0.5\linewidth]{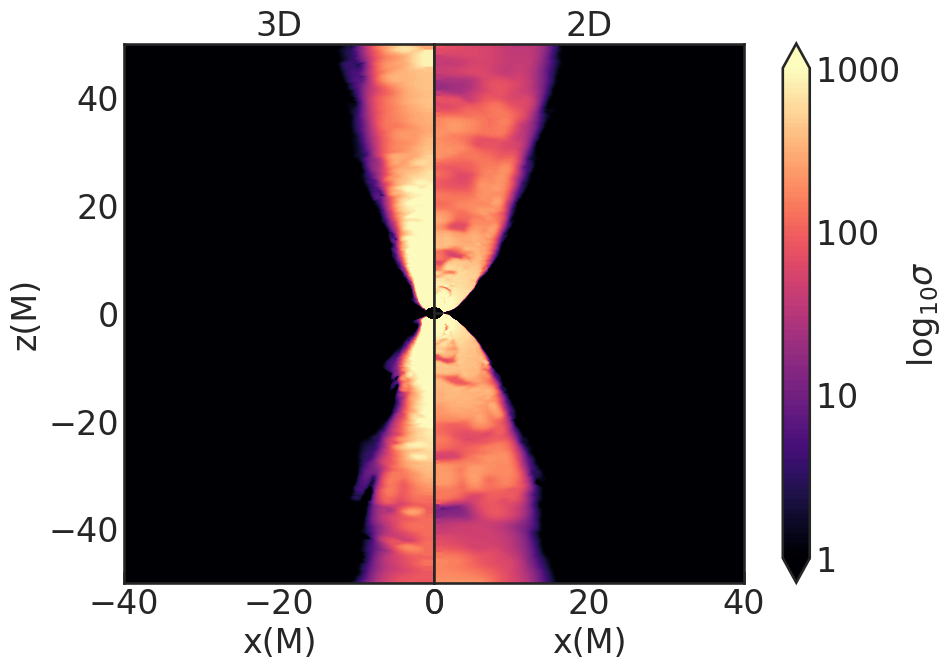}
   } 
    \caption{The left panel shows Lorentz factor $\Gamma$ in the $x-z$ plane for the 3DS-9375 and 2DS-9375 runs at $t=6000M$, while the right panel shows magnetisation $\rm{log}_\mathrm{10}\sigma$ for the same. We see that the overall properties are almost similar in the plots irrespective of the difference in $\dot{M}_\mathrm{BH}$ between the two runs. However, the jet relatively looks narrower in the 3DS-9375 case.}
    \label{fig:jet_diag_3Dsane}
\end{figure}
Despite this reduction, the jet does not shut down completely in 3DS-9375. The Poynting flux in the jet at $r=50\ M$ remains non-zero (see \Fig{fig:poynting_3Dsane}). Both magnetisation $\sigma$ and Lorentz factor $\Gamma$, stay at modest but non-negligible values, comparable to that of 2DS-9375 as seen in \Fig{fig:jet_diag_3Dsane}. This indicates the presence of a weak but persistent jet throughout the evolution, in contrast to the shutting down of jets in the sub-SANE cases. This can be attributed to the relatively coherent field at the horizon which is reflected in coherence parameter, $\mathcal{C}_\mathrm{BH}$ \Eq{ratio_BH}. $\mathcal{C}_\mathrm{BH}$ remains close to $\sim 0.9$ during most of the evolution(See \Fig{fig:ratio-signed}). Such a high value implies that the magnetic field retains a relatively ordered, coherent structure. We understand this as the continuing effect of the initial strong field development at the horizon which remains largely unaffected by the developing turbulence and accretion of more random fields at later times.


\bibliographystyle{aasjournal}
\bibliography{jetdynrefs}

\end{document}